\documentclass{article}
\usepackage{amsmath,amssymb}
\usepackage[utf8]{inputenc}
\usepackage[margin=1in,headheight=13.6pt]{geometry}
\usepackage[framemethod=tikz]{mdframed}
\usepackage{footnote}
\usepackage{wasysym} 
\usepackage{enumitem}

\usepackage{url}
\usepackage{pythontex}

\usepackage{pbox} 

\usepackage{placeins}

\usepackage{cellspace}
\usepackage{tikz}
\usetikzlibrary{arrows, arrows.meta}

\tikzset{
BasicNode/.style={circle, draw= black!50, fill=colora!40, thin, minimum size
  = 5mm, inner sep = 0mm}
  }
\tikzset{
BasicNode/.style={circle, draw= black!50, fill=colora!40, thin, minimum size
  = 5mm, inner sep = 0mm},
SusCirc/.style={BasicNode,fill=colorS!40},
InfCirc/.style={BasicNode,fill=colorI!40},
RecCirc/.style={BasicNode,fill=colorR!40},
DefCirc/.style={BasicNode}
}

\definecolor{colorS}{RGB}{0,154,128}
\definecolor{colorI}{RGB}{255,32,0}
\definecolor{colorR}{RGB}{205,10,179}
\definecolor{colorE}{RGB}{240,228,66} 

\makesavenoteenv{tabular}
\makesavenoteenv{table} 

\newtheorem{exercisex}{Exercise}
\newtheorem{example}{Example}
\newtheorem{property}{Property}
\newtheorem{thm}{Theorem}
\newtheorem{cor}{Corollary}
\newtheorem{solutionx}{Solution} 

\let\oldexercisex\exercisex
\let\oldendexercisex\endexercisex
\def\exercisex{\begingroup%
\ifsolns
\oldexercisex\begin{mdframed}[
  linecolor=red,
  topline=false,
  bottomline=false,
  rightline=false,
  skipabove=\topsep,
  skipbelow=\topsep,
  linewidth=1pt,
  innertopmargin=0pt,
  innerbottommargin=0pt,
  leftmargin = -\leftmargin,
  innerleftmargin=0.4em,
  rightmargin = 0,
  innerrightmargin=0, 
]
\else
\oldexercisex\begin{mdframed}[
  topline=false,
  bottomline=false,
  rightline=false,
  skipabove=\topsep,
  skipbelow=\topsep,
  linewidth=1pt,
  innertopmargin=0pt,
  innerbottommargin=0pt,
  leftmargin = -\leftmargin,
  innerleftmargin=0.4em,
  rightmargin = 0,
  innerrightmargin=0, 
]
\fi
\renewcommand{\theenumi}{{\bf\alph{enumi}}}
}
\def\endexercisex{\end{mdframed}\oldendexercisex%
\endgroup}

\let\oldsolutionx\solutionx
\let\oldendsolutionx\endsolutionx
\def\solutionx{\begingroup%
\oldsolutionx\begin{mdframed}[
  linecolor=cyan,
  topline=false,
  bottomline=false,
  rightline=false,
  skipabove=\topsep,
  skipbelow=\topsep,
  linewidth=1pt,
  innertopmargin=0pt,
  innerbottommargin=0pt,
  leftmargin = -\leftmargin,
  innerleftmargin=0.4em,
  rightmargin = 0,
  innerrightmargin=0, 
]\renewcommand{\theenumi}{{\bf\alph{enumi}}}
}
\def\endsolutionx{\end{mdframed}\oldendsolutionx%
\endgroup}

\let\oldexample\example
\let\oldendexample\endexample
\def\example{\begingroup \oldexample\begin{mdframed}[
  linecolor=cyan,
  topline=false,
  bottomline=false,
  rightline=false,
  skipabove=\topsep,
  skipbelow=\topsep,
  linewidth=1pt,
  innertopmargin=0pt,
  innerbottommargin=0pt,
  leftmargin = -\leftmargin,
  innerleftmargin=0.4em,
  rightmargin = 0,
  innerrightmargin=0, 
]
}
\def\endexample{\end{mdframed}\oldendexample\endgroup}

\let\oldthm\thm
\let\oldendthm\endthm
\def\thm{\begingroup \oldthm\begin{mdframed}[
  linecolor=cyan,
  roundcorner=5pt,
  skipabove=\topsep,
  skipbelow=\topsep,
  linewidth=1pt,
  innertopmargin=0.4em,
  innerbottommargin=0.4em,
  leftmargin = -\leftmargin,
  innerleftmargin=0.4em,
  rightmargin = 0,
  innerrightmargin=0.4em, 
]
}
\def\endthm{\end{mdframed}\oldendthm\endgroup}

\let\oldcor\cor
\let\oldendcor\endcor
\def\cor{\begingroup \oldcor\begin{mdframed}[
  linecolor=cyan,
  roundcorner=5pt,
  skipabove=\topsep,
  skipbelow=\topsep,
  linewidth=1pt,
  innertopmargin=0.4em,
  innerbottommargin=0.4em,
  leftmargin = -\leftmargin,
  innerleftmargin=0.4em,
  rightmargin = 0,
  innerrightmargin=0.4em, 
]
}
\def\endcor{\end{mdframed}\oldendcor\endgroup}

\let\oldproperty\property
\let\oldendproperty\endproperty
\def\property{\begingroup \oldproperty\begin{mdframed}[
  linecolor=cyan,
  roundcorner=5pt,
  skipabove=\topsep,
  skipbelow=\topsep,
  linewidth=1pt,
  innertopmargin=0.4em,
  innerbottommargin=0.4em,
  leftmargin = -\leftmargin,
  innerleftmargin=0.4em,
  rightmargin = 0,
  innerrightmargin=0.4em, 
]
}
\def\endproperty{\end{mdframed}\oldendproperty\endgroup}

\newcounter{myenumi}

\newenvironment{myenumerate}{%
\setcounter{myenumi}{0}
\begin{enumerate} %
\setcounter{enumi}{\value{myenumi}}%
}
{%
\setcounter{myenumi}{\value{enumi}}%
\end{enumerate}
}

\newenvironment{continuemyenumerate}{
\begin{enumerate}%
\setcounter{enumi}{\value{myenumi}}%
}
{%
\setcounter{myenumi}{\value{enumi}}%
\end{enumerate}
}

\usepackage{todonotes}

\newcommand{\Tree}{\mathcal{T}}
\newcommand{\Seq}{\mathcal{S}}

\newcommand{\littleo}{
{
  \mathchoice
    {{\scriptstyle\mathcal{O}}}
    {{\scriptstyle\mathcal{O}}}
    {{\scriptscriptstyle\mathcal{O}}}
    {\scalebox{.7}{$\scriptscriptstyle\mathcal{O}$}}
  }
}
\newcommand{\ave}[1]{\left \langle #1 \right \rangle}
\newcommand{\Ro}{\ensuremath{\mathcal{R}_0}}

\newcommand{\diff}[2]{\frac{\mathrm{d} #1}{\mathrm{d} #2}}

\newcommand{\pd}[2]{\ensuremath{\frac{\partial #1}{\partial #2}}}
\newcommand{\pds}[2]{\ensuremath{\frac{\partial^2 #1}{\partial #2^2}}}

\usepackage[misc]{ifsym}
\newcommand{\die}[1]{{\raisebox{-2pt}{\scalebox{0.9}{\Cube{#1}}}}}

\def\CC{\mbox{C\hspace{-.05em}\raisebox{.4ex}{\tiny\bf ++}}}

\numberwithin{exercisex}{section}

\numberwithin{solutionx}{section}

\numberwithin{example}{section}

\numberwithin{thm}{section}

\numberwithin{cor}{section}

\numberwithin{property}{section}

\usepackage{atbegshi}
\newcommand{\handlethispage}{}
\AtBeginShipout{\handlethispage}

\usepackage{answers}

\newif\ifsolns 
\ifsolns
\Newassociation{exercise}{exercise_aux}{solnfile}
\Newassociation{solution}{solution_aux}{solnfile}

\Opensolutionfile{solnfile}
\usepackage{etoolbox}
\makeatletter
\pretocmd\@startsection{\Writetofile{solnfile}{\string\csname\space #1\string\endcsname}}{}{} 
\pretocmd\@sect{\Writetofile{solnfile}{[#7]{#8}}}{}{}
\pretocmd\@ssect{\Writetofile{solnfile}{*{#5}}}{}{}
\AtBeginDocument{%
\Writetofile{solnfile}{\string \noindent{\Large\bf Solutions to: {\it A primer
      on the use of probability generating functions in infectious
      disease modeling }\\[12pt]  Joel C. Miller}\\[24pt]}
\let\mtrefstepcounter\refstepcounter
\let\refstepcounter\@gobble
\let\handlethispage\AtBeginShipoutDiscard
\let\solnappendix\appendix \renewcommand{\appendix}{\Writetofile{solnfile}{\string\appendix}}}
\AtEndDocument{%
\let\appendix\solnappendix
\let\refstepcounter\mtrefstepcounter
\clearpage\pagenumbering{arabic}
\let\handlethispage\relax
\Closesolutionfile{solnfile}
\Readsolutionfile{solnfile}}
\makeatother
\else
\newenvironment{exercise}{\begin{exercisex}}{\end{exercisex}}
\Newassociation{solution}{solutionx}{solnfile}
\Opensolutionfile{solnfile}
\AtEndDocument{%
\Closesolutionfile{solnfile}}
\fi

\definecolor{colora}{RGB}{0,115,179}
\definecolor{colorb}{RGB}{230,154,0} 
\definecolor{colorc}{RGB}{0,154,128} 
\definecolor{colord}{RGB}{205,10,179}
\definecolor{colore}{RGB}{255,32,0}
\definecolor{colorf}{RGB}{240,228,66}
\definecolor{colorg}{RGB}{90,179,230}
\definecolor{colorh}{RGB}{205,154,179}

\input{PythontexFile}

\author{Joel C. Miller}
\title{A primer on the use of probability generating functions in infectious disease modeling}

\graphicspath{ {./figures/} }

\begin{document}
\maketitle
\begin{abstract}
  We explore the application of probability generating functions
  (PGFs) to invasive processes, focusing on infectious disease
  introduced into large populations.  Our goal is to acquaint the
  reader with applications of PGFs, moreso than to derive new results.
  PGFs help predict a number of properties about early outbreak
  behavior while the population is still effectively infinite,
  including the probability of an epidemic, the size distribution
  after some number of generations, and the cumulative size
  distribution of non-epidemic outbreaks.  We show how PGFs can be
  used in both discrete-time and continuous-time settings, and discuss
  how to use these results to infer disease parameters from observed
  outbreaks.  In the large population limit for
  susceptible-infected-recovered (SIR) epidemics PGFs lead to
  survival-function based models that are equivalent the the usual
  mass-action SIR models but with fewer ODEs.  We use these to explore
  properties such as the final size of epidemics or even the dynamics
  once stochastic effects are negligible.  We target this primer at
  biologists and public health researchers with mathematical modeling
  experience who want to learn how to apply PGFs to invasive diseases,
  but it could also be used in an applications-based mathematics course on
  PGFs.  We include many exercises to help demonstrate concepts and to
  give practice applying the results.  We summarize our main results
  in a few tables.  Additionally we provide a small python package
  which performs many of the relevant calculations.
\end{abstract}

\ifsolns
\else
\tableofcontents
\fi

\section{Introduction}
The spread of infectious diseases remains a public health challenge.
Increased interaction between humans and wild animals leads to
increased zoonotic introductions, and modern travel networks allows
these diseases to spread quickly.
Many mathematical approaches have been developed to give us insight
into the early behavior of disease outbreaks.
An important tool for understanding the stochastic behavior of an outbreak soon after introduction is the probability generating function (PGF)~\cite{gf,allen2010introduction,yan2008distribution}. 

Specifically, PGFs frequently give insight the statistical behavior of
outbreaks before they are large enough to be affected by the
finite-size of the population.  In these cases, both
susceptible-infected-recovered (SIR) disease (for which nodes recover
with immunty) and susceptible-infected-susceptible (SIS) disease (for
which nodes recover and can be reinfected immediately) are equivalent.
In the case of SIR disease they can be used to study the dynamics of
disease once an epidemic is established in a large population.  

We can
investigate properties such as the early growth rate of the disease,
the probability the disease becomes established, or the distribution
of final sizes of outbreaks that fail to become established.  Similar
questions also arise in other settings where some introduced agent can
reproduce or die, such as invasive species in ecological
settings~\cite{lewis2016mathematics}, early within-host pathogen
dynamics~\cite{conway2011stochastic}, and the accumulation of
mutations in precancerous and cancerous
cells~\cite{durrett2015branching,antal2011exact} or in pathogen evolution~\cite{volz2017phylodynamic}.  These are all
examples of \emph{branching processes}, and PGFs are a central tool
for the analysis of branching
processes~\cite{bartlett1949some,kendall1949stochastic,kimmel2002branching}.
Except for Section~\ref{sec:large-time} where we develop deterministic
equations for later-time SIR epidemics, based on~\cite{volz:cts_time,
  miller:volz, miller:ebcm_overview}, the approaches we describe here
have direct application in these other branching processes as well.

Before proceeding, we define what a PGF is.  Let $r_i$ denote the probability of drawing the value $i$ from a given distribution of non-negative integers.  Then
\[
f(x) = \sum_i r_i x^i 
\]
is the PGF of this distribution.  We should address a potential confusion
caused by the name.  A ``generating function'' is a function which is
defined from (or ``generated by'') a sequence of numbers $a_i$ and takes the form $\sum_i
a_i x^i$.  So a ``probability generating function'' is a generating
function defined from a probability distribution on integers.  It is
\emph{not} a function that generates probabilities when values are
plugged in for $x$.  There are other generating functions, including
the ``moment generating function'', defined to be $\sum_m \ave{i^m}
x^m$ where $\ave{i^m} = \sum_ir_i i^m$ (the moment and probability
generating functions turn out to be closely related).

PGFs have a number of useful properties which we derive in Appendix~\ref{app:properties}.  We have structured this paper so that a reader can skip ahead now and read Appendix~\ref{app:properties} in its entirety to get a self-contained introduction to PGFs, or wait until a particular property is referenced in the main text and then read that part of the appendix.  

As we demonstrate in Table~\ref{tab:example_PGFs}, for many important
distributions the PGF takes a simple form.  We derive this for the
Poisson distribution.
\begin{example}
\label{example:poisson2PGF}
Consider the Poisson distribution with mean $\lambda$
\[
r_i = \frac{e^{-\lambda}\lambda^i}{i!} \, .
\]
For this we find
\begin{align*}
f(x) &= \sum_i \frac{e^{-\lambda} \lambda^i}{i!}x^i= e^{-\lambda} \sum_i \frac{(\lambda x)^i}{i!}= e^{-\lambda}e^{\lambda x}\\
&= e^{\lambda(x-1)} \, .
\end{align*}
\end{example}

\begin{table}
\begin{center}
\begin{tabular}{|Sc|Sc|}
\hline
\textbf{Distribution} & \textbf{PGF $f(x)=\sum_i r_ix^i$} \\
\hline\hline
Poisson, mean $\lambda$: $r_i = \frac{e^{-\lambda}\lambda^i}{i!}$ & $e^{\lambda(x-1)}$\\
\hline
Uniform: $r_\lambda = 1$ &  $x^\lambda$\\
\hline
Binomial: $n$ trials, with success probability $p$: $r_i = \binom{n}{i}p^iq^{n-i}$ for $q=1-p$ & $[q + px]^n$\\
\hline
Geometric\footnote{Another definition of the geometric
  distribution with different indexing, $r_i = q^{i-1}p$ for $i=1,2, \ldots$, gives a different PGF.}: $r_i =
  q^ip$ for $q=1-p$ and $i=0, 1, \ldots$
& $p/(1-qx)$\\
\hline
Negative binomial\footnote{Typically the negative binomial is
  expressed in terms of a parameter $r$ which is the number of
  failures at which the experiment stops, assuming each with success
  probability $p$.  For us $r_i$ plays an important role, so to help
  distinguish these, we use $\hat{r}$ rather than $r$.  Then $r_i$ is the probability of $i$ successes.}: $r_i = \binom{i+\hat{r}-1}{i} q^{\hat{r}}p^i$ for $q=1-p$ & $\left(\frac{q}{1-px}\right)^{\hat{r}}$\\
\hline
\end{tabular}
\end{center}
\caption{A few common probability distributions and their PGFs.}
\label{tab:example_PGFs}
\end{table}

In this primer, we explore the application of PGFs to the study of disease spread.  We will use PGFs to answer questions about the early-time behavior of an outbreak (neglecting depletion of susceptibles):
\begin{itemize}
\item What is the probability an outbreak goes extinct within $g$ generations (or by time $t$) in an arbitrarily large population?
\item What is the probability an index case causes an epidemic?
\item What is the final size distribution of small outbreaks?
\item What is the size distribution of outbreaks at generation $g$ (or time $t$)?
\item How fast is the initial growth for those outbreaks that do not go extinct?
\end{itemize}
Although we present these early-time results in the context of SIR outbreaks they also apply to SIS outbreaks and many other invasive processes.  

We can also use PGFs for some questions about the full behavior
accounting for depletion of susceptibles. Specifically:
\begin{itemize}
\item In a continuous-time Markovian SIR or SIS outbreak spreading in a finite population, what is the distribution of possible system states at time $t$?
\item In the large-population limit of an SIR epidemic, what fraction of the population is eventually infected?
\item In the large-population limit of an SIR epidemic, what fraction of the population is infected or recovered at time $t$?
\end{itemize}

\begin{table}
\begin{center}
\begin{tabular}{|Sc|Sc|}
\hline
\textbf{Function/variable name} & \textbf{Interpretation}\\
\hline
\hline
\parbox{0.3\textwidth}{%
$f(x)=\sum_i p_i x^i$\\[3pt]
$g(x)=\sum_i q_i x^i$} 
& \parbox{0.65\textwidth}{Arbitrary PGFs.}\\
\hline
\parbox{0.3\textwidth}{%
$\mu(y)= \sum_i p_i y^i$\\[3pt]
$\hat{\mu}(y) = (\beta y^2 +\gamma)/(\beta+\gamma)$\\[3pt]
$\hat{\mu}(y,z)=(\beta y^2 +\gamma z)/(\beta+\gamma)$} &
\parbox{0.65\textwidth}{\textbf{Without hats:} The PGF for the
offspring distribution in discrete time.\\
\textbf{With hats:} The PGF for the outcome of an unknown event in a
  continuous-time Markovian outbreak: $y$ accounts for  active
  infections and $z$ accounts for  completed
  infections.}\\
\hline
\parbox{0.3\textwidth}{$\alpha$,  $\alpha_g$, $\alpha(t)$} & 
\parbox{0.65\textwidth}{Probability of either eventual extinction, extinction by generation
$g$, or by time $t$ in an infinite population.}\\
\hline
\parbox{0.3\textwidth}{%
$\Phi_g(y) = \sum_i \phi_i(g) y^i$\\[3pt]
$\Phi(y,t) = \sum_i \phi_i(t) y^i$} 
& \parbox{0.65\textwidth}{PGF for the number of active infections in generation $g$ or at time $t$ in an infinite population.}\\
\hline
\parbox{0.3\textwidth}{%
$\Omega_\infty(z) = \sum_{r<\infty} \omega_r z^r+ \omega_\infty z^\infty$\\[3pt]
$\Omega_g(z) = \sum_{r} \omega_r(g) z^r$\\[3pt]
$\Omega(z,t)= \sum_{r} \omega_r(t) z^r$}
& 
\parbox{0.65\textwidth}{
The PGF for the distribution of completed infections at  the end of a small outbreak, in generation $g$, or at time $t$ in an infinite population.  If $\Ro>1$, then one of the terms in the expansion of $\Omega_\infty(z)$ is $\omega_\infty z^\infty$ where $\omega_\infty$ is the probability of an epidemic.}\\
\hline 
\parbox{0.3\textwidth}{%
$\Pi_g(y,z)=\sum_{i,r} \pi_{i,r}(g)y^iz^r$\\[3pt]
$\Pi(y,z,t)=\sum_{i,r} \pi_{i,r}(t)y^iz^r$} 
& 
\parbox{0.65\textwidth}{The PGF for the joint distribution of current infections and completed infections either at generation $g$ or time $t$ in an infinite population.}\\
\hline
\parbox{0.3\textwidth}{$\Xi(x,y,t)=\sum_{s,i} \xi_{s,i}(t)x^sy^i$} & 
\parbox{0.65\textwidth}{%
The PGF for the joint distribution of susceptibles and current infections at time $t$ in a finite population of size $N$ (used for continuous time only). In the SIR case we can infer the number recovered from this and  the total population size.}\\
\hline
\parbox{0.3\textwidth}{$\chi(x)=\sum_ip_i x^i$} & 
\parbox{0.65\textwidth}{PGF for the ``ancestor distribution'', analogous to the offspring distribution.}\\
\hline
\parbox{0.3\textwidth}{$\psi(x) = \sum_\kappa P(\kappa) x^\kappa$} & 
\parbox{0.65\textwidth}{PGF for the distribution of susceptibility for the continuous time model where rate of receiving transmission is proportional to $\kappa$.}\\
\hline
\parbox{0.3\textwidth}{$\beta$, $\gamma$} & \parbox{0.65\textwidth}{The individual transmission and recovery rates for the Markovian continuous time model.}\\
\hline
\end{tabular}
\end{center}
\caption{Common function and variable names.  When we use a PGF for the number of susceptible individuals, active infections, and/or completed infections $x$ and $s$ correspond to susceptible individuals, $y$ and $i$ to active infections, and $z$ and $r$ to completed infections.}
\label{tab:functions}
\end{table}

We will consider both discrete-time and Markovian continuous-time
models of disease.   In the discrete-time case each infected individual
transmits to some number of ``offspring'' before recovering.  In the
continuous-time case each infected individual trasmits with a rate
$\beta$ and recovers with a rate $\gamma$.

In Section~\ref{sec:discrete} we begin our study
investigating properties of epidemic emergence in a discrete-time,
generation-based framework, focusing on the probability of extinction
and the sizes of outbreaks assuming that the disease is invading a
sufficiently large population with enough mixing that we can treat the
infections caused by any one infected individual as independent of the
others.  We also briefly discuss how we might use our observations to
infer disease parameters from observed small outbreaks.  In Section~\ref{sec:cts}, we repeat this analysis for a
continuous-time case treating transmission and recovery as Poisson
processes, and then adapt the analysis to a population with finite
size $N$.  Next in Section~\ref{sec:large-time} we use PGFs to derive
simple models of the large-time dynamics of SIR disease spread, once
the infection has reached enough individuals that we can treat the
dynamics as deterministic.  Finally, in Section~\ref{sec:multitype} we
explore multitype populations in which there are different types of infected individuals, which may produce different distributions of infections.
 We provide three appendices.  In Appendix~\ref{app:properties}, we
 derive the relevant properties of PGFs, in
 Appendix~\ref{app:power_magic} we provide elementary (\emph{i.e.},
 not requiring Calculus) derivations of two important theorems, and in
 Appendix~\ref{app:software} we provide details of a Python package
 \texttt{Invasion\_PGF} available at
 \url{https://github.com/joelmiller/Invasion_PGF} that implements most
 of the results described in this primer.  Python code that uses this
 package to implement the figures of Section~\ref{sec:discrete} is
 provided in the supplement.

Our primary goal here is to provide modelers with a useful PGF-based
toolkit, with derivations that focus on developing intuition and
insight into the application rather than on providing fully rigorous
proofs.  Throughout, there are exercises designed to increase
understanding and help prepare the reader for applications.  This
primer (and Appendix~\ref{app:properties} in particular) could serve
as a resource for a mathematics course on PGFs.  For readers wanting to take a deep dive into the underlying theory, there are resources that provide a more technical look into PGFs in general~\cite{gf} or specifically using PGFs for infectious disease~\cite{yan2008distribution}.

\ifsolns
\else
\subsection{Summary}
\fi
Before presenting the analysis, we provide a collection of tables that summarize our main results. Table~\ref{tab:functions} summarizes our notation.
Tables~\ref{tab:discrete} and~\ref{tab:continuous} summarize our main results for the discrete-time and continuous-time models.  Table~\ref{table:infinite_proportion} shows applications of PGFs to the continuous-time dynamics of SIR epidemics once the disease has infected a non-negligible proportion of a large population, effectively showing how PGFs can be used to replace most common mass-action models.  Finally, Table~\ref{table:final_probs} provides the probability of each finite final outbreak size assuming a sufficiently large population that susceptible depletion never plays a role.  
\begin{table}[h]
\begin{center}
\begin{tabular}{|Sc|Sc|Sc|}
\hline
\textbf{Question} & \textbf{Section} &\textbf{Solution}\\
\hline\hline
\parbox{0.4\columnwidth}{Basic Reproductive Number $\Ro$ [the
  average number of transmissions an infected individual causes early
  in an outbreak].} & Intro to~\ref{sec:discrete} & \parbox{0.4\columnwidth}{$\Ro=\mu'(1)$.}\\
\hline
\parbox{0.4\columnwidth}{Probability of extinction, $\alpha$, given a single introduced infection.}
                  &\ref{sec:extinction} & \parbox{0.4\columnwidth}{$\alpha = \lim_{g\to\infty}\mu^{[g]}(0)$
                    or, equivalently, the smallest $x$ in $[0,1]$ for which $x=\mu(x)$.}\\
\hline
\parbox{0.4\columnwidth}{Probability of extinction within $g$
  generations} &\ref{sec:extinction_generation} & \parbox{0.4\columnwidth}{$\alpha_g=\mu^{[g]}(0)$.}\\
\hline
\parbox{0.4\columnwidth}{PGF of the distribution of the number of
  infected individuals in the $g$-th generation.} 
&
\ref{sec:current_size} &
\parbox{0.4\columnwidth}{$\Phi_g(y)$ where $\Phi_g$ solves $\Phi_g(y) = \mu^{[g]}(y)$.}\\
\hline 
 \parbox{0.4\columnwidth}{Average number of active infections in
  generation $g$ and average number if the outbreak has not yet gone extinct.} &
                                                          \ref{sec:current_size}                                   
&
\parbox{0.4\columnwidth}{$\Ro^g$, and $\frac{\Ro^g}{1-\alpha_g}$.}\\
\hline
\parbox{0.4\columnwidth}{PGF of the number of completed cases at
  generation $g$ in an infinite population.} & \ref{sec:cum_size}
                                     &\parbox{0.4\columnwidth}{$\Omega_g(z)$ where $\Omega_g$ solves $\Omega_g(z) = z\mu(\Omega_{g-1}(z))$ with $\Omega_0(z)=1$.}\\
\hline
\parbox{0.4\columnwidth}{PGF of the joint distribution of the number of current and completed cases at generation $g$ in an infinite
  population.} & \ref{sec:joint_size}
                                     &\parbox{0.4\columnwidth}{$\Pi_g(y,z)$ where $\Pi_g$ solves $\Pi_g(y,z) = z\mu(\Pi_{g-1}(y,z))$ with $\Pi_0(y,z) = y$.}\\
\hline
\parbox{0.4\columnwidth}{PGF of the final size distribution.} 
&
  \ref{sec:discrete_small_final_size}
&
\parbox{0.4\columnwidth}{$\Omega_\infty(z)$ where $\Omega_\infty$ solves $\Omega_\infty(z)=\lim_{g\to\infty}\Omega_g(z)$.  It also solves $\Omega_\infty(z) = z\mu(\Omega_\infty(z))$.  This has a discontinuity at $|z|=1$ if epidemics are possible.}\\
\hline
\parbox{0.4\columnwidth}{Probability an outbreak infects exactly $j$ individuals} & \ref{sec:discrete_small_final_size} & \parbox{0.4\columnwidth}{$\frac{p_{j-1}^{(j)}}{j}$ where $p_i^{(j)}$ is the coefficient of $y^i$ in the expansion of $[\mu(y)]^j$.}\\
\hline
\parbox{0.4\columnwidth}{Probability a disease has a particular set of
  parameters $\Theta$ given a set of observed independent outbreak
  sizes $X=(j_1,\ldots, j_\ell)$ and a prior belief $P(\Theta)$.} & \ref{sec:inference}
& \parbox{0.4\columnwidth}{$P(\Theta|X) = \frac{P(j_1|\Theta)\cdots
  P(j_{\ell}|\Theta) P(\Theta)}{\sum_{\Theta'}P(j_1|\Theta')\cdots P(j_{\ell}|\Theta')}$,
which can be solved numerically using our prior knowledge $P(\Theta)$
and our knowledge of the probability of each $j_i$ given $\Theta$.}\\
\hline
\end{tabular}
\end{center}
\caption{A summary of our results for application of PGFs to
 discrete-time SIS and SIR disease processes in the  infinite population limit.  The function $\mu(x)$ is the PGF for the offspring distribution.  The notation $[g]$ in the exponent denotes function composition $g$ times. For example, $\mu^{[2]}(y) = \mu(\mu(y))$.}
\label{tab:discrete}
\end{table}

\begin{table}
\begin{tabular}{|Sc|Sc|Sc|}
\hline
\textbf{Question} & \textbf{Section} & \textbf{Solution}\\
\hline \hline
\parbox{0.4\columnwidth}{Probability of eventual extinction $\alpha$ given a single introduced infection.} 
& \ref{sec:cts_extinction} &\parbox{0.4\columnwidth}{$\alpha = \min(1,\gamma/\beta)$}\\
\hline
\parbox{0.4\columnwidth}{Probability of extinction by time $t$, $\alpha(t)$.} 
& \ref{sec:cts_extinction_time}
& \parbox{0.4\columnwidth}{$\alpha(t)$ where $\dot{\alpha} = (\beta+\gamma)[\hat{\mu}(\alpha) - \alpha]$ and $\alpha(0)=0$.}\\
\hline
\parbox{0.4\columnwidth}{PGF of the distribution of number of infected individuals at time $t$ (assuming one infection at time $0$).} 
& \ref{sec:cts_infection_PGF}
&\parbox{0.4\columnwidth}{$\Phi(y,t)$ where $\Phi(y,0)=y$ and $\Phi$ solves either 
\[
\pd{}{t} \Phi = (\beta+\gamma) [\hat{\mu}(y)-y]\pd{}{y} \phi
\]
or 
\[
\pd{}{t}\Phi=(\beta+\gamma)[\hat{\mu}(\Phi)-\Phi]\,.
\]
}\\
\hline
\parbox{0.4\columnwidth}{PGF of the number of completed cases at time $t$.} 
& \ref{sec:cts_final_size}
& \parbox{0.4\columnwidth}{$\Omega(z,t)$ where $\Omega(z,0)=1$ and $\Omega$ solves 
\[
\pd{}{t}\Omega = (\beta+\gamma) \big[\hat{\mu}(\Omega,z)-\Omega\big]
\]
}\\
\hline
\parbox{0.4\columnwidth}{PGF of the joint distribution of the number of current and completed cases at time $t$ (assuming one infection at time $0$).} 
& \ref{sec:cts_joint_size}
& \parbox{0.4\columnwidth}{$\Pi(y,z,t)$ where $\Pi(y,z,0)=y$ and $\Pi$ solves either 
\[
\pd{}{t} \Pi = (\beta+\gamma) \big[\hat{\mu}(y,z) - y \big] \pd{}{y}\Pi
\]
or 
\[
\pd{}{t} \Pi = (\beta+\gamma) \big[ \hat{\mu}(\Pi, z) - \Pi \big]\, .
\]}\\
\hline
\parbox{0.4\columnwidth}{PGF of the final size distribution.}
& \ref{sec:cts_final_size}
& \parbox{0.4\columnwidth}{$\Omega_\infty(z)=\lim_{t\to\infty} \Omega(z,t)$.  This also solves $\Omega_\infty(z) = \hat{\mu}(\Omega_\infty(z),z)$.  If epidemics are possible this has a discontinuity at $|z|=1$.}\\
\hline
\parbox{0.4\columnwidth}{Probability an outbreak infects exactly $j$
  individuals} & \ref{sec:cts_final_size} & \parbox{0.4\columnwidth}{$\frac{1}{j} \frac{\beta^{j-1}\gamma^j}{(\beta+\gamma)^{2j-1}}\binom{2j-2}{j-1}$.}\\
\hline
\parbox{0.4\columnwidth}{PGF for the joint distribution of the number susceptible and infected at time $t$ for SIS dynamics in a population of size $N$.}&\ref{sec:cts_full_SIS} & \parbox{0.4\columnwidth}{$\Xi(x,y,t)$ where 
$\Xi$ solves
\[
\pd{}{t}\Xi = \frac{\beta}{N}(y^2-xy)\pd{}{x}\pd{}{y}\Xi + \gamma(x-y)\pd{}{y} \Xi
\]}\\
\hline
\parbox{0.4\columnwidth}{PGF for the joint distribution of the number susceptible and infected at time $t$ for SIR dynamics in a population of size $N$.}&\ref{sec:cts_full_SIR} & \parbox{0.4\columnwidth}{$\Xi(x,y,t)$ where 
$\Xi$ solves
\[
\pd{}{t}\Xi = \frac{\beta}{N}(y^2-xy)\pd{}{x}\pd{}{y}\Xi + \gamma(1-y)\pd{}{y} \Xi
\]}\\
\hline
\end{tabular}
\caption{A summary of our results for application of PGFs to the continuous-time disease process.  We assume individuals transmit with rate $\beta$ and recover with rate $\gamma$.  The functions $\hat{\mu}(y)=(\beta y^2 + \gamma)/(\beta+\gamma)$ and $\hat{\mu}(y,z)=(\beta y^2 + \gamma z)/(\beta+\gamma)$ are given in System~\eqref{sys:muhat}.}
\label{tab:continuous}
\end{table}

\begin{table}
\begin{tabular}{|Sc|Sc|Sc|}
\hline
\textbf{Question} & \textbf{Section} &\textbf{Solution}\\
\hline\hline
\parbox{0.45\columnwidth}{Final size relation for an SIR epidemic assuming a vanishingly small fraction $\rho$ randomly infected initially with $\rho N \gg 1$.} &
\ref{sec:final_size} & \parbox{0.4\columnwidth}{$r(\infty) = 1-\chi(1-r(\infty))$.  [For standard assumptions, including the usual continuous-time assumptions, $\chi(x) = e^{-\Ro(1-x)}$.]}\\
\hline
\parbox{0.45\columnwidth}{Discrete-time number susceptible, infected, or recovered in a population with homogeneous susceptibility and given $\Ro$, assuming an initial fraction $\rho$ is randomly infected with $\rho N \gg 1$.} & \ref{sec:large_time_discrete_SIR} 
&
\parbox{0.4\columnwidth}{For $g>0$:\begin{align*}S(g) &= N(1-\rho)e^{-\Ro(1-S(g-1)/N)}\\
I(g) &= N-S(g)-R(g)\\ R(g) &= R(g-1) + I(g-1)\end{align*}
with the initial condition $S(0)=(1-\rho)N$, \ $I(0)=\rho N$, and $R(0)=0$.}\\
\hline
\parbox{0.45\columnwidth}{Discrete-time number susceptible, infected, or recovered in a population with heterogeneous susceptibility for SIR disease after $g$ generations with an initial fraction $\rho$ randomly infected where $\rho N \gg 1$.} 
& \ref{sec:large_time_discrete_SIR}
& \parbox{0.4\columnwidth}{For $g>0$:\begin{align*}S(g) &= N(1-\rho)\chi(S(g-1)/N)\\
I(g) &= N-S(g)-R(g)\\ R(g) &= R(g-1) + I(g-1)\end{align*}
with the initial condition $S(0)=(1-\rho)N$, \ $I(0)=\rho N$, and $R(0)=0$.}\\
\hline
\parbox{0.45\columnwidth}{Continuous time number susceptible, infected, or recovered for SIR disease as a function of time with an initial fraction $\rho$ randomly infected where $\rho N \gg 1$.  Assumes $u$ receives infection at rate $\beta I \kappa_u/N\ave{K}$} 
& \ref{sec:large_time_cts_SIR}
& \parbox{0.4\columnwidth}{For $t>0$:\begin{align*}S(t) &= (1-\rho)N\psi(\theta(t))\\
I(t) &= N-S(t)-R(t)\\ 
R(t) &= \frac{\gamma N\ave{K}}{\beta} 
\ln \theta(t)\\
\dot{\theta}(t) &= -\frac{\beta}{N\ave{K}}I\theta(t)
\end{align*}
with the initial condition
$\theta(0)=1$.}\\
\hline
\end{tabular}
\caption{A summary of our results for application of PGFs to the final
  size and large-time dynamics of SIR disease.  The PGFs $\chi$ and
  $\psi$ encode the heterogeneity in susceptibility.  The PGF $\chi$
  is the PGF of the ancestor distribution (an ancestor of $u$ is any
  individual who, if infected, would infect $u$).  The PGF
  $\psi(x)=\sum_\kappa p(\kappa)x^\kappa$ encodes the distribution of the contact rates.}
\label{table:infinite_proportion}
\end{table}

\FloatBarrier

\begin{table}[t!]
\begin{center}
\begin{tabular}{|Sc|Sc|Sc|Sc|}
\hline
\textbf{Distribution} &\textbf{PGF}
  & \parbox{0.2\textwidth}{\textbf{Probability of\\ $j$ infections}} &
  \parbox{0.3\textwidth}{\textbf{Log-Likelihood of\\
Parameters given $j$}}\\
\hline
\hline
Poisson & $e^{\lambda(y-1)}$ & $\frac{(j\lambda)^{j-1}}{j!}
                               e^{-j\lambda}$ 
& \parbox{0.3\textwidth}{$-j\lambda + (j-1)
                                                \log(j\lambda) - \log(j!)$}\\
\hline
Uniform & $y^\lambda$ & $\begin{cases} 1 & j=1, \ \lambda = 0\\
0 & \text{otherwise} \end{cases}$ & \parbox{0.3\textwidth}{$\begin{cases} 0 & j=1, ] \lambda=
0\\
-\infty & \text{otherwise} \end{cases}$}\\ 
\hline
Binomial & $(q+py)^n$ & $ \frac{1}{j} \binom{nj}{j-1}
                        p^{j-1}q^{nj-j+1}$ & \parbox{0.3\textwidth}{$\log((nj)!) -
                                             \log((nj-j+1)!) -
                                             \log(j!) + (j-1) \log p +
                                             (nj-j+1) \log q$} \\
\hline
Geometric & $p/(1-qy)$ & $\frac{1}{j} \binom{2j-2}{j-1}p^jq^{j-1}$ &
                                                                     \parbox{0.3\textwidth}{$\log((2j-2)!)
                                                                     -
  \log ((j-1)!) - \log(j!) + j \log p + (j-1) \log q$}\\
\hline
Negative Binomial & $\left(\frac{q}{1-py}\right)^{\hat{r}}$ &
                                                              $\frac{1}{j}
                                                              \binom{\hat{r}j+j-2}{j-1}
                                                              q^{\hat{r}j}p^{j-1}$
                                                      
&\parbox{0.3\textwidth}{$\log ((\hat{r}j+j-1)!) - \log ((\hat{r}j - 1)!) - \log (j!) +
  \hat{r}j\log q + (j-1) \log p$}
  \\ 
\hline
\end{tabular}
\end{center}
\caption{The probability of $j$ total infections in an infinite
  population for different offspring distributions, derived using
  Theorem~\ref{thm:power_magic} and the corresponding log-likelihoods.
  For any one of these, if we sum the probability of $j$ over (finite)
  $j$, we get the probability that the outbreak remains finite in an
  infinite population.  This is particularly useful when inferring
  disease parameters from observed outbreak sizes
  (Section~\ref{sec:inference}).   The parameters' interpretations are given in
  Table~\ref{tab:example_PGFs}. }
\label{table:final_probs}
\end{table}

\ifsolns
\else
\subsection{Exercises}
\fi
We end each section with a collection of exercises.  We have designed these exercises to give the reader more experience applying PGFs and to help clarify some of the more subtle points.


\begin{exercise}
Except for the Poisson distribution handled in Example~\ref{example:poisson2PGF}, derive the PGFs shown in Table~\ref{tab:example_PGFs} directly from the definition $f(x) = \sum_i r_i x^i$.

For the negative binomial, it may be useful to use the binomial series:
\[
(1+\delta)^\eta = 1+ \eta \delta + \frac{\eta(\eta -1)}{2!} \delta^2 + \cdots +
\frac{\eta(\eta-1)\cdots(\eta-i+1)}{i!}\delta^i + \cdots
\]
using $\eta = -\hat{r}$ and $\delta = -px$.
\end{exercise}
\begin{solution}
\mbox{}
\begin{myenumerate}
\item Uniform: $p(\lambda)=1$, so the PGF is simply $0 + 0 + \cdots + 1 x^\lambda + 0 + 0 + \cdots = x^\lambda$.
\item Binomial: We have $r_i = \binom{n}{i} p^i q^{n-i}$.  So
\begin{align*}
\sum_i r_i x^i &= \sum_i \binom{n}{i} p^i q^{n-i}x^i\\
&= (px+q)^n
\end{align*}
by the binomial theorem.
\item Geometric: We have $r_i = q^ip$.  So
\begin{align*}
\sum_i r_i x^i &= \sum_i pq^ix^i\\
&= p \sum_i (qx)^i\\
&= \frac{p}{1-qx}
\end{align*}
by the sum of a geometric series.
\item Negative binomial: We have $r_i = \binom{i + \hat{r}-1}{i} q^{\hat{r}}p^i$.  So
\begin{align*}
\sum_i r_i x^i &= \sum_i \binom{i + \hat{r}-1}{i} q^{\hat{r}}p^ix^i\\
&= q^{\hat{r}} \sum_i \binom{i + \hat{r}-1}{i}p^i x^i\\
&=q^{\hat{r}}(1-px)^{-\hat{r}}\\
&= \left( \frac{q}{1-px}\right)^{\hat{r}}
\end{align*}
\end{myenumerate}
\end{solution}

\begin{exercise}
Consider the binomial distribution with $n$ trials, each having
success probability $p=\lambda/n$.  Using
Table~\ref{tab:example_PGFs}, show that the PGF for the binomial distribution
converges to the PGF for the Poisson distribution in the limit $n \to
\infty$, if $\lambda$ is fixed.
\end{exercise}
\begin{solution}
We have $(px+q)^n$ where $p = \lambda/n$ and $q=1-p$.  Taking $n$ to be large gives
\begin{align*}
\lim_{n \to \infty} (q+px)^n &= \lim_{n\to\infty} \left( 1 + \frac{\lambda (x-1)}{n}\right)^n\\
&= e^{\lambda(x-1)}
\end{align*}
\end{solution}

\section{Discrete-time spread of a simple disease: early time}
\label{sec:discrete}
We begin with a simple model of disease transmission using a
discrete-time setting.  In the time step after becoming infected, an
infected individual causes some number of additional cases and then
recovers.  We let $p_i$ denote the probability of causing exactly $i$
infections (referred to as ``offspring'') before recovering.  It will be useful to define  the PGF for the offspring distribution
\begin{equation}
\label{eqn:mu}
\mu(y) = \sum_{i=0}^\infty p_i y^i \, .
\end{equation}
For results related to early extinction or early-time dynamics, we
will assume that the population is large enough and sufficiently
well-mixed that the transmissions in successive generations are all
independent events and unaffected by depletion of susceptible
individuals.  Before deriving our results for the
early-time behavior of our discrete-time model, we offer a summary in table~\ref{tab:discrete}.

\begin{figure}
\begin{center}
\includegraphics[width=0.5\textwidth]{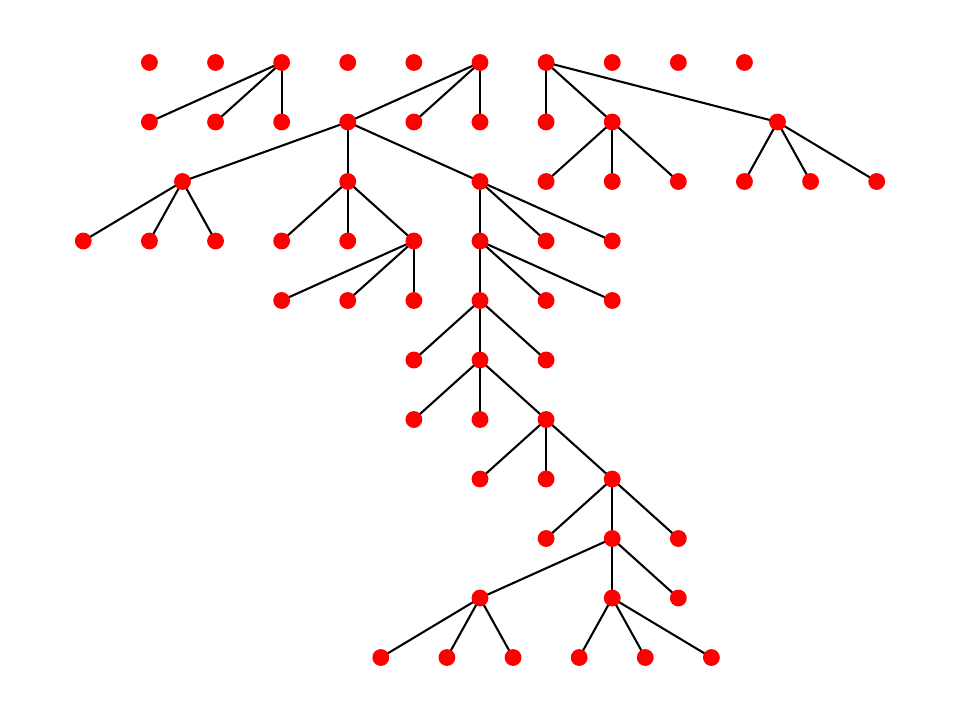}
\end{center}
\caption{A sample of $10$ outbreaks starting with a bimodal distribution having $\Ro = 0.9$ in which $3/10$ of the population causes $3$ infections and the rest cause none.  The top row denotes the initial states, showing each of the $10$ initial infections.  An edge from one row to the next denotes an infection from the higher node to the lower node.  Most outbreaks die out immediately.}
\label{fig:sample_outbreaks}
\end{figure}

Often in disease spread we are interested in the expected number of infections caused by an infected individual early in an outbreak, which we define to be $\Ro$.  
\begin{equation}
\label{eqn:Ro}
\Ro = \sum_i i p_i = \mu'(1) 
\end{equation}
where $\mu'(x) = \diff{}{x}\mu(x)$.  The value of $\Ro$ is related to disease dynamics, but it is not the only important property of $\mu$.
\begin{example}
We demonstrate a few sample outbreaks in Fig.~\ref{fig:sample_outbreaks}.  Here we take a bimodal case with $\Ro = 0.9$ such that a proportion $0.3$ of the population cause $3$ infections and the remaining $0.7$ cause none.  Most of the outbreaks die out immediately, but some persist, surviving multiple generations before extinction.
\end{example}

\begin{figure}
\begin{center}
\begin{tabular}{ccc}
& Poisson & Bimodal\\
$\Ro=0.75$ 
& \raisebox{-0.5\height}{\includegraphics[width=0.4\textwidth]{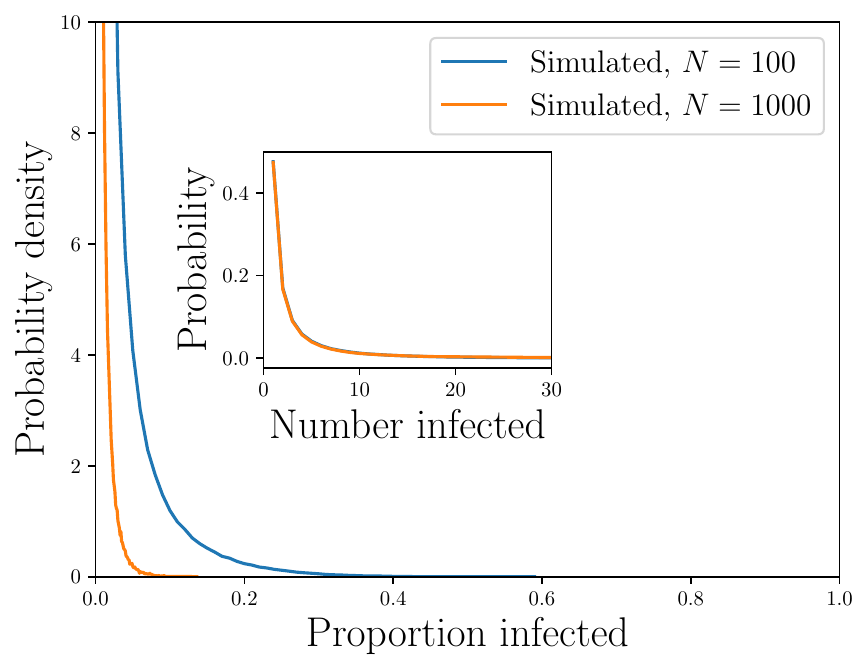}}
&\raisebox{-0.5\height}{\includegraphics[width=0.4\textwidth]{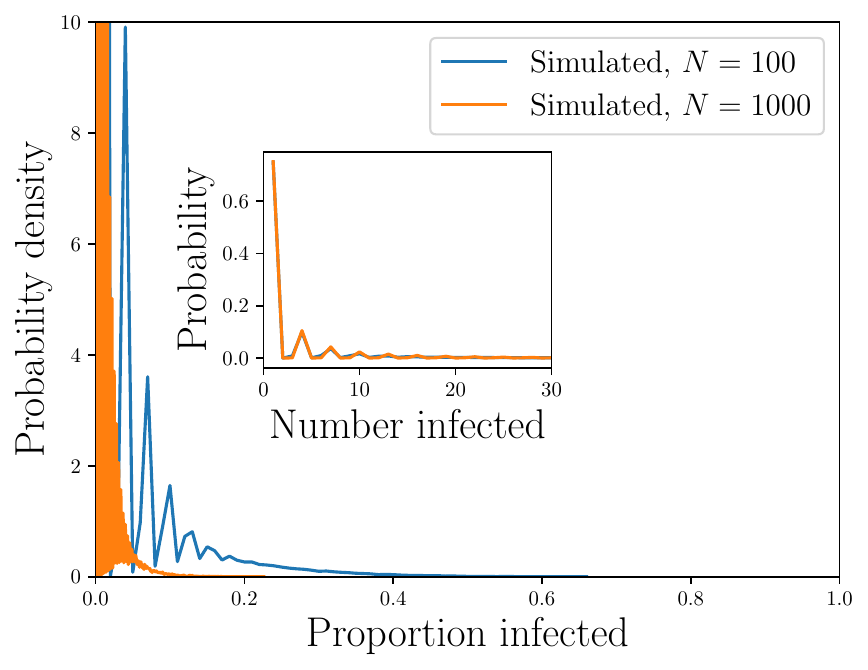}}\\
$\Ro = 2$ 
&\raisebox{-0.5\height}{\includegraphics[width=0.4\textwidth]{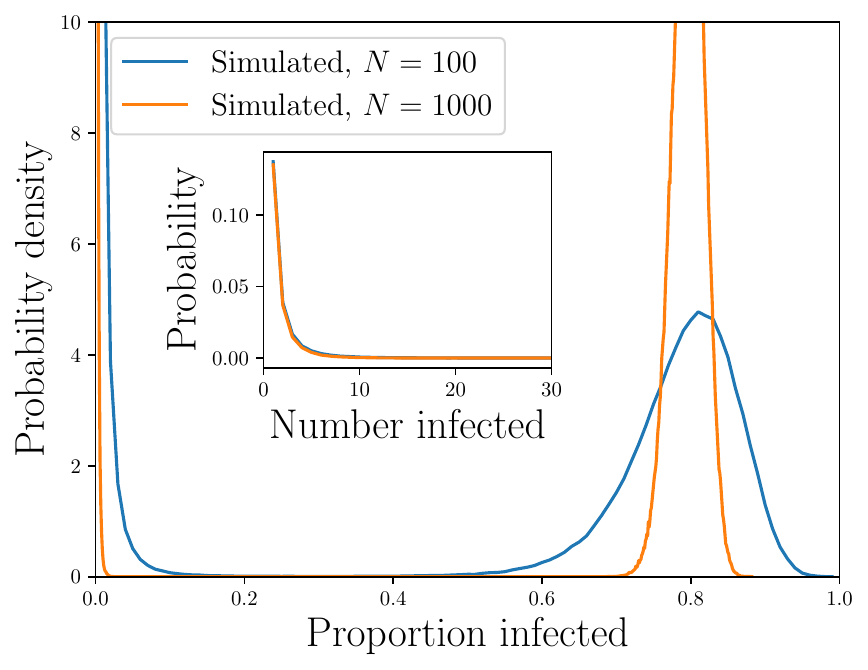}}
&\raisebox{-0.5\height}{\includegraphics[width=0.4\textwidth]{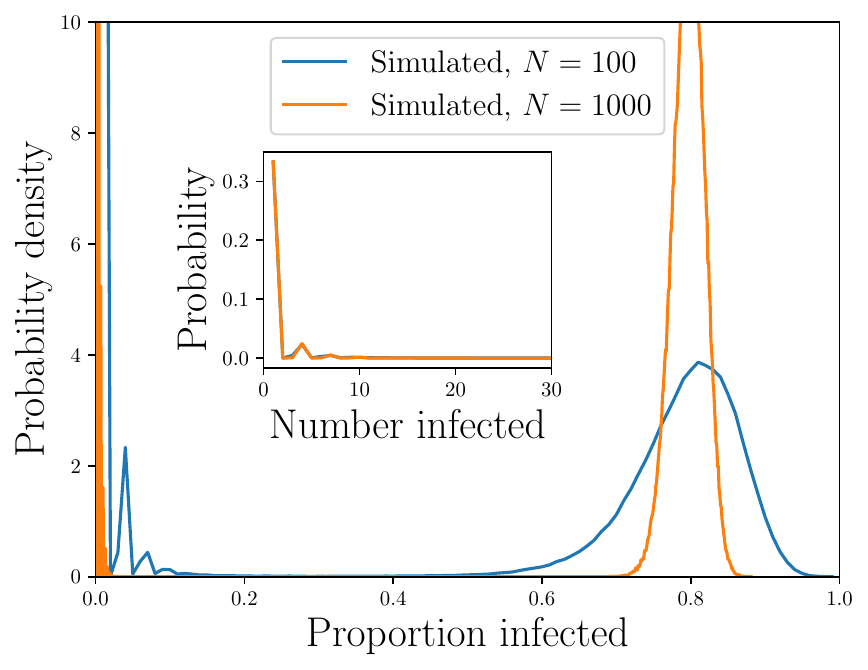}}
\end{tabular}
\end{center}
\caption{Simulated outcomes of SIR outbreaks in populations as described in example~\ref{example:model_description}.  Outbreaks tend to be either small or large.  
The typical \emph{number infected} in small outbreaks (insets) is affected by the details of the offspring distribution, but not the population size.  The typical \emph{proportion infected} in large outbreaks (epidemics) appears to depend on the average number of transmissions an individual causes, but not the population size or the offspring distribution.  These observations will be explained later.  These simulations are reused throughout this section to show how PGFs capture different properties of the distributions.} 
\label{fig:basic_observations}
\end{figure}
\begin{example}
\label{example:model_description}
Throughout Section~\ref{sec:discrete} we compare simulated SIR
outbreaks with the theoretical predictions which we calculate using
the Python package \texttt{Invasion\_PGF} described in Appendix~\ref{app:software}.  We assume that all individuals are equally likely to be infected by any transmission, and we focus on $\Ro = 0.75$ and $\Ro = 2$.  For each $\Ro$, we consider two distributions for the number of new infections an infected individual causes:
\begin{itemize}
\item a Poisson-distributed number of infections with mean $\Ro$, or
\item a bimodal distribution with either $0$ or $3$ infections, with the proportion chosen to give
  a mean of $\Ro$.  The probabilities are $p_0 = 1-\Ro/3$ and $p_3 =
  \Ro/3$ ($\Ro>3$ is impossible).
\end{itemize}
The bimodal distribution is similar to that of
Fig.~\ref{fig:sample_outbreaks}, but with different probabilities of
$0$ or $3$.  After an individual chooses the number of infections to
cause, the recipients are selected uniformly at random (with
replacement) from the population.  If they are susceptible, an
infection occurs at the next time step, otherwise nothing happens.  We
use $5\times10^5$ simulations for $N=100$ and $N=1000$.

Figure~\ref{fig:basic_observations} looks at the final size
distribution.  The distribution of the \emph{number} infected in small
outbreaks (insets) is not significantly affected by the total
population size.  This is because they do not grow large enough to
``see'' the system size.  They would die out even in an infinite
population.  Large outbreaks, or \emph{epidemics}, on the other hand
would grow without bound in an infinite population, and their growth
is limited by the finiteness of the population.  We will see that
(assuming homogeneous susceptibility and the large population
limit), the \emph{proportion} infected in an SIR epidemic depends only on $\Ro$.
\end{example}

\ifsolns
\else
\subsection{Early extinction probability}
\fi
\label{sec:extinction}
A common misconception is that if $\Ro>1$ an epidemic is inevitable.  In fact, if we are lucky an outbreak can die out  stochastically before  the number infected is large.  Conversely, if we are not lucky it may initially grow faster than our deterministic models predict.

In any finite population a disease will eventually go extinct because
the disease interferes with its own spread.  Our observations show
that the typical final outcomes of an outbreak are either an
``epidemic'' which grows until the number infected is limited by the
finiteness of the population or a small outbreak which dies out before
it can see the system size.  One of our first questions about a
possible disease emergence is ``what is the probability that an
outbreak will grow into an epidemic?''  We focus on the equivalent
question, ``what is the probability the outbreak goes extinct before
causing an epidemic?''.  We aim to calculate the probability that the
disease would go extinct if it never interferes with its own spread,
or in other words, if it were spreading through an unlimited
population.  Throughout we assume that disease is introduced with a
single randomly chosen index case.

The theory for the extinction probability in an unbounded population has been developed extensively in the context of Galton--Watson processes~\cite{watson1875probability}.  It has been applied to infectious disease many times, e.g., \cite[section 21.8]{easley2010networks} and~\cite{getz2006basic,lloyd2005superspreading}.

\ifsolns
\else
\subsubsection{Derivation as a fixed point equation}
\fi

We present two derivations of the extinction probability.  Our first is quicker, but gives less insight.  We start with the \emph{a priori} observation that the extinction probability takes some value between $0$ and $1$ inclusive.  Our goal is to filter out the vast majority of these options by finding a property of the extinction probability that most values between $0$ and $1$ do not have.

Let $\alpha$ be the probability of extinction if the spread starts from a single infected individual.  Then from Property~\ref{property:notclothesline} of Appendix~\ref{app:properties} we have $\alpha = \sum_ip_i \hat{\alpha}^i = \mu(\hat{\alpha})$ where $\hat{\alpha}$ is the probability that, in isolation, an offspring of the initial infected individual would not cause an epidemic.  Because we assume that the offspring distribution of later cases is the same as for the index case, we must have $\hat{\alpha} = \alpha$ and so the extinction probability solves $\alpha = \mu(\alpha)$.

 We have established:
\begin{thm}
\label{thm:weak_extinction}
Assuming that each infected individual produces an independent number of offspring $i$ chosen from a distribution having PGF $\mu(y)$,  then $\alpha$, the probability an outbreak starting from a single infected individual goes extinct, satisfies
\begin{equation}
\alpha = \mu(\alpha) \, .
\label{eqn:discrete_extinction_naive}
\end{equation}
Not all solutions to $x=\mu(x)$ must give the extinction probability.
\end{thm}

There can be more than one $x$ solving $x=\mu(x)$.  In fact $1=\mu(1)$
is always a solution, and from Property~\ref{property:PGFcobweb} it
follows that there is another solution if and only if
$\Ro=\mu'(1) >1$.  In this case, our derivation of
Theorem~\ref{thm:weak_extinction} does not tell us which of the
solutions is correct.  However,
Section~\ref{sec:extinction_generation} shows that the correct
solution is the smaller solution when it exists.  More specifically
the extinction probability is $\alpha = \lim_{g\to\infty} \alpha_g$ where
$\alpha_g = \mu(\alpha_{g-1})$ starting with $\alpha_0=0$.  This gives
a condition for a nonzero epidemic probability.  Namely
$\Ro = \mu'(1) = \sum_i ip_i > 1$.

\begin{figure}
\begin{center}
\begin{tabular}{ccc}
& Poisson & Bimodal\\
$\Ro=0.75$ 
& \raisebox{-0.5\height}{\includegraphics[width=0.4\textwidth]{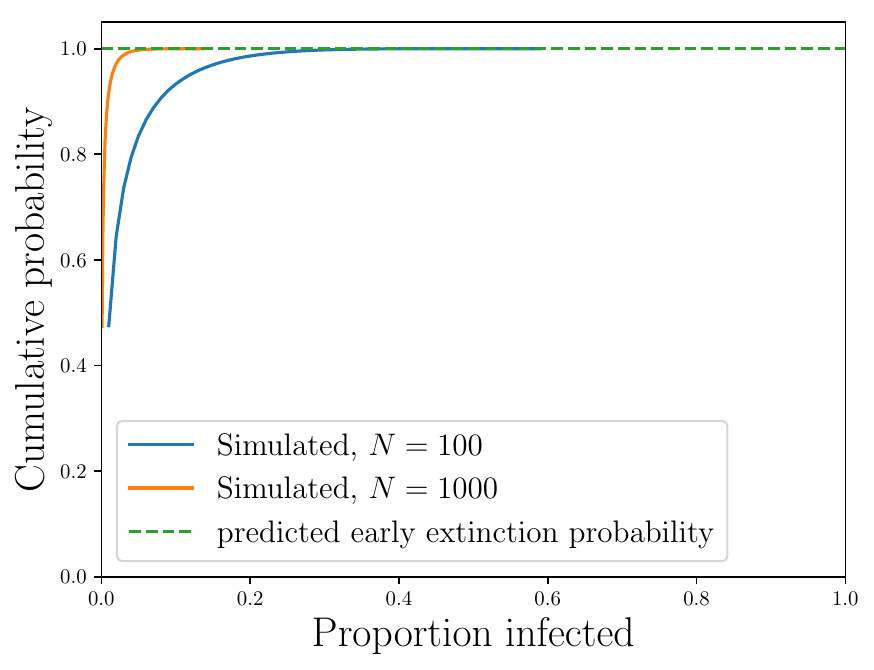}}
& \raisebox{-0.5\height}{\includegraphics[width=0.4\textwidth]{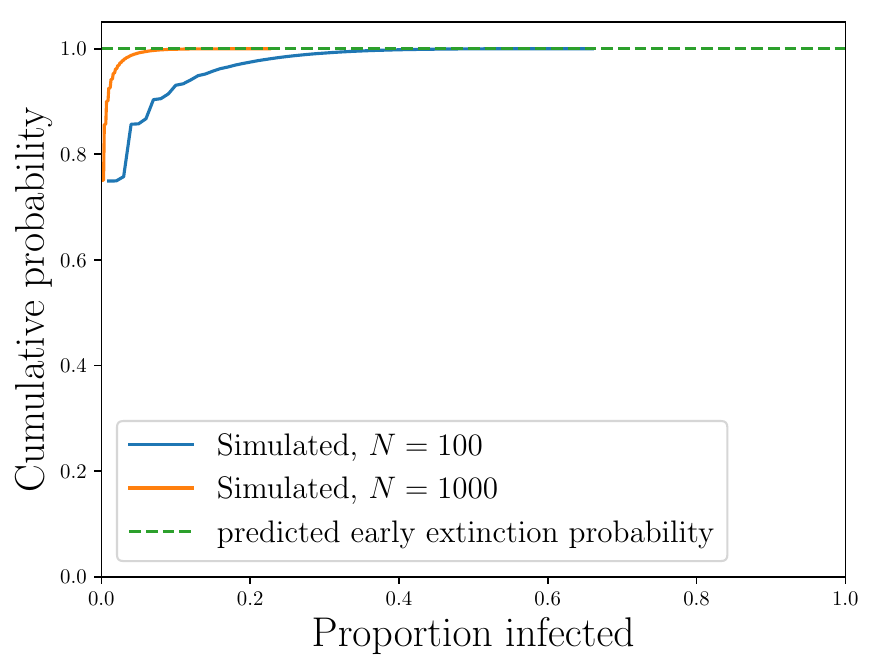}}\\
$\Ro = 2$ 
&\raisebox{-0.5\height}{\includegraphics[width=0.4\textwidth]{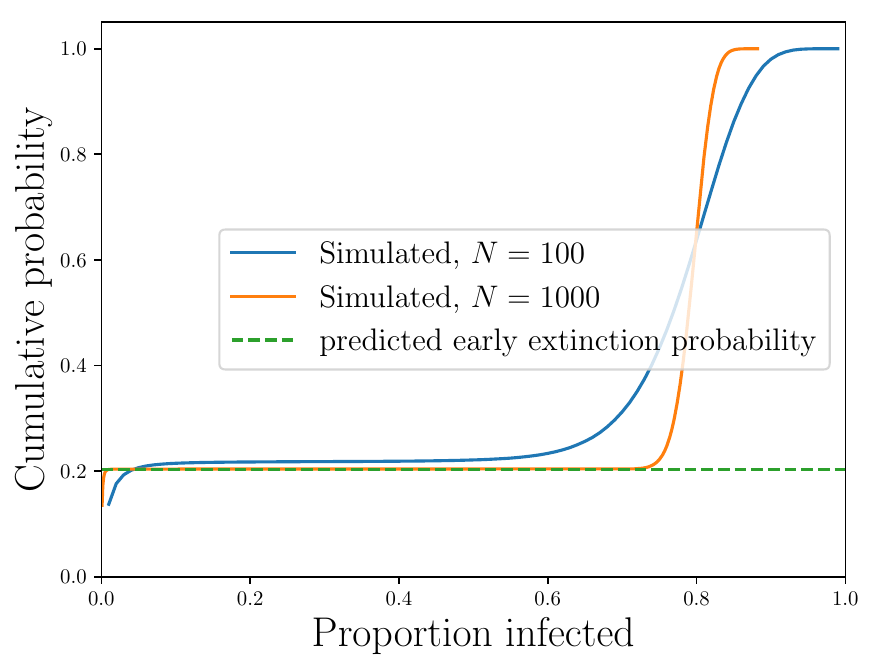}}
& \raisebox{-0.5\height}{\includegraphics[width=0.4\textwidth]{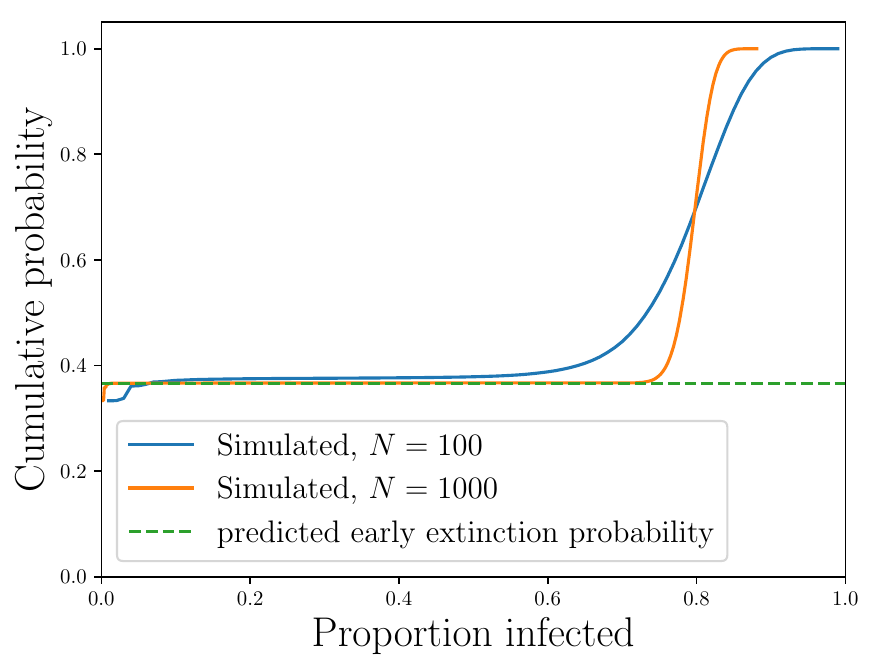}}
\end{tabular}
\end{center}
\caption{\textbf{Illustration of Theorem~\ref{thm:weak_extinction}.} The cumulative density function (cdf) for the total proportion ever infected (effectively the integral of Fig.~\ref{fig:basic_observations}). 
For small $\Ro$, all outbreaks die out without affecting a sizable portion of the population.  For larger $\Ro$, there are many small outbreaks and many large outbreaks, but very few outbreaks in between, so the cdf is flat in this range.  The height of this plateau is the probability the outbreak dies out while small.  This is approximately the predicted extinction probability for an infinite population (dashed).   The probability of a small outbreak is different for the different distributions, but the proportion infected corresponding to epidemics is the same (for given $\Ro$).}
\label{fig:epiprob}
\end{figure}
\begin{example}
\label{example:cdf}
We now consider the Poisson and bimodal offspring distributions
described in Example~\ref{example:model_description}. We saw that
typically an outbreak either affects a small proportion of the population (a vanishing fraction in the infinite population limit) or a large number (a nonzero fraction in the infinite population limit).

By plotting the cumulative density function (cdf) of proportion infected in Fig.~\ref{fig:epiprob}, we extend our earlier
observations.  The cdf is steep near zero (becoming vertical in the
infinite population limit).  Then it is effectively flat for a while.
Finally if $\Ro > 1$ it again grows steeply at some proportion
infected well above $0$ (the size of epidemic outbreaks).

The plateau's height is the probability that an outbreak dies out
while small.  Fig.~\ref{fig:epiprob} shows that this is well-predicted by choosing the smaller of the solutions to $x = \mu(x)$.

For a fixed $\Ro>1$, the the plateau's height (i.e., the early
extinction probability) depends on the details of the offspring
distribution and not simply $\Ro$.  However, the critical value at
which the cdf increases for the second time depends only on $\Ro$.
This suggests that even though the probability of an epidemic depends
on the details of the offspring distribution, the proportion infected in an SIR epidemic depends only on $\Ro$, the reproductive number.  We explore this in more detail in Section~\ref{sec:final_size}.
\end{example}

\ifsolns
\else
\subsubsection{Derivation from an iterative process}
\fi
\label{sec:extinction_generation}

In our second derivation, we calculate the probability that the
outbreak dies out within $g$ ``generations''.  Then the probability
the outbreak would die out after a finite number of steps in an
infinite population is simply the limit of this as $g \to \infty$.  In
our counting of ``generations'', we consider the index case to be
generation $0$.  An individual's generation is equal to the number of transmissions occurring in the chain from the index case to that individual.

We define $\alpha_g$ to be the probability that the longest chain an
index case will initiate has fewer than  $g$ transmissions.  So
because there are always at least $0$ transmissions, $\alpha_0 =
0$. The probability that there is no transmission is by definition
$\alpha_1$.  Recalling that the probability the index case causes zero infections is $p_0$, we have
\[
\alpha_1=p_0 = \mu(0) = \mu(\alpha_0)
\]
is the probability that the index case does not cause a chain of $1$
or more transmissions.  The probability that all chains die out after
at most $1$ transmission (that is, there are no second generation
cases) is the probability that the index case causes $i$ infections,
$p_i$, times the probability none of those $i$ individuals causes
further infections, $\alpha_1^i$, summed over all $i$.  We introduce
the notation $\mu^{[g]}(x)$ to be the result of iterative applications
of $\mu$ to $x$ $g$ times, so $\mu^{[1]}(x) = \mu(x)$ and for $g>1$, \
$\mu^{[g]}(x) = \mu(\mu^{[g-1]}(x))$.  Then following
Property~\ref{property:notclothesline} we have
\[
\alpha_2 = p_0 + p_1 \alpha_1 + p_2 \alpha_1^2 + \cdots =
\mu(\alpha_1) = \mu^{[2]}(0)
\]
We generalize this by stating that the probability an initial infection fails to initiate any length $g$ chains is equal to the probability that all of its $i$ offspring fail to initiate a chain of length $g-1$.  
\[
\alpha_g = \sum_i p_i \alpha_{g-1}^i = \mu(\alpha_{g-1}) = \mu^{[g]}(0) \, .
\]
So the probability of not starting a chain of length at least $g$ is found by iteratively applying the function $\mu$ \ $g$ times to $x=0$.  Taking $g \to \infty$ gives the extinction probability~\cite{getz2006basic}:
\begin{equation}
\label{eqn:extinction}
\alpha = \lim_{g\to\infty} \mu^{[g]}(0) \, .
\end{equation}

The fact that there is a biological interpretation of $\alpha_g$
starting with $\alpha_0=0$ is important.  It effectively guarantees
that the iterative process converges and that the speed of convergence
reflects the typical speed of extinction.  Iteration appears to be an
efficient way to solve $x = \mu(x)$ numerically and because of the
biological interpretation, we can avoid questions that might arise
about whether there are multiple solutions of $x=\mu(x)$ and, if so,
which of them corresponds to the biological problem.  Instead we
simply iterate starting from $0$ and the result must converge to the probability that in an infinite population the outbreak would go extinct in finite time, regardless of what other solutions $x=\mu(x)$ might have.

Exercise~\ref{exercise:alpha_g} shows that if $\mu(0)\neq 0$ then the limit of the sequence $\alpha_g$ is $1$ if $\Ro\leq 1$ and some $\alpha<1$ satisfying $\alpha = \mu(\alpha)$ if $\Ro >1$.  This proves:
\begin{thm}
\label{thm:strong_extinction}
Assume that each infected individual produces an independent number of offspring $i$ chosen from a distribution having PGF $\mu(y)$.  Then
\begin{itemize}
\item The probability an outbreak goes extinct within $g$ generations is
\begin{equation}
\alpha_g = \mu^{[g]}(0)\, . 
\label{eqn:discrete_extinct}
\end{equation}
\item The probability of extinction in an infinite population is 
\[
\alpha = \lim_{g\to\infty} \alpha_g \, .
\]
\item If $\Ro=\mu'(1) \leq 1$ and $\mu(0) \neq 0$
then $\alpha=1$.  If $\Ro>1$ extinction occurs with probability $\alpha<1$.
\end{itemize}
\end{thm}


\begin{figure}
\begin{center}
\parbox{0.095\textwidth}{Poisson\\$\Ro=0.75$}
\raisebox{-0.5\height}{\includegraphics[width=0.85\textwidth]{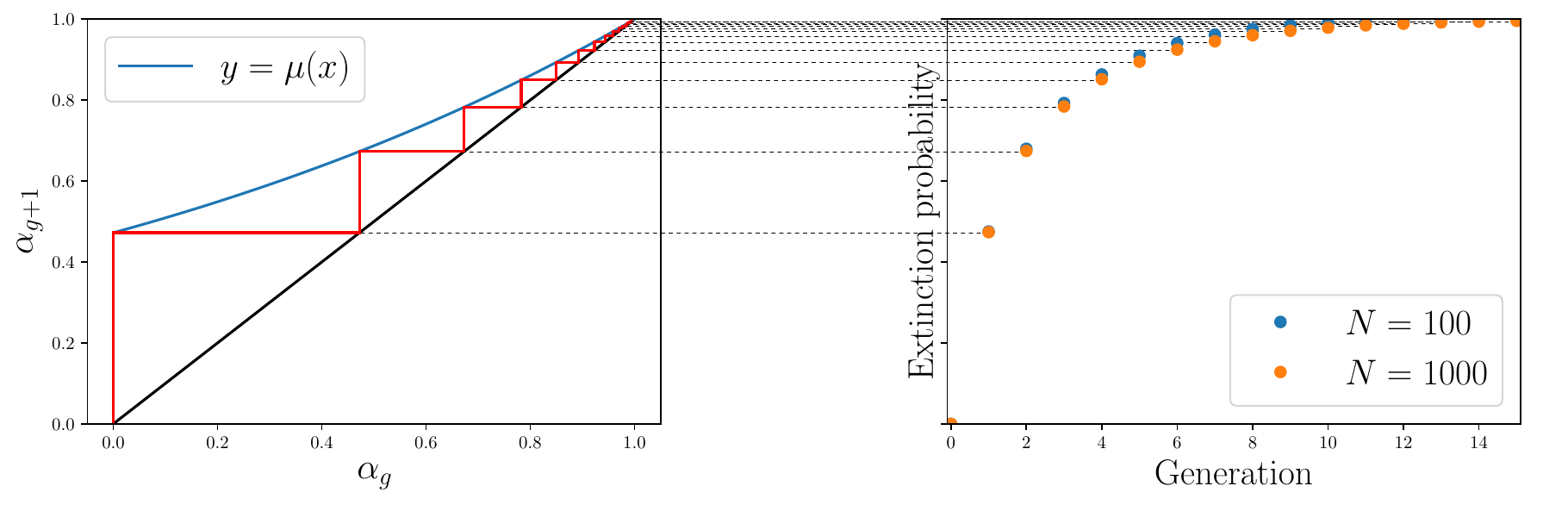}}
\parbox{0.095\textwidth}{Bimodal\\$\Ro=0.75$}
\raisebox{-0.5\height}{\includegraphics[width=0.85\textwidth]{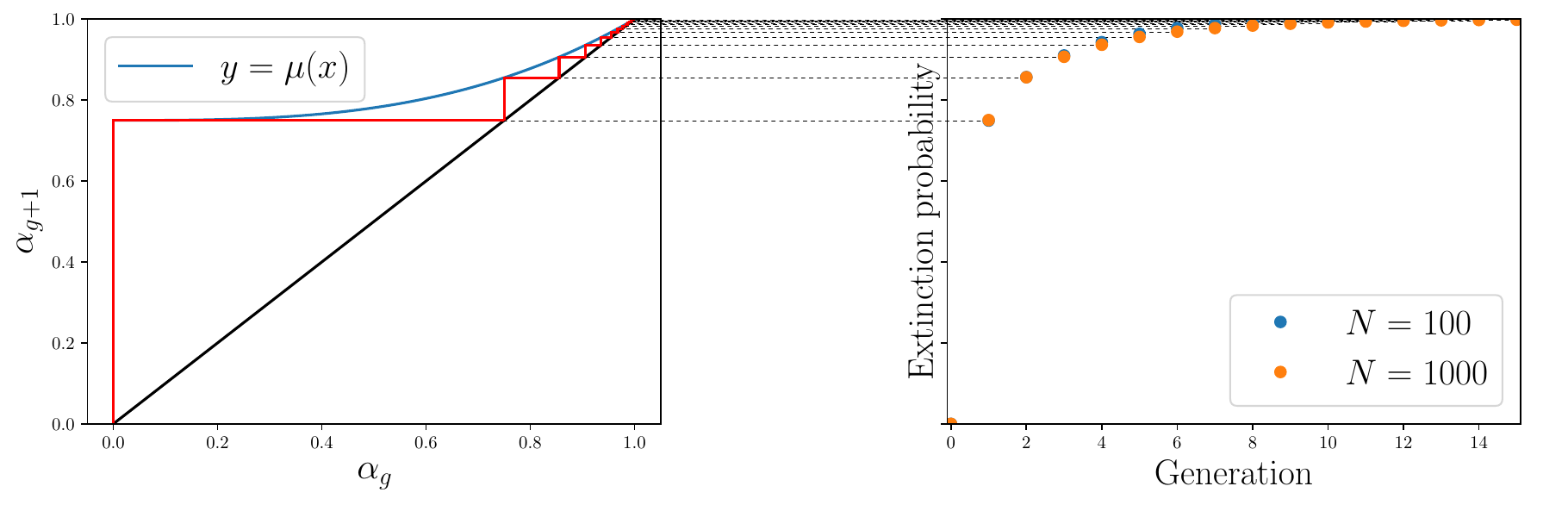}}
\parbox{0.095\textwidth}{Poisson\\$\Ro=2$}
\raisebox{-0.5\height}{\includegraphics[width=0.85\textwidth]{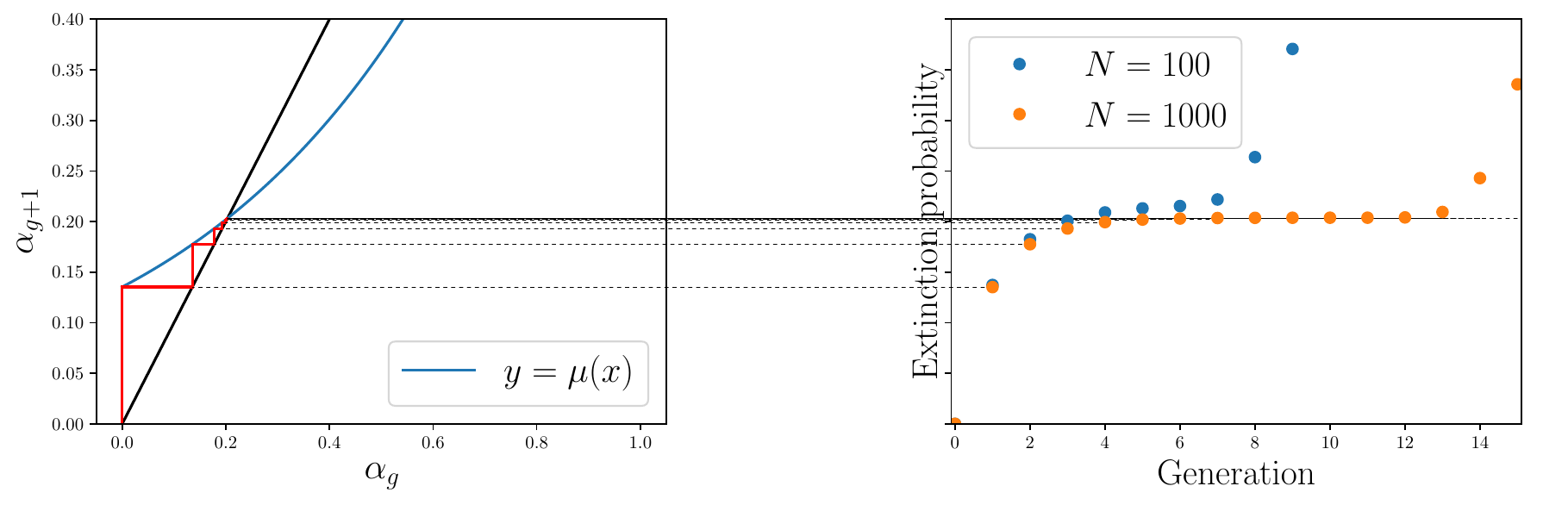}}
\parbox{0.095\textwidth}{Bimodal\\$\Ro=2$}
\raisebox{-0.5\height}{\includegraphics[width=0.85\textwidth]{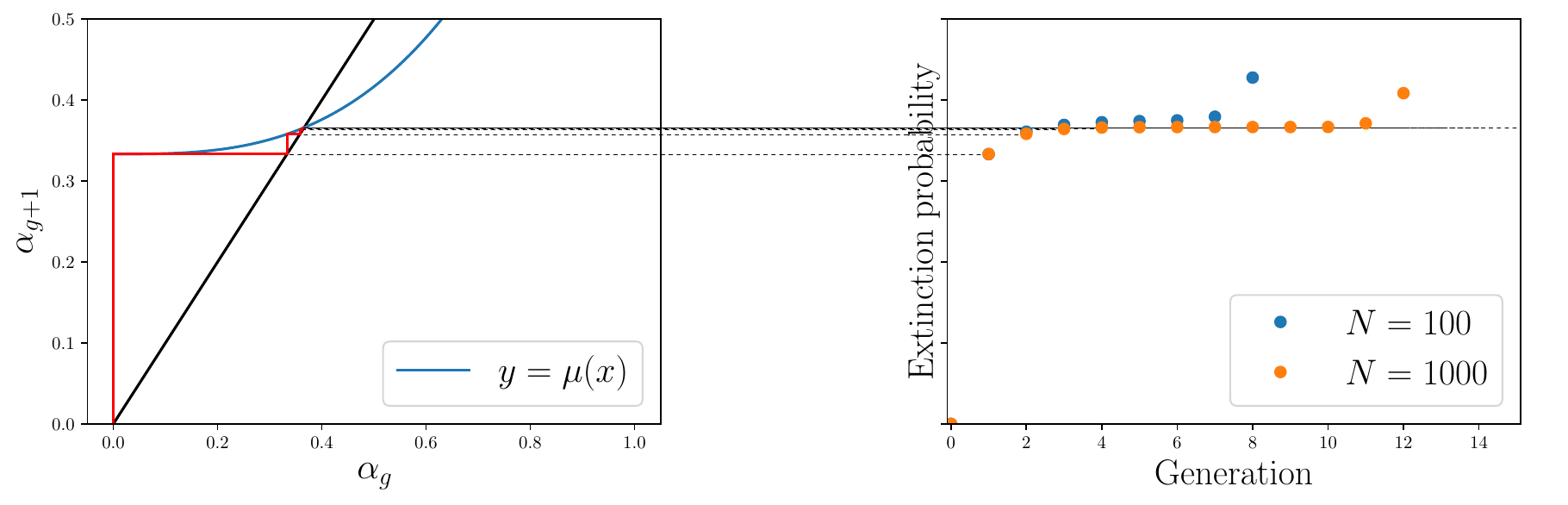}}
\end{center}

\caption{\textbf{Illustration of Theorem~\ref{thm:strong_extinction}.}
   \textbf{Left:} Cobweb diagrams showing convergence of iterations to
   the predicted outbreak extinction probability (see
   Fig.~\ref{fig:cobweb}).  \textbf{Right:} Observed probabilities of
   no infections remaining after each generation for simulations of
   Fig.~\ref{fig:basic_observations} showing the probability of
   extinction by generation $g$.  Thin lines show the relation between
   the cobweb diagram and the extinction probabilities.  The simulated
   probability tends to rise quickly representing outbreaks that die
   out early on, then it remains steady at a level representing the
   probability of outbreaks dying out while small.  For $\Ro>1$ it increases again
   because the epidemics burn through the finite population (and so
   the infinite population theory breaks down).  The values match the corresponding iteration of the cobweb diagrams.}
   \label{fig:cobweb_comp}
\end{figure}

\begin{example}
We now consider the Poisson and bimodal offspring distributions described in Example~\ref{example:model_description}

Figure~\ref{fig:cobweb_comp} shows that starting with $\alpha_0=0$ and
defining $\alpha_g = \mu(\alpha_{g-1})$, the  values of $\alpha_g$ emerging from
the iterative process correspond to the observed probability outbreaks
have gone extinct by generation $g$ for early values of $g$.

In the infinite population limit, this provides a match for all $g$.
So this gives the probability the outbreak goes extinct by generation
$g$ assuming it has not grown large enough to see the finite-size of
the population (i.e., assuming it has not become an epidemic).  For
SIR epidemics in the
finite populations we use for simulations, the plateaus eventually
give way to extinction because eventually there are not enough
remaining susceptibles.
\end{example}

\ifsolns
\else
\subsection{Early-time outbreak dynamics}
\fi
\label{sec:current_size}
We now explore the number of active infections present in generation $g$.  Setting $\phi_i(g)$ to be the probability $i$ active infections exist at generation $g$, we define the PGF $\Phi_g(y)= \sum_i \phi_i(g)y^i$.  Assuming at generation $0$ there is a single infection ($\phi_1(0)=1$) then the initial condition is $\Phi_0(y)=y$. From inductive application of Property~\ref{prop:composition} for composition of PGFs (exercise~\ref{exercise:discrete_backward_and_forward}) it is straightforward to conclude that for $g>0$, \ $\Phi_g(y) = \mu^{[g]}(y)$ where $\mu(y)$ is the PGF for the offspring distribution.

\begin{thm}
\label{thm:generation_size}
Assuming that each infected individual produces an independent number of offspring $\ell$ chosen from a distribution with PGF $\mu(y)$, the number infected in the $g$-th generation has PGF
\begin{equation}
\Phi_g(y) = \sum_\ell \phi_\ell(g)y^\ell = \mu^{[g]}(y) 
\label{eqn:discrete_early_PGF}
\end{equation}
where $\phi_i(g)$ is the probability there are $\ell$ active infections in generation $g$.
This does not provide information about the cumulative number infected.
\end{thm}

It is worth highlighting that for general distributions, the
calculation of coefficients of $\Phi_g(y)$ may seem quite challenging.
Luckily, it is not so difficult.  
Property~\ref{prop:cauchy_coefficient} states (taking $i = \sqrt{-1}$)
\[
\phi_\ell(g) \approx \frac{1}{M} \sum_{m=1}^M \frac{\Phi_g(Re^{2\pi i m/M})}{R^\ell e^{2\ell \pi i m/M}}
\]
for large $M$ and any $R \leq 1$.  For each $y_m = R e^{2\pi i m/M}$ we can calculate $\Phi_g(y_m) = \mu^{[g]}(y_m)$ by numerically iterating $\mu$ $g$ times.  Then for large enough $M$, this gives a remarkably accurate and efficient approximation to the individual coefficients.  

\begin{figure}
\begin{center}
\begin{tabular}{ccc}
& Poisson & Bimodal\\
$\Ro=0.75$ 
& \raisebox{-0.5\height}{\includegraphics[width=0.4\textwidth]{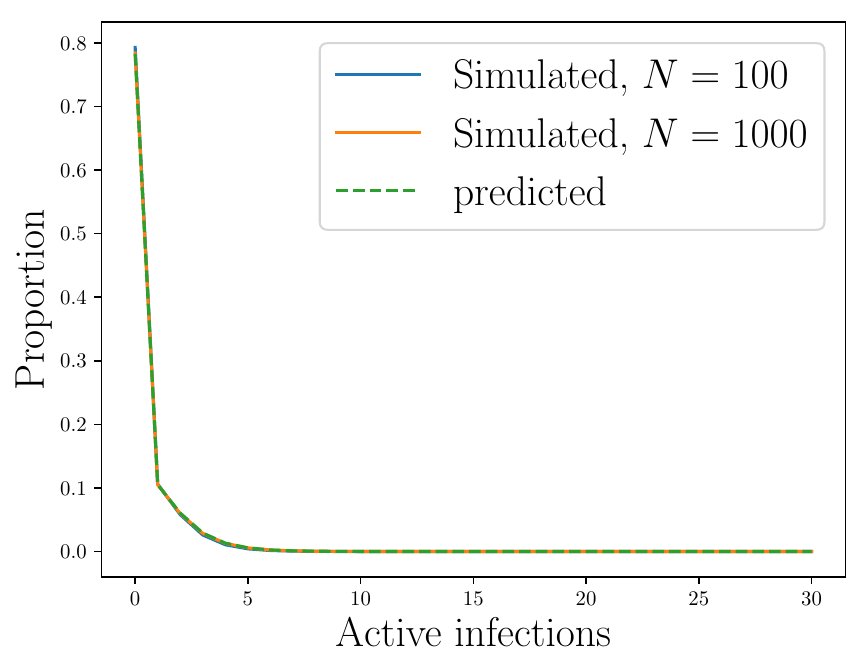}}
& \raisebox{-0.5\height}{\includegraphics[width=0.4\textwidth]{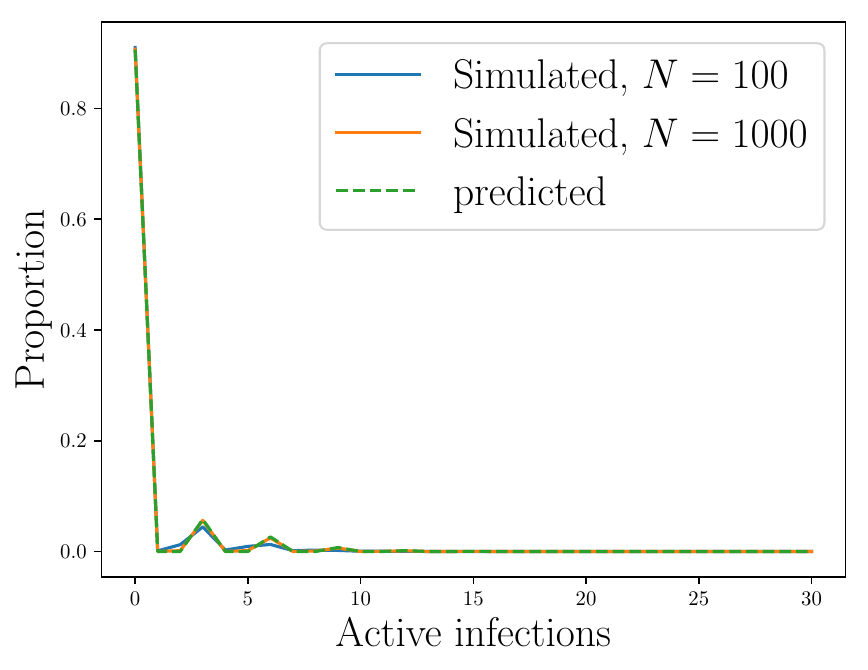}}\\
$\Ro=2$ 
& \raisebox{-0.5\height}{\includegraphics[width=0.4\textwidth]{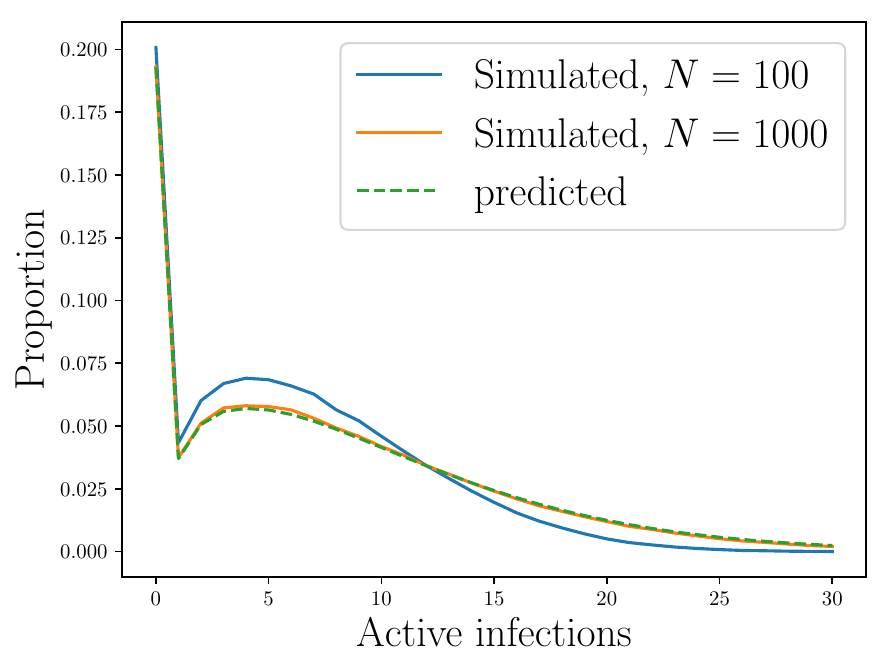}}
& \raisebox{-0.5\height}{\includegraphics[width=0.4\textwidth]{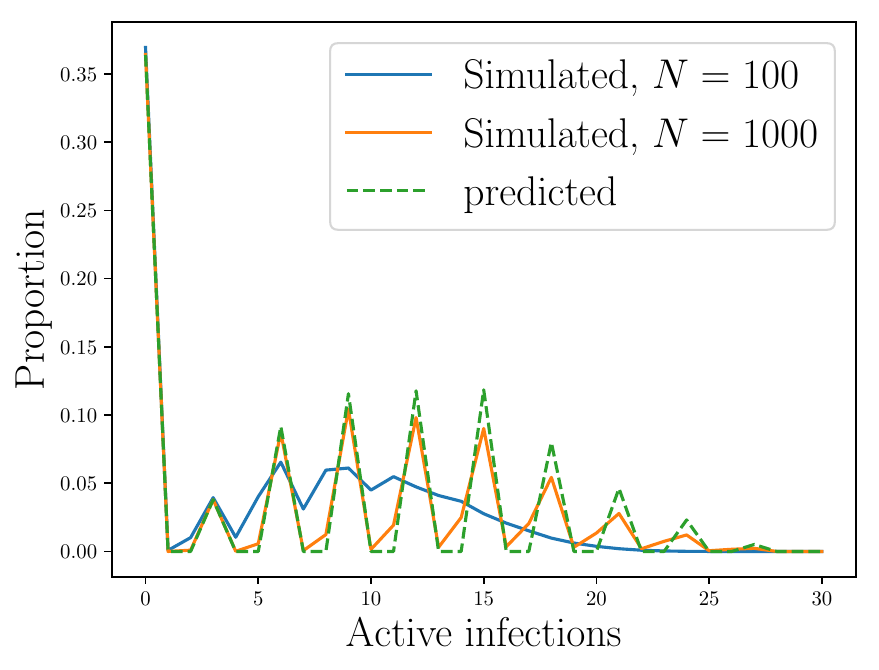}}
\end{tabular}
\end{center}
\caption{\textbf{Illustration of Theorem~\ref{thm:generation_size}.}
  Comparison of predictions and the simulations from
  Fig.~\ref{fig:basic_observations} for the number of active
  infections in the third generation.  The bimodal case with $N=100$
  shows a clear impact of population size as a sizable number of
  transmissions fail because the population is finite.  The
  predictions were made numerically using the summation in Property~\ref{prop:cauchy_coefficient}. }
\label{fig:current_size}
\end{figure}
\begin{example}
We demonstrate Theorem~\ref{thm:generation_size} in
Fig.~\ref{fig:current_size}, using the simulations from Example~\ref{example:model_description}.  Simulations and predictions are in excellent agreement.

There is a mismatch noticeable for the bimodal distribution with $\Ro=2$ particularly with $N=100$, which is a consequence of the fact that the population is finite.  In stochastic simulations, occasionally an individual receives multiple transmissions even early in the outbreak, but in the PGF theory this does not happen.

\end{example}

We are often interested in the expected number of active infections in generation $g$, \ $\sum_\ell \ell \phi_\ell(g)$ (however, as seen below this is not the most relevant measure to use if $\Ro > 1$).  Property~\ref{property:mean} shows that this is given by $\left.\pd{}{y}\Phi_g(y)\right|_{y=1}$. To calculate this we use  $\Phi_g(1)=1$ for all $g$ (Property~\ref{property:1}) and $\mu'(1)=\Ro$.  Then through induction and the chain rule we show that $\pd{}{y}\left. \Phi_g(y)\right|_{y=1} = \Ro^g$:
\begin{align*}
\left. \pd{}{y} \Phi_g(y)\right|_{y=1}
&= \left.\pd{}{y} \mu(\Phi_{g-1}(y))\right|_{y=1}\\
&= \left. \left( \mu'(\Phi_{g-1}(y)) \times \pd{}{y} \Phi_{g-1}(y) \right)\right|_{y=1}\\
&=  \mu'(1) \times \Ro^{g-1}\\
&= \Ro^g \, .
\end{align*}
we initialized the induction with the case $g=1$ which is the definition of $\Ro$.  If $\Ro<1$, this shows that we expect decay.  

If $\Ro>1$, there is a more relevant measure.  On average we see
growth, but a sizable fraction of outbreaks may go extinct, and these
zeros are included in the average, which alters our prediction.  This
is closely related to the ``push of the past'' effect observed in
phylodynamics~\cite{nee1994extinction}.  For policy purposes, we are more
interested in the expected size if the outbreak is not yet extinct
because a response that is scaled to deal with the average size
including those that are extinct is
either too big (if the disease has gone extinct) or too small (if the disease has become established)~\cite{miller2010epidemics}.  It is very unlikely to be just right.  The expected number infected in generation $g$ conditional on the outbreaks not dying out by generation $g$ is $\Ro^g/(1-\alpha_g)$.  This has an important consequence.  We can have different extinction probabilities for different offspring distributions with the same $\Ro$.  The disease with a higher extinction probability tends to have considerably more infections in those outbreaks that do not go extinct.  

We have
\begin{cor}
\label{cor:early_expectation}
In the infinite population limit, the expected number infected in generation $g$ starting from a single infection is
\begin{equation}
[I]_g = \Ro^g
\end{equation}
and the expected number starting from a single infection conditional
on the disease persisting to generation $g$ is
\begin{equation}
\langle I \rangle_g = \frac{\Ro^g}{1-\alpha_g}
\end{equation}
\end{cor}
We can explore higher moments through taking more derivatives of $\Phi_g(y)$ and evaluating at $y=1$.

\ifsolns
\else
\subsection{Cumulative size distribution}
\fi
We now look at the total number infected while the outbreak is small. 
There are multiple ways to calculate how the cumulative size of small
outbreaks is distributed.  We look at two of these.    The first
focuses just on the number of completed infections by generation $g$.
The second calculates the joint distribution of the number of
completed infections and the number of active infections at generation
$g$.  Later we address the distribution of final sizes.

\ifsolns
\else
\subsubsection{Focused approach to find the cumulative size distribution}
\fi
\label{sec:cum_size} 
We begin by calculating just the number of completed infections at generation $g$.  We define $\omega_j(g)$ to be the probability that there are $j$ completed infections at generation $g$ (by ``completed'' we only include individuals who are no longer infectious in generation $g$).  We will use PGFs of the variable $z$ when focusing on completed infections.

We define 
\[
\Omega_g(z)= \sum_j \omega_j(g) z^j
\]
to be the PGF for the number of completed infections $j$ at generation
$g$.  Although we use $j$ to represent recoveries, this model is still
appropriate for SIS disease because we are interested in small
outbreak sizes in a well-mixed infinite population for which we can
assume no previously infected individuals have been reexposed.  If the
outbreak begins with a single infection, then
\[
\Omega_0(z) = 1 \quad \text{and} \quad \Omega_1(z) = z
\]
showing that the first individual (infectious during generation $0$) completes his infection at the start of generation $1$.  For generation $2$ we have the initial individual and his direct offspring, so $\Omega_2(z)=z\mu(z)$.  

More generally, to calculate for $g>1$, the completed infections
consist of
\begin{itemize}
\item  the initial infection
\item the active infections in generation $1$.
\item any descendants of those active infections in generation $1$
  that will have recovered by generation $g$.
\end{itemize}
The distribution of the number of descendants of a generation $1$
individual (including that individual) who have recovered by generation $g$ is given by
$\Omega_{g-1}(z)$.  That is each generation $1$ individual and its 
descendants for the following $g-1$ infections have the same
distribution as an initial infection and its descendants after $g-1$
generations.  

From Property~\ref{prop:composition} the number of descendants by generation $g$ (not counting the initial infection) that have recovered is distributed like $\mu(\Omega_{g-1}(z))$. Accounting for the initial individual requires that we increment the count by $1$ which requires increasing the exponent of $z$ by $1$.  So we multiply by $z$.  This yields
\[
\Omega_g(z) = z \mu(\Omega_{g-1}(z))
\]

To sustain an outbreak up to generation $g$ there must be at least one infection in each generation from $0$ to  $g-1$. So any outbreak with fewer than $g$ completed infections at generation $g$ must be extinct.  So the coefficient of $z^j$ does not change once $g>  j$.  Thus we have shown
\begin{thm}
\label{thm:cum_size}
Assuming a single initial infection in an infinite population, the PGF $\Omega_g(z) = \sum_j \omega_j(g) z^j$ for the distribution of the number of completed infections at
generation $g> 1$ is given by
\begin{equation}
\Omega_g(z) = z \mu(\Omega_{g-1}(z))
\label{eqn:discrete_completed}
\end{equation}
with $\Omega_1(z)=z$.  Once $g > j$, the coefficient $\omega_j(g)$ is constant.
\end{thm}

\begin{figure}
\begin{center}
\begin{tabular}{ccc}
& Poisson & Bimodal\\
$\Ro=0.75$ 
& \raisebox{-0.5\height}{\includegraphics[width=0.4\textwidth]{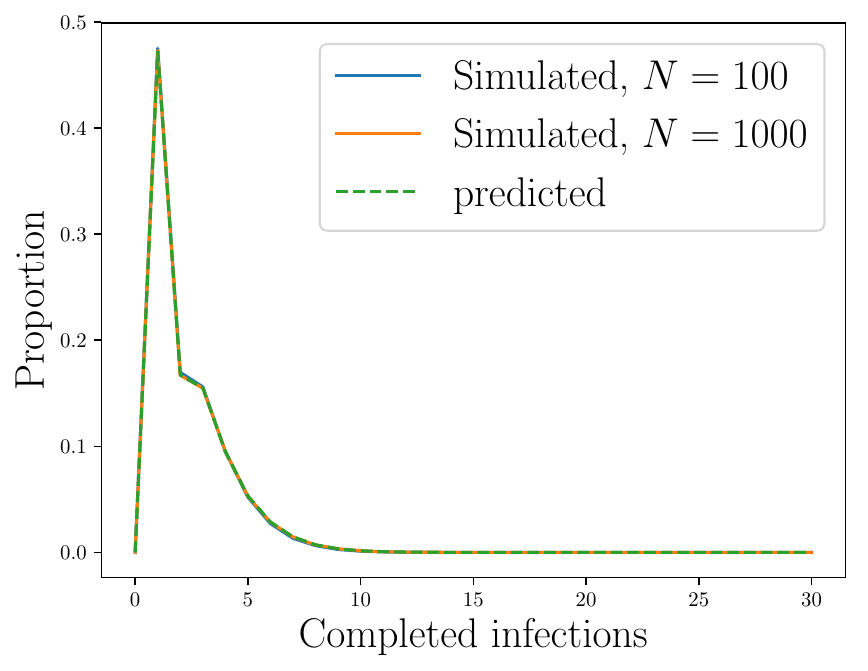}}
& \raisebox{-0.5\height}{\includegraphics[width=0.4\textwidth]{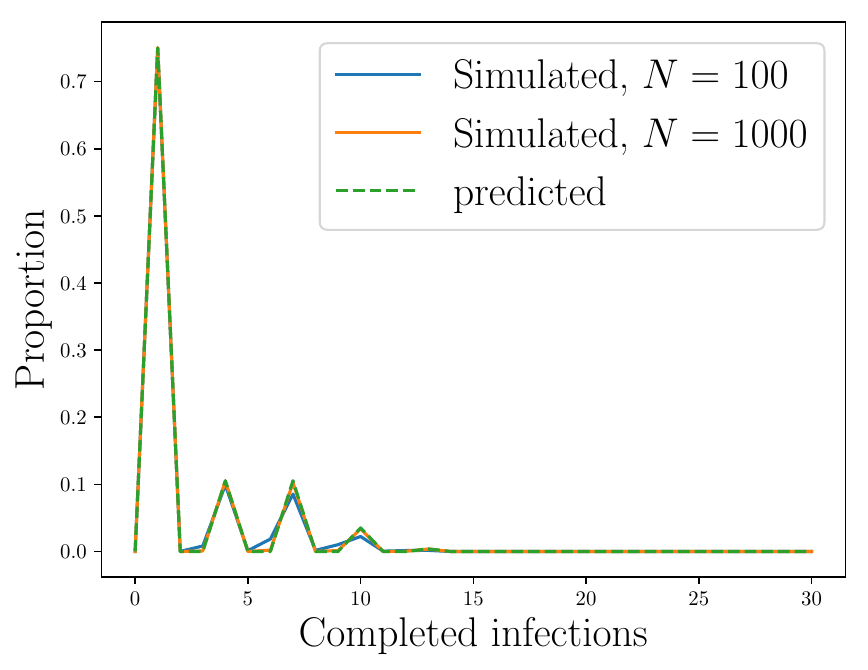}}\\
$\Ro=2$ 
& \raisebox{-0.5\height}{\includegraphics[width=0.4\textwidth]{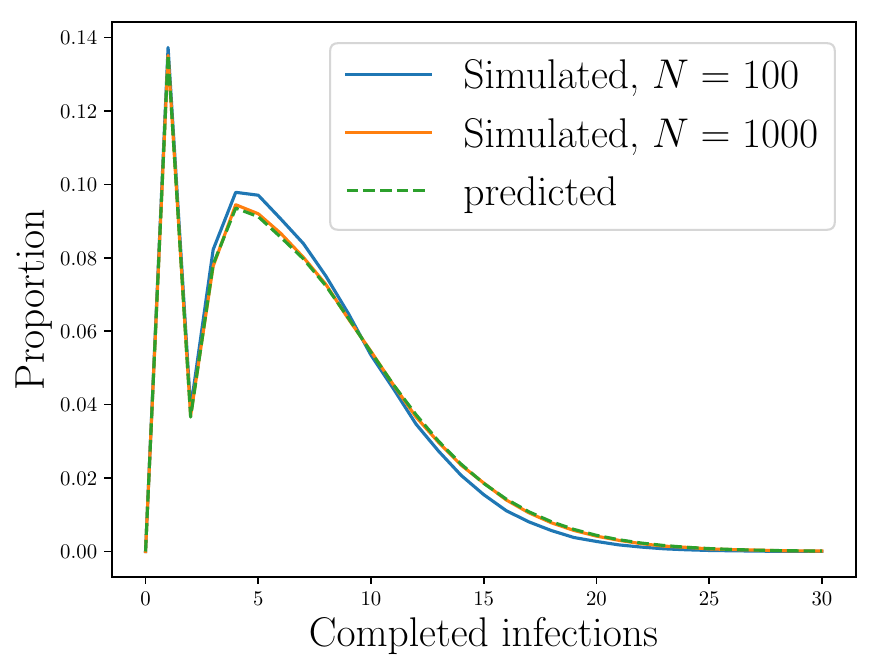}}
& \raisebox{-0.5\height}{\includegraphics[width=0.4\textwidth]{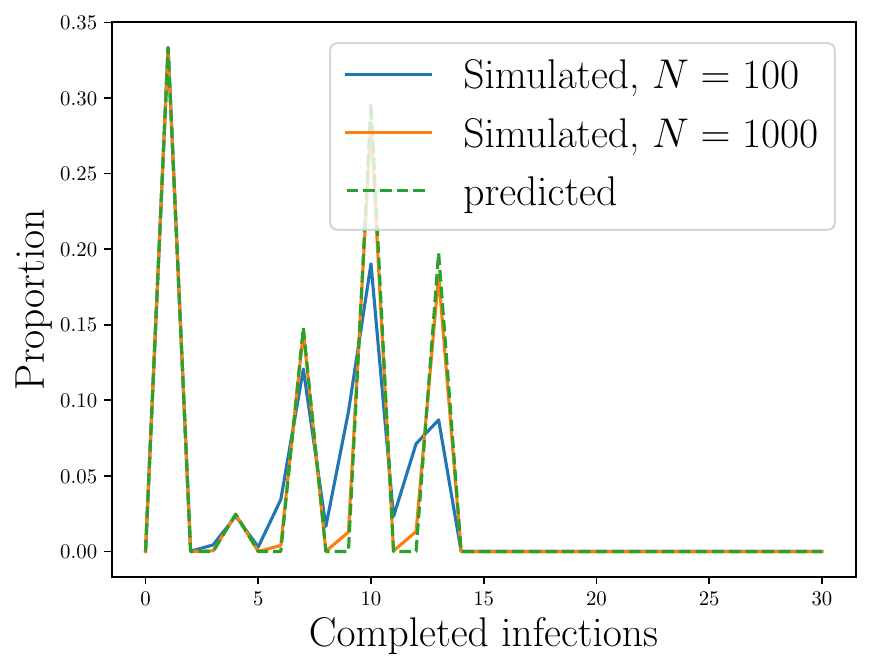}}
\end{tabular}
\end{center}
\caption{\textbf{Illustration of Theorem~\ref{thm:cum_size}}
  Comparison of predictions with the simulations from
  Fig.~\ref{fig:basic_observations} for the number of completed
  infections at the start of the third generation. The predictions
  were calculated using Property~\ref{prop:cauchy_coefficient}.  }
\label{fig:cum_size}
\end{figure}
\begin{example}
We test Theorem~\ref{thm:cum_size} in Fig.~\ref{fig:cum_size}, using
the simulations from Example~\ref{example:model_description}.  Simulations and predictions are in excellent agreement.

\end{example}

\begin{example}
\textbf{Expected cumulative size}
It is instructive to calculate the expected number of completed infections at generation $g$.  
Note that $\left. \Omega_g(z)\right|_{z=1}=1$, \  $\mu(1)=1$, and $\mu'(1) = \Ro$.   We use
induction to show that for $g\geq 1$ the expected number of completed infections is
$\sum_{j=0}^{g-1} \Ro^j$:
\begin{align*}
\left . \pd{}{z} \Omega_g(z)\right|_{z=1}
              &= \left . \pd{}{z} z \mu(\Omega_{g-1}(z)) \right |_{z=1}\\
              &=  \left.\mu(\Omega_{g-1}(z))+ z \mu'(\Omega_{g-1}(z))
                \pd{}{z} \Omega_{g-1}(z) \right|_{z=1}\\
              &= \mu(1) + \mu'(1) \sum_{j=0}^{g-2} \Ro^j\\
              &= 1 + \Ro \sum_{j=0}^{g-2} \Ro^j\\
              &= \sum_{j=0}^{g-1} \Ro^j 
\end{align*}
This is in agreement with our earlier result that the expected number that are infected in generation $j$ is $\Ro^j$.  

This is
\[
\left.\pd{}{z}\Omega_g'(z) \right|_{z=1}= \begin{cases}
\frac{1-\Ro^{g}}{1-\Ro} & \Ro \neq 1\\
g & \Ro = 1 
\end{cases} 
\]
As with our previous results, the sum shows a threshold behavior at $\Ro=1$.  If $\Ro<1$, then in the limit $g\to \infty$, the expected cumulative outbreak size converges to the finite value $1/(1-\Ro)$.  If $\Ro \geq 1$, it diverges.
\end{example}
This example shows
\begin{cor}
\label{cor:expected_cum_size}
In the infinite population limit the expected number of completed infections at the start of generation $g$ assuming a single randomly chosen initial infection is
\begin{subequations}
\label{sys:expected_cum_size}
\begin{equation}
\left.\pd{}{z}\Omega_g'(z) \right|_{z=1}= \begin{cases}
\frac{1-\Ro^{g}}{1-\Ro} & \Ro \neq 1\\
g & \Ro = 1 
\end{cases} 
\end{equation}
For $\Ro \geq 1$ this diverges as $g\to\infty$.  Otherwise it
converges to $1/(1-\Ro)$.
\end{subequations}
\end{cor}

\ifsolns
\else
\subsubsection{Broader approach}
\fi
\label{sec:joint_size}
An alternate approach calculates both the current and cumulative size at generation $g$.  We let $\pi_{i,r}(g)$ be the probability that there are $i$ currently infected individuals and $r$ completed infections in generation $g$.  We define $\Pi_g(y,z) = \sum_{i,r}\pi_{i,r}(g)y^iz^r$, so $y$ represents the active infections and $z$ the completed infections.

Assume we know the values $i_{g-1}$ and $r_{g-1}$ for generation $g-1$.  Then $r_g$ is simply $i_{g-1}+r_{g-1}$ and $i_g$ is distributed according to $\mu(y)^{i_{g-1}}$.  So given those known $i_{g-1}$ and $r_{g-1}$, the distribution for the next generation would be $[z \mu(y)]^{i_{g-1}}z^{r_{g-1}}$.  Summing over all possible $i_{g-1}$ and $r_{g-1}$ yields
\begin{align*}
\Pi_{g}(y,z) &= \sum_{i,r} \pi_{i,r}(g-1) [z\mu(y)]^i z^r\\
&= \Pi_{g-1}(z\mu(y), z)
\end{align*}
with the initial condition
\[
\Pi_0(y,z)=y
\]
The first few iterations are
\begin{align*}
\Pi_1(y,z) &= z \mu(y)\\
\Pi_2(y,z) &= z \mu(z\mu(y))
\end{align*}
and we can use induction on this to show that in general
\[
\Pi_g(y,z) = z \mu(\Pi_{g-1}(y,z))
\]
\begin{thm}
\label{thm:joint_size}
Given a single initial infection in an infinite population, the PGF $\Pi_g(y,z) = \sum_{i,r} \pi_{i,r}(g)y^iz^r$ for the joint distribution of the
number of active $i$ and completed infections $r$ in generation $g$ is given
by 
\begin{equation}
\Pi_g(y,z) = z \mu(\Pi_{g-1}(y,z))
\label{eqn:discrete_joint_dist}
\end{equation}
with $\Pi_0(y,z)=y$.
\end{thm}

\begin{figure}
\begin{center}
\includegraphics[width=0.8\textwidth]{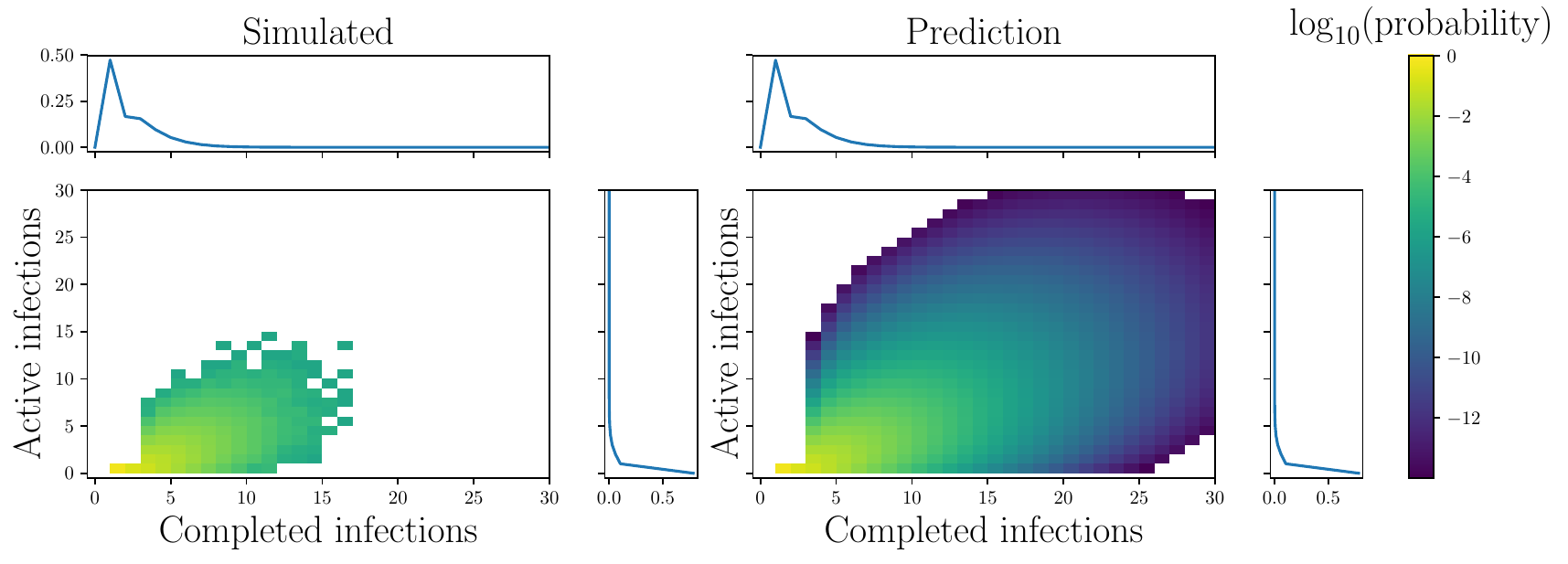}\\
\includegraphics[width=0.8\textwidth]{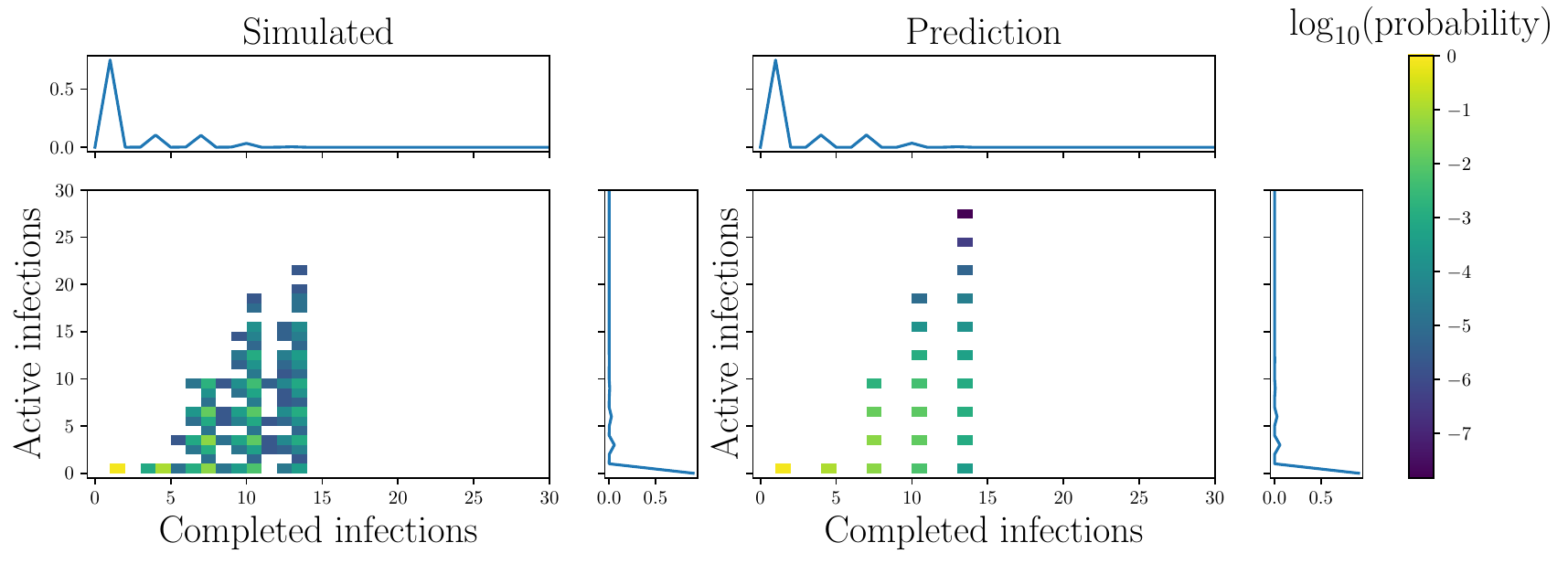}\\
\includegraphics[width=0.8\textwidth]{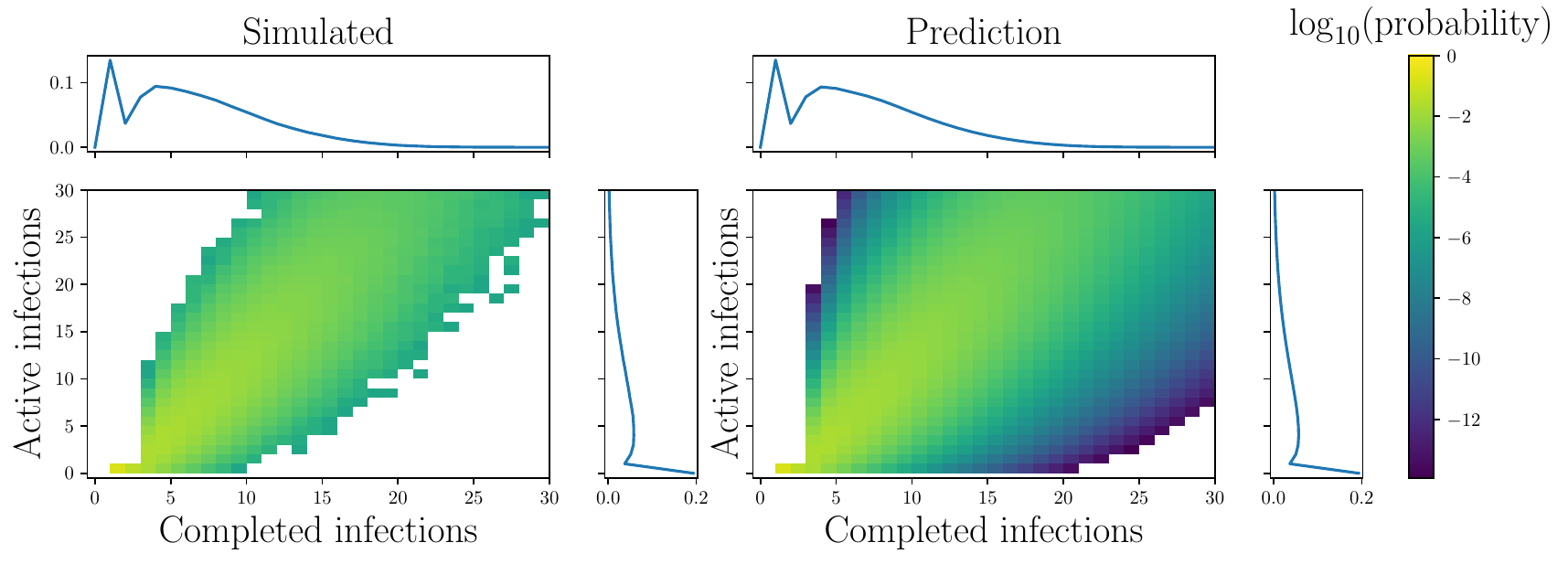}\\
\includegraphics[width=0.8\textwidth]{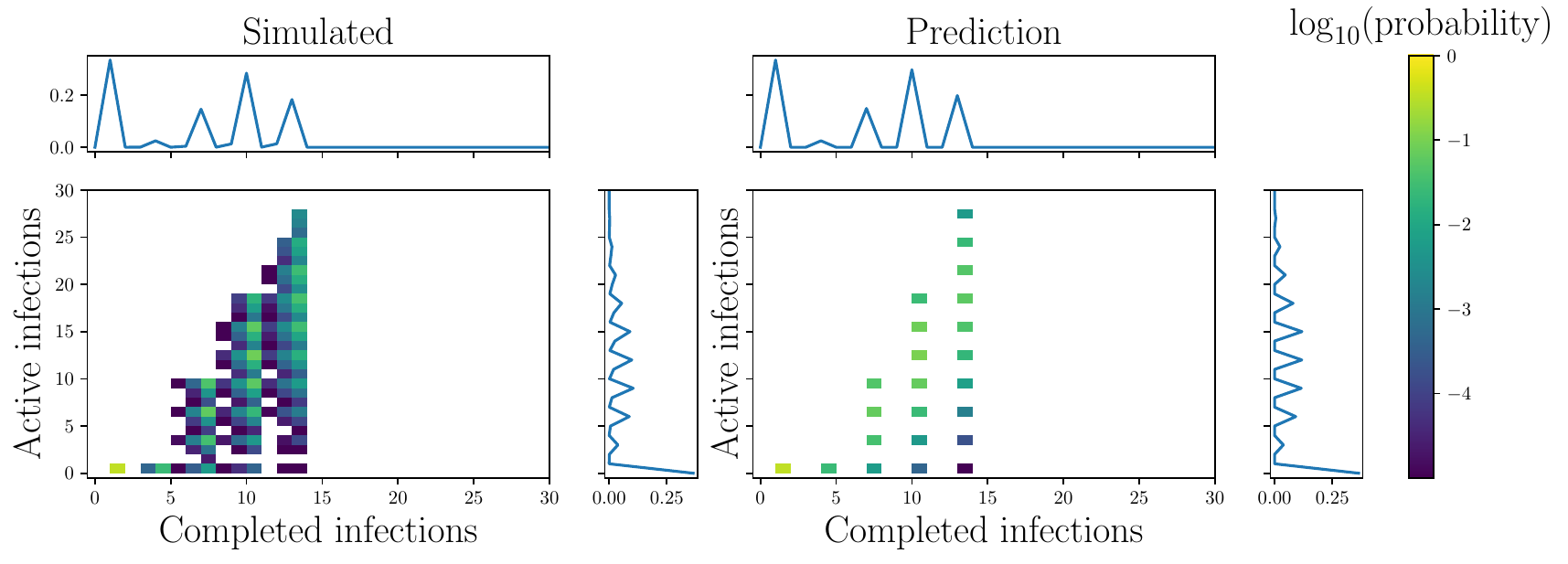}
\end{center}
\caption{\textbf{Illustration of Theorem~\ref{thm:joint_size}.}
  Comparison of predictions and simulations for the joint distribution
  of the number of current and completed infections at generation
  $g=3$. The predictions were calculated using
  Property~\ref{prop:cauchy_coefficient}.  \textbf{Left:} simulations
  from Fig.~\ref{fig:basic_observations} for $N=1000$ and
  \textbf{Right:} predictions (note vertical scales on left and right
  are the same).  \textbf{Top to Bottom:}
  Poisson $\Ro = 0.75$, \ Bimodal $\Ro = 0.75$, \ Poisson $\Ro = 2$,
  and Bimodal $\Ro = 2$. The predictions match our observations, with
  some difference for two reasons: 1) because $5\times10^5$
  simulations cannot resolve events with probabilities as small as $10^{-12}$, but the PGF
  approach can, and 2) due
  to finite-size effects as occasionally an individual receives multiple transmissions even early on.  The plots also show the marginal distributions, matching Figs.~\ref{fig:current_size} and~\ref{fig:cum_size}.}
\label{fig:joint_size}
\end{figure}
\begin{example}
We demonstrate Theorem~\ref{thm:joint_size} in Fig.~\ref{fig:joint_size}, using the same simulations as in Example~\ref{example:model_description}.  Simulations and predictions are in excellent agreement.

\end{example}


\ifsolns
\else
\subsection{Small outbreak final size distribution}
\fi
\label{sec:discrete_small_final_size}

There are many diseases for which there have been multiple small
outbreaks in recent years but no large-scale epidemics (such as Nipah,
H5N1 avian influenza, Pneumonic Plague, Monkey pox, and --- prior to
2013 --- Ebola).  A natural question emerges: what can we infer about
the epidemic potential of these diseases?    The size distribution may
help us to infer properties of the disease and in particular to
estimate the probability that $\Ro>1$~\cite{blumberg2013inference,kucharski2015characterizing,nishiura2012estimating}.

We have found that $\Omega_g(z)$ gives the PGF for the number of completed infections by generation $g$.  We noted earlier that for a given $r$, once $g > r$, the coefficient of $z^r$ in $\Omega_g(z)$ is fixed and equal to the probability that the outbreak goes extinct after exactly $r$ infections.  Motivated by this, we look for the limit as $g \to \infty$.
We define 
\[
\Omega_\infty(z) = \lim_{g\to\infty} \Omega_g(z)
\]
We expect this to be the PGF for the final size of the outbreaks.

We can express the pointwise limit\footnote{Although this converges
  for any given $z$ in $[0,1]$, it does not do so ``uniformly'' if
  $\Ro>1$.  That is, for $\Ro >1$ no matter how large $g$ is, there
  are always some values of $z<1$,  but sufficiently close to $1$, which are far from converged.} as
\[
\Omega_\infty(z) = \sum_r \omega_r z^r + \omega_\infty z^\infty 
\]
where for $r<\infty$ the coefficient $\omega_r$ is the probability an outbreak causes exactly $r$ infections in an infinite population.  We use $\omega_\infty$ to denote the probability that the outbreak is infinite in an infinite population (i.e., that it is an epidemic), and we interpret $z^\infty$ as $1$ when $z=1$ and $0$ for $0\leq z<1$.  So if epidemics are possible, $\Omega_\infty(z)$ has a discontinuity at $z=1$, and the limit as $z \to 1$ from below gives $\sum_{r<\infty} \omega_r = 1-\omega_\infty$ which is the extinction probability $\alpha$.

We now look for a recurrence relation for $\Omega_\infty(z)$ in the infinite population limit.  Each offspring of the initial infection independently causes a set of infections. The distribution of the these new infections (including the original offspring) also has PGF $\Omega(z)$.  So the distribution of the number of descendants of the initial infection (but not including the initial infection) has PGF $\mu(\Omega_\infty(z))$.  To include the initial infection, we must increase the exponent of $z$ by one, which we do by multiplying by $z$. We conclude that $\Omega_\infty(z) = z\mu(\Omega_\infty(z))$.  Although we have shown that $\Omega_\infty(z)$ solves $f(z) = z \mu(f(z))$, we have not shown that there is only one function that solves this.


We may be interested in the outbreak size distribution conditional on the outbreak going extinct.  For this we are looking at $\Omega_\infty(z)/\alpha$ for any $z<1$, and at $z=1$, this is simply $1$.  Note that if $\Ro<1$ then $\alpha=1$.

Summarizing this we have
\begin{thm}
\label{thm:discrete}
Given a single initial infection in an infinite population, consider $\Omega_\infty(z)$, the PGF for the final size distribution: $\Omega_\infty(z)=\left(\sum_{r<\infty} \omega_r z^r \right)+ \omega_\infty z^\infty$ where $z^\infty=0$ if $|z|<1$ and $1$ if $|z|=1$.
\begin{itemize}
\item Then
\begin{equation}
\Omega_\infty(z) = \begin{cases}
    z \mu(\Omega_\infty(z)) & z \neq 1\\
    1 & z=1
    \end{cases} \, .
\end{equation}

\item We have $\lim_{z\to1^{-}} \Omega_\infty(z)=\alpha=1-\omega_\infty$.  If $\Ro>1$ then $\Omega_\infty(z)$ is discontinuous at $z=1$, with a jump discontinuity of $\omega_\infty$, the probability of an epidemic.

\item The PGF for outbreak size distribution conditional on the outbreak being finite  is
\[
\begin{cases}
\Omega_\infty(z)/\alpha & 0<z<1\\
1 & z=1
\end{cases}
\]
\end{itemize}
\end{thm}

\begin{figure}
\begin{center}
\begin{tabular}{ccc}
& Poisson & Bimodal\\
$\Ro=0.75$ 
& \raisebox{-0.5\height}{
\includegraphics[width=0.4\textwidth]{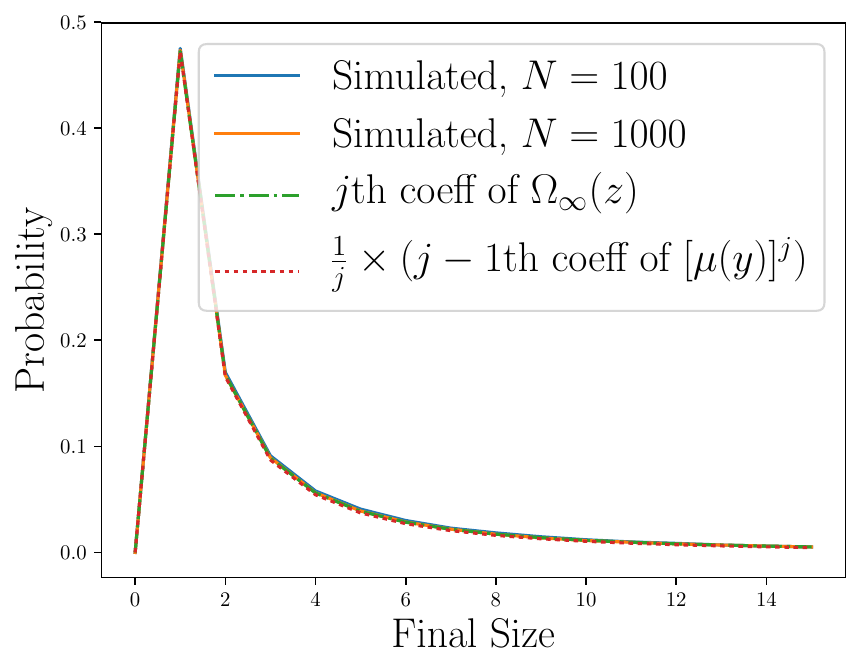}}
& \raisebox{-0.5\height}{\includegraphics[width=0.4\textwidth]{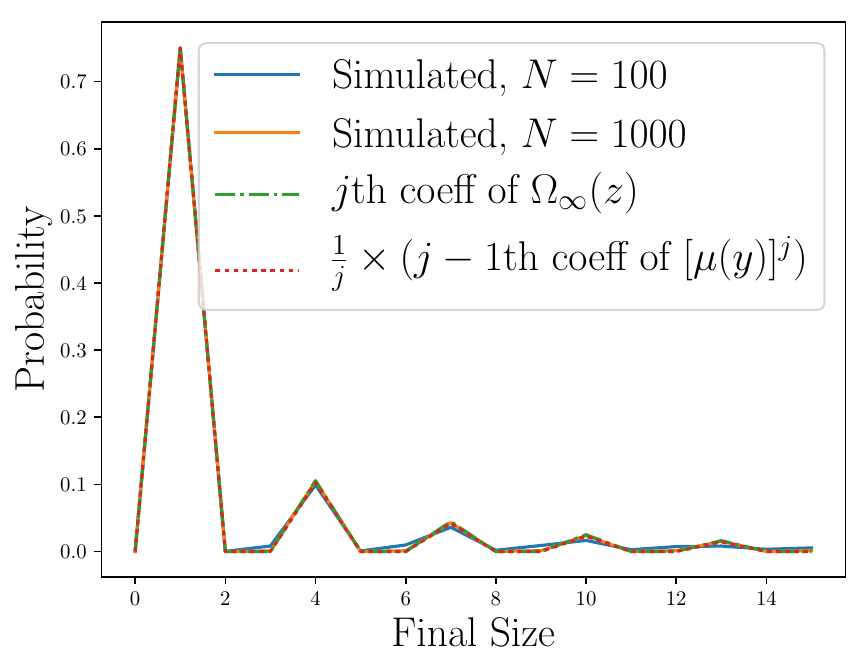}}\\
$\Ro=2$ 
& \raisebox{-0.5\height}{\includegraphics[width=0.4\textwidth]{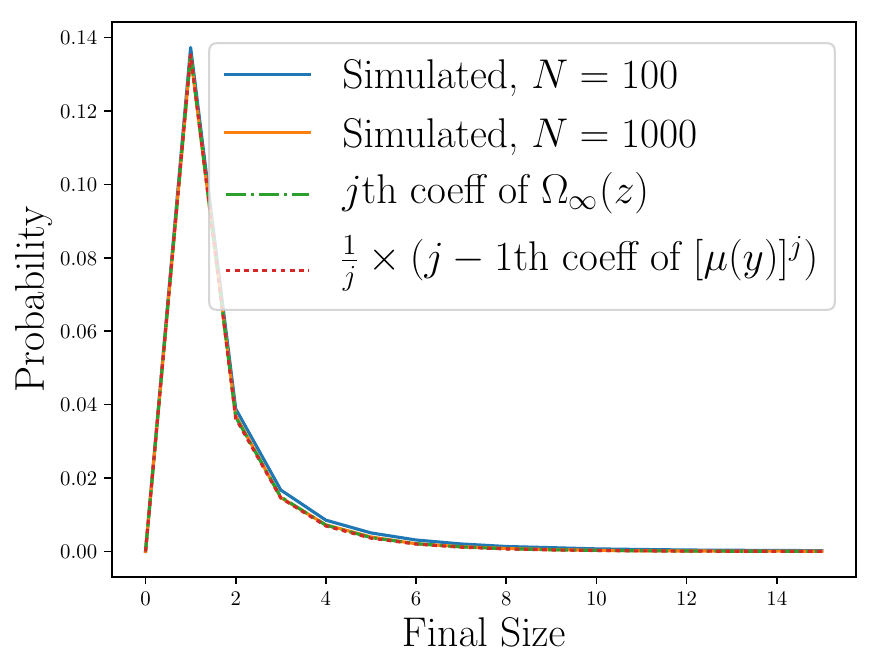}}
& \raisebox{-0.5\height}{\includegraphics[width=0.4\textwidth]{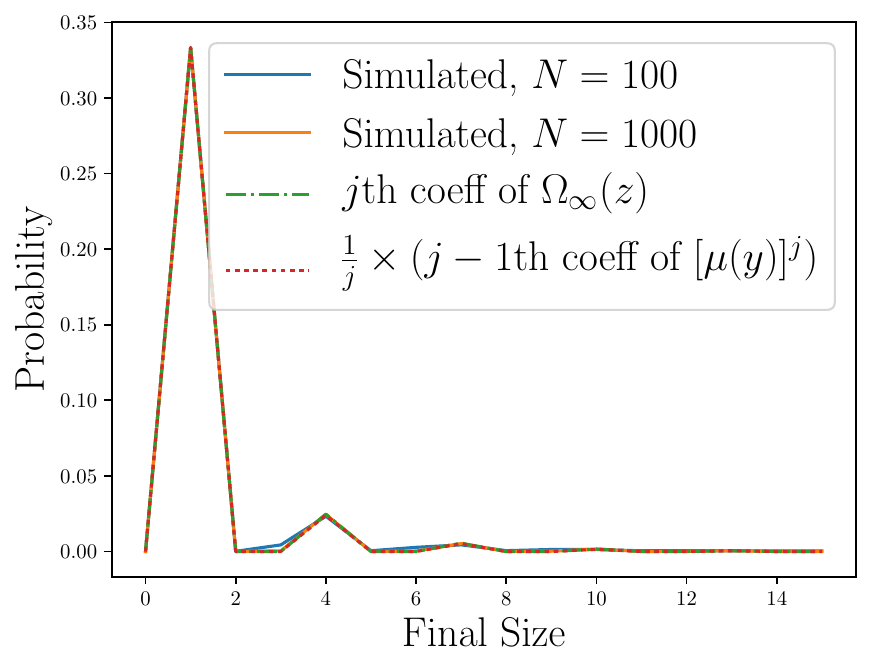}}
\end{tabular}
\end{center}
\caption{\textbf{Illustration of Theorems~\ref{thm:discrete} and~\ref{thm:power_magic}}. The final size of small outbreaks predicted by Theorem~\ref{thm:discrete} and by Theorem~\ref{thm:power_magic} as calculated using Property~\ref{prop:cauchy_coefficient} matches observations  from the simulations in Fig.~\ref{fig:basic_observations} (see also insets of Fig.~\ref{fig:basic_observations}).}
\label{fig:discrete}
\end{figure}

Perhaps surprisingly we can often find the coefficients of $\Omega_\infty(z)$ analytically if $\mu(y)$ is known.  We use a remarkable result showing that the probability of infecting exactly $n$ individuals is equal to the coefficient of $z^{n-1}$ in $[\mu(z)]^n$~\cite{blumberg2013inference,dwass1969total,van2008elementary,wendel1975left}.  The theorem is
\begin{thm}
\label{thm:power_magic}
Given an offspring distribution with PGF $\mu(y)$, for $j<\infty$ the coefficient of $z^j$ in $\Omega_\infty(z)$ is $\frac{1}{j} p_{j-1}^{(j)}$ where $[\mu(y)]^j = \sum_ip_{i}^{(j)} y^i$.

That is, the probability of having exactly $j<\infty$ infections in an outbreak starting from a single infection is $\frac{1}{j}$ times the coefficient of $y^{j-1}$ in $[\mu(y)]^j$.
\end{thm}

We prove this theorem in Appendix~\ref{app:power_magic}.  The proof is based on observing that if we draw a sequence of $j$ numbers from the offspring distribution, the probability they sum to $j-1$ (corresponding to $j-1$ transmissions and hence $j$ infected individuals including the index case) is the coefficient of $z^{j-1}$ in $[\mu(z)]^j$.  A fraction $1/j$ of these satisfy additional constraints needed to correspond to a valid transmission tree\footnote{If the index case causes 0 infections and its first offspring causes 1 infection, we have a sequence of two numbers that sum to 1, but it is biologically meaningless because it does not make sense to talk about the first offspring of an individual who causes no infections.} and thus the probability of a valid transmission tree with exactly $j-1$ transmissions  is $1/j$ times $p_{j-1}^{(j)}$.

Because the coefficient of $y^{j-1}$ in $[\mu(y)]^j$ is $\frac{1}{(j-1)!} \left(\diff{}{y}\right)^{j-1} \left. [\mu(y)]^j \right|_{y=0}$ (by Property~\ref{prop:coefficient}), we have that the probability of an outbreak of size $j$ is 
\[
\frac{1}{j!} \left. \left(\diff{}{y}\right)^{j-1}[\mu(y)]^j \right|_{y=0}
\]

It is enticing to think there may be a similar theorem for coefficients of $\Pi(y,z)$, but we are not aware of one.  The theorem has been generalized to models having multiple types of individuals~\cite{kucharski2015characterizing}.  

\begin{example}
We demonstrate Theorems~\ref{thm:discrete} and~\ref{thm:power_magic} in Fig.~\ref{fig:discrete}, using the same simulations as in Example~\ref{example:model_description}.  
\end{example}

\begin{example}
\label{example:neg_binom}
The PGF for the negative binomial distribution with parameters $p$ and $\hat{r}$ (with $q=1-p$) is 
\[
\mu(y) = \left( \frac{q}{1-py}\right)^{\hat{r}}
\]
We can rewrite this as
\[
\mu(y) = q^{\hat{r}}(1-py)^{-\hat{r}}
\]
We will use this to find the final size distribution.
We expand $[\mu(y)]^j=q^{\hat{r}j}(1-py)^{-\hat{r}j}$  using the
binomial series 
\[
(1+\delta)^\eta = 1+ \eta \delta + \frac{\eta(\eta -1)}{2!} \delta^2 + \cdots +
\frac{\eta(\eta-1)\cdots(\eta-i+1)}{i!}\delta^i + \cdots
\]
which holds for integer or non-integer $\eta$.  Then with $-py$,
$-\hat{r}j$, and $j-1$ playing the role of $\delta$, $\eta$, and $i$:
\begin{align*}
[\mu(y)]^j &= q^{\hat{r}j}(1-py)^{-\hat{r}j}\\
&= q^{\hat{r}j}\left (1 +\hat{r}j p y +
\frac{\hat{r}j(\hat{r}j+1)}{2!}p^2y^2 + \cdots + \frac{\hat{r}j(\hat{r}j+1)\cdots(\hat{r}j+j-2)}{(j-1)!} p^{j-1}y^{j-1}+\cdots\right)
\end{align*}
[the negatives all cancel].
So the coefficient of $y^{j-1}$ is $q^{\hat{r}j} p^{j-1}
\frac{(\hat{r}j+j-2)!}{(\hat{r}j-1)!(j-1)!} = \binom{\hat{r}j+j-2}{j-1}q^{\hat{r}j}p^{j-1}$ (assuming $\hat{r}$ is an integer).
Looking at $1/j$ times this, we conclude that the
probability an outbreak infects exactly $j$ individuals is
\[
\frac{1}{j}
\binom{\hat{r}j+j-2}{j-1}q^{\hat{r}j}p^{j-1}
\]
A variation of this result for non-integer $\hat{r}$ is commonly used
in work estimating disease parameters~\cite{blumberg2013inference,nishiura2012estimating}. Exercise~\ref{exercise:neg_binom} generalizes the formula for this.
\end{example}

Applying Theorem~\ref{thm:power_magic} to several different families
of distributions yields Table~\ref{table:final_probs} for the
probability of a final size $j$.

\ifsolns
\else
\subsubsection{Inference based on outbreak sizes}
\fi
\label{sec:inference}
A major challenge in infectious disease modeling is inferring
parameters of an infectious disease.  In
Section~\ref{sec:discrete_small_final_size} we alluded to the use of
PGFs to infer disease properties from observations of the size
distribution of small outbreaks.  In this section we describe how to
do this using a Bayesian approach, using the probabilities given in
Table~\ref{table:final_probs}.  A number of researchers have used this
approach to estimate disease parameters~\cite{blumberg2013inference,nishiura2012estimating,kucharski2015characterizing}

We assume that we know what type of distribution the offspring
distribution, but that there are some unknown parameters.  
We also assume that we have some prior belief about the probability of various
parameters.  For practical purposes, we will assume that we have some
finite number of possible parameter values, each with a probability.

We use Bayes' Theorem~\cite{hoff2009first}:
\begin{equation}
\label{eqn:bayes}
P(\Theta|X) = \frac{P(\Theta,X)}{P(X)} = \frac{P(X|\Theta) P(\Theta)}{P(X)}
\end{equation}
Here we think of $\Theta$ as the specific parameter values and $X$ as
the observed data (typically the observed size of an outbreak or sizes
of multiple independent outbreaks, in which case $P(X|\Theta)$ comes
from Theorem~\ref{thm:power_magic} or Table~\ref{table:final_probs}).
In our calculations we can simply use the fact that $P(\Theta|X)
\propto P(X|\Theta) P(\Theta)$ with a normalization constant which can
be dealt with at the end.

The \emph{prior} for $\Theta$ is the probability distribution we assume for
the parameter values before observing the data, given by $P(\Theta)$.
We often simply assume that all parameter values are equally probable initially.

The \emph{likelihood} of the parameters $\Theta$ is defined to be $P(X|\Theta)$, the
probability that we would observe $X$ for the given parameter values.
If we are choosing between two sets of parameter values $\Theta_1$ and $\Theta_2$ and
the observations have consistently higher likelihood for $\Theta_2$, then
we intuitively expect that $\Theta_2$ is the more probable parameter value.

In practice the likelihood may be very small which can lead to
numerical error.  It is often useful to
instead look at \emph{log-likelihood}\footnote{Throughout this section, we assume that
  $\log$ is taken with base $e$.}, $\log P(X|\Theta)$.  For example, if
we have many observed outbreak sizes, the likelihood $P(X|\Theta)$ under
independence is the
product of the probabilities of each individual outbreak size.  The
likelihood is thus quite small (perhaps less than machine precision),
while the log-likelihood is simply the sum of the
log-likelihoods  of
each individual observation.

We know that
\[
\log P(\Theta|X) - C= \log P(X|\Theta) + \log P(\Theta)
\]
where $C$ is the logarithm of the proportionality constant $1/P(X)$ in
Equation~\eqref{eqn:bayes}. If we have
a prior and the likelihood, the right hand side can be calculated.  It
is often possible (and advisable) to calculate the log likelihood $\log P(X|\Theta)$ 
directly rather than calculating $P(X|\Theta)$ and then taking the
logarithm.

Exponentiating the right hand side and then finding the appropriate
normalization constant will yield $P(\Theta|X)$.  Numerically the numbers
may be very small when we exponentiate, so to prior to exponentiating
it is advisable to add a constant value to all of the expressions.
This constant is corrected for in the final normalization step.

We now provide the steps for a numerical calculation of $P(\Theta|X)$ given
the prior $P(\Theta)$, the observations $X$, and the log likelihood
$\log P(X|\Theta)$.

\begin{enumerate}
\item For each $\Theta$, calculate $f(\Theta) = \log P(X|\Theta) + \log P(\Theta)$.
\item Find the maximum $X_{\text{max}}$ over all $\Theta$ and subtract it
  to yield $\hat{f}(\Theta) = \log P(X|\Theta) + \log P(\Theta) - X_{\text{max}}$.
  Note that $X_{\text{max}}\leq 0$, and this brings all of our numbers
  closer to zero.
\item Calculate $g(\Theta) = e^{\hat{f}(\Theta)}$.  This will be
  proportional to $P(\Theta|X)$.  Note that by using $e^{\hat{f}(\Theta)}$ rather
  than $e^{f(\Theta)}$ we have reduced the impact of roundoff error.
\item Find the normalization constant $\sum_{\Theta'} g(\Theta')$.  Then
\[
P(\Theta|X) = \frac{g(\Theta)}{\sum_{\Theta'} g(\Theta')}
\]
\end{enumerate}
Note that if $\Theta$ comes from a continuous distribution rather than
a discrete distribution, then the same approach works, except that $P$
is a probability density and the summation in the final step becomes
an integral.
\begin{example}
\label{example:inference}
  A frequent assumption is that the offspring distribution is negative
  binomial.  Let us make this assumption with unknown $p$ and $\hat{r}$.  

  To artificially simplify the problem, we assume that we know that
  there are only two possible pairs of $\Theta=(p, \hat{r})$, namely
  $\Theta_1=(p_1,\hat{r}_1)=(0.02,40)$ or $\Theta_2=(p_2,\hat{r}_2)=(0.03, 20)$, and that our \emph{a priori} belief is
  that they are equally probable.

After observing $2$ independent outbreaks, with total sizes $j_1=8$
  and $j_2=7$, we want to use our observations to update $P(\Theta)$.

  From Table~\ref{table:final_probs}, the likelihood of a given
  $\Theta$ given the two independent observations is
\begin{align*}
f(\Theta) &= \left(\log 
\prod_{j=7,8}\frac{1}{j} \binom{\hat{r}j+j-2}{j-1}q^{\hat{r}j}
p^{j-1}\right) + \log 0.5\\
&= \left( \sum_{j=7,8} \log \frac{1}{j}
  \binom{\hat{r}j+j-2}{j-1}q^{\hat{r}j}p^{j-1}\right) + \log 0.5\\
&= \left(\sum_{j=7,8} \log ((\hat{r}j+j-2)!) - \log(j!) - \log
  ((\hat{r}j-1)!) +\hat{r}j \log q + (j-1) \log p\right) + \log 0.5
\end{align*}
In problems like this, we will often encounter logarithms of
factorials.  Many programming languages provide this, typically using
Stirling's approximation.  For example,  Python, R, and \CC{} all have
a special function \texttt{lgamma} which calculates the natural log of
the absolute value of the gamma function\footnote{The Gamma function
  is an analytic function that satisfies $\Gamma(n) = (n+1)!$ for
  positive integer values so to calculate $\log (n!)$ we use $\texttt{lgamma}(n+1)$.}  We find
\begin{align*}
f(\Theta_1) &\approx -8.495\\
f(\Theta_2) &\approx -9.135 
\end{align*}
So $\hat{f}(\Theta_1)  = 0$ and $\hat{f}(\Theta_2) \approx -0.640$.
Exponentiating, we have
\begin{align*}
g(\Theta_1) & = 1\\
g(\Theta_2) &\approx 0.5277
\end{align*}
So now
\begin{align*}
P(\Theta_1|X) &\approx \frac{1}{1.5277} \approx 0.6546 \\
P(\Theta_2|X) &\approx \frac{0.5277}{1.5277} \approx 0.3454 
\end{align*}
So rather than the two parameter sets
being equally probable, $\Theta_2$ is now about half as likely as $\Theta_1$
given the observed data.
\end{example}

\ifsolns
\else
\subsection{Generality of discrete-time results}
\fi
\label{sec:discrete_general}

Thus far we have measured time in generations.  However, many models measure time differently and different generations may overlap.  For both SIS and SIR disease, our results above about final size distribution or extinction probability still apply.  To see this, we note first that our results have been derived assuming that the population is infinite and well-mixed so no individuals receive multiple transmissions.  Regardless of the clock time associated with transmission and recovery, there is still a clear definition of the length of the transmission chain to an infected individual.  Once we group individuals by length of the transmission chain, we get the generation-based model used above.  This equivalence is studied more in~\cite{yan2008distribution,ludwig}.


\ifsolns
\else
\subsection{Exercises}
\fi
\begin{exercise}
\label{exercise:alpha_g}
\textbf{Monotonicity of $\alpha_g$}
\begin{myenumerate}
\item By considering the biological interpretation of $\alpha_g$,
  explain why the sequence of inequalities $0=\alpha_0 \leq \alpha_1
  \leq \cdots \leq 1$ should hold.  That is, explain why $\alpha_0=0$,
  why the $\alpha_i$ form a monotonically increasing sequence, and why
  all of them are at most $1$.
\item Show that $\alpha_g$ therefore converges to some non-negative limit
  $\alpha$ that is at most $1$ and that $\alpha = \mu(\alpha)$.
\item Use Property~\ref{property:PGFcobweb} to show that if $\mu(0) \neq 0$ there exists a unique
  $\alpha<1$ solving $\alpha = \mu(\alpha)$ if and only if $\Ro = \mu'(1)>1$.
\item Assuming $\mu(0)\neq 0$, use Property~\ref{property:PGFcobweb} to show that if $\Ro>1$ then $\alpha_g$ converges to the unique $\alpha<1$ solving $\alpha=\mu(\alpha)$, and otherwise $\alpha_g$ converges to $1$.
\end{myenumerate}
\end{exercise}
\begin{solution}\mbox{}
\begin{myenumerate}
\item In generation $0$ there is an infected individual.  So the disease is not extinct.  Thus $\alpha_0=0$.  If the outbreak is extinct at generation $g$, it remains extinct at later generations.  So $\alpha_g \leq \alpha_{g-1}$.  The eventual extinction probability is at most $1$.
\item Any sequence of numbers that is increasing and bounded from above must have a limit that is at most that bound.  So it converges to some limit $\alpha$.  Since $\mu(\alpha_g) = \alpha_{g+1}$, we conclude that $\mu(\alpha_g)$ must converge to the same limit.  Thus since $\mu$ is continuous, $\mu(\alpha)=\alpha$.
\item The assumptions of Property~\ref{property:PGFcobweb} hold, with $\Ro$ playing the role of $f'(1)$.  The property states that if $\Ro\leq 1$ the only solution is $\alpha=1$, while if $\Ro>1$ there is exactly one other solution in $(0,1)$.
\item The second part of Property~\ref{property:PGFcobweb} states that for $\Ro>1$, the solution must converge to this unique $\alpha<1$.  If $\Ro \leq 1$, the above observations show that it must converge to a solution to $\alpha = \mu(\alpha)$, and the only possible choice is $\alpha=1$.
\end{myenumerate}
\end{solution}

\begin{exercise}
Use Theorem~\ref{thm:strong_extinction} to prove Theorem~\ref{thm:weak_extinction}.
\end{exercise}

\begin{solution}
We only need the first two parts of Theorem~\ref{thm:strong_extinction}.  We have $\alpha = \lim_{g\to\infty}\alpha_g = \lim_{g\to\infty}\mu(\alpha_g)$.  Because $\mu$ is continuous we have $\lim_{g\to\infty} \mu(\alpha_g) = \mu(\lim_{g\to\infty}\alpha_g) = \mu(\alpha)$.  

So $\alpha = \mu(\alpha)$.
\end{solution}

\begin{exercise}
\label{exercise:mu0eq0}
Show that if $\mu(0)=0$, then $\lim_{g\to\infty} \alpha_g = 0$.  By
referring to the biological interpretation of $\mu(0)=0$, explain this result.
\end{exercise}
\begin{solution}
Let $\mu(0)=0$ and consider the smallest $g\geq 0$ such that $\alpha_{g+1} \neq 0$ (if it exists).  Then $\alpha_{g+1} = \mu(\alpha_g)=\mu(0)=0$.  So no such $g$ exists.

Biologically, $\mu(0)=0$ means that every individual has at least one offspring.  Thus if we start with one individual, we can never end up with zero.
\end{solution}

\begin{exercise}
Find all PGFs $\mu(y)$ with $\Ro \leq 1$ and $\mu(0)=0$.  Why were these excluded from Theorem~\ref{thm:strong_extinction}?
\end{exercise}
\begin{solution}
If $\mu(y) = \sum_{i=0}^\infty p_i x^i$ is a PGF and $\mu(0)=0$, then $p_0=0$ and $\mu(y) = \sum_{i=1}^\infty p_i x^i$ so $\Ro=\mu'(1) = \sum_{i=1}^\infty i p_i$.  This is $\geq \sum_{i=1}^\infty p_i=1$, with equality only if $p_i=0$ for $i>1$.  Thus the only such function is $\mu(y)=y$.

This corresponds to each individual having exactly one offspring.  So starting with one infection, at each generation there will remain exactly one infection, and there is no chance for extinction.  All other cases with $\Ro \leq 1$ have a nonzero chance of having zero offspring, and thus extinction is possible (and in fact inevitable).
\end{solution}

\begin{exercise}
\textbf{Larger initial conditions}

Assume that disease is introduced with $m$ infections rather than just $1$, or that it is not observed by surveillance until $m$ infections are present.  Assume that the offspring distribution PGF is $\mu(y)$.
\begin{myenumerate}
\item If $m$ is known, find the extinction probability.
\item If $m$ is unknown but its distribution has PGF $h(y)$, find the extinction probability.
\end{myenumerate}
\end{exercise}
\begin{solution}
Let $\alpha$ be the extinction probability from one individual.
\begin{myenumerate}
\item The extinction probability given $m$ initial infections is $\alpha^m$.  
\item Let $h(x) = \sum_m q_m x^m$.  The probability of extinction is $\sum_m q_m \alpha^m = h(\alpha)$.
\end{myenumerate}
\end{solution}

\begin{exercise}
\textbf{Extinction probability}

Consider a disease in which $p_0 = 0.1$, \ $p_1 = 0.2$, \ $p_2 =
0.65$, and $p_3=0.05$ with a single introduced infection.
\begin{myenumerate}
\item Numerically approximate the probability of extinction within $0$, $1$, $2$, $3$, $4$, or $5$ generations up to five significant digits (assuming an infinite population).
\item Numerically approximate the probability of eventual extinction up to five significant digits (assuming an infinite population).
\item A surveillance program is being introduced, and detection will
  lead to a response.  But it will not be soon enough to affect the
  transmissions from generations $0$ and $1$.  From then on
  $p_0=0.3$, \ $p_1= 0.4$, \ $p_2 = 0.3$, and $p_3=0$.   Numerically approximate the
  new probability of eventual extinction after an introduction in an unbounded population [be careful
  that you do the function composition in the right order -- review Properties~\ref{property:notclothesline} and~\ref{prop:composition}].
\end{myenumerate}
\end{exercise}
\begin{solution}
Define $\mu_1(y) = \sum_i p_i y^i = 0.1 + 0.2y + 0.65y^2+0.05y^3$.
\begin{myenumerate}
\item The extinction probability after $0$ generations is $\alpha_0=0$ and for $g>0$, the extinction probability after $g$ generations is $\alpha_g=\mu_1(\alpha_{g-1})$.  So by iteratively applying $\mu_1$ to $0$ we get
\begin{align*}
\alpha_0&=0\\
\alpha_1 &= \mu_1(\alpha_0) = 0.1\\
\alpha_2 &= \mu_1(\alpha_1) = 0.12655\\
\alpha_3 &= \mu_1(\alpha_2) \approx 0.13582\\
\alpha_4 &= \mu_1(\alpha_3) \approx 0.13928\\
\alpha_5 &= \mu_1(\alpha_4) \approx 0.14060
\end{align*}
\item Repeatedly applying $\mu_1$ to the result quickly yields convergence to $\alpha \approx 0.14143$.
\item Let $\mu_2(y) = 0.3+0.4y+0.3y^2$.  Assuming that there are still infected individuals when the intervention is introduced.  For each of them, the probability that all offspring die out is $\lim_{g\to\infty} \mu_2^{[g]}(0)\approx 0.9688$.

The distribution of the number infected when the intervention is introduced is $\mu_1(\mu_1(y))$.  So the probability of extinction is $\mu_1(\mu_1(0.9688)) \approx 0.9185$.
\end{myenumerate}
\end{solution}

\begin{exercise}
\label{exercise:discrete_backward_and_forward}
We look at two inductive derivations of $\Phi_g(y) = \mu^{[g]}(y)$.  They are similar, but when adapted to the continuous-time dynamics we study later, they lead to two different models.  We take as given that $\Phi_{g-1}(y)$ gives the distribution of the number of infections caused after $g-1$ generations starting from a single case.  One argument is based on discussing the results of outcomes attributable to the infectious individuals of generation $g-1$ in the next generation.  The other is based on the outcomes indirectly attributable to the infectious individuals of generation $1$ through their descendants after another $g-1$ generations.
\begin{myenumerate}
\item Explain why Property~\ref{prop:composition} shows that $\Phi_g(y) = \Phi_{g-1}(\mu(y))$.\label{item:forward_derivation}
\item (without reference to~\ref{item:forward_derivation}) Explain why Property~\ref{prop:composition} shows that $\Phi_g(y) = \mu(\Phi_{g-1}(y))$.
\end{myenumerate}
\end{exercise}
\begin{solution}
\begin{myenumerate}
\item The PGF for the number infected in generation $g-1$ is
  $\Phi_{g-1}(y)$.  Each of the individuals who are infected in
  generation $g-1$ make some sort of contribution (possibly 0) to
  generation $g$.  The PGF for this contribution is $\mu(y)$.  Thus by
  Property~\ref{prop:composition}, the PGF for the combination of
  steps from generation $0$ to $g-1$ and from $g-1$ to $g$ is given by
  $\Phi_{g-1}(\mu(y))$.
\item The PGF for the number infected in generation $1$ is $\mu(y)$.
  If we trace the contributions of these individuals to generation
  $g$, we find that the distribution is the same as for the
  contribution of a generation $0$ individual to generation $g-1$.  So
  by Property~\ref{prop:composition}, the PGF for the  combination of steps from
  generation $0$ to $1$ and from $1$ to $g$ is $\mu(\Phi_{g-1}(y))$.
\end{myenumerate}
\end{solution}

\begin{exercise}
Use Theorem~\ref{thm:generation_size} to prove the first part of Theorem~\ref{thm:strong_extinction}.
\end{exercise}
\begin{solution}
The probability that there are no infections at generation $g$ is
given by $\Phi_g(0)$ (The expansion eliminates all terms except the
coefficient of $y^0$).

Simply observing $\Phi_g(0) = \mu^{[g]}(0)$ finishes the proof.
\end{solution}

\begin{exercise}
How does Corollary~\ref{cor:early_expectation} change if we start with $k$ infections?
\end{exercise}
\begin{solution}
The first part is simply multiplied by $k$.

For  $\langle I \rangle_g$, the numerator is multiplied by $k$.  The
denominator is the probability that the disease persists to generation
$g$.  This becomes $1-\alpha_g^k$.  This is because to be extinct, all $k$
introductions must independently go extinct, which occurs with
probability $\alpha_g^k$.
\end{solution}

\begin{exercise}
Assume the PGF of the offspring size distribution is $\mu(y) = (1+y +y^2)/3$.
\begin{myenumerate}
\item What  offspring size distribution yields this PGF?
\item Find the PGF $\Omega_g(z)$ for the number of completed infections at $0$, $1$, $2$, $3$, and $4$ generations [it may be helpful to use a symbolic math program once $g>2$.].
\item Check that for these cases, once $g > r$, the coefficient of $z^r$ does not change.
\end{myenumerate}
\end{exercise}
\begin{solution}\mbox{}
\begin{myenumerate}
\item Each individual causes 0, 1, or 2 transmissions with equal
  probability.
\item \begin{enumerate}
\renewcommand\labelenumii{(\roman{enumii})}
\renewcommand\theenumii\labelenumii
\item We have $\Omega_0(z)=1$ [there are no completed infections at generation $0$].
\item We have $\Omega_1(z)=z\mu(1) = z$ [which means that there is a single completed infection at generation $1$].
\item We have
\begin{align*}
\Omega_2(z)&= z\mu(\Omega_1(z))\\
&= z\mu(z)\\
&= \frac{z+z^2+z^3}{3}
\end{align*}
\item We have 
\begin{align*}
\Omega_3(z) &= z \mu(\Omega_2(z))\\
&= z \frac{1+\Omega_2(z) + \Omega_2(z)^2}{3}\\
&= \frac{9z + 3z^2 + 4 z^3  + 5 z^4+ 3z^5 + 2 z^6 +z^7 }{27}\\
&= \frac{z}{3} + \frac{z^2}{9} + \cdots
\end{align*}
\item and finally
\begin{align*}
\Omega_4(z) &= z\mu(\Omega_3(z))\\
&= z \frac{1+\Omega_3(z) + \Omega_3(z)^2 + }{3}\\
&= 
\frac{1}{2187} \big(729z  + 243z^2 + 162 z^3 +  162 z^4 +  216 z^5 +
  105 z^6 +  154 z^7\\
&\qquad\qquad +  121 z^8 +  79 z^9 +  52 z^{10} +  37 z^{11} +  22 z^{12} +  10 z^{13} +  4 z^{14} +  z^{15}\big)\\
&= \frac{z}{3} + \frac{z^2}{9} + \cdots
\end{align*}
\end{enumerate}
\item 
\begin{itemize}
\item For $r=0$, we see that from $\Omega_1(z)$ onwards, the coefficient of $z^0$ is $0$.
\item For $r=1$, we see that from $\Omega_2(z)$ onwards, the coefficient of $z$ is $1/3$
\item For $r=2$, we see that from $\Omega_3(z)$ onwards, the coefficient of $z^2$ is $1/9$.
\end{itemize}
\end{myenumerate}
\end{solution}

\begin{exercise}\mbox{}
By setting $y=1$, use Theorem~\ref{thm:joint_size} to prove Theorem~\ref{thm:cum_size}.  
\end{exercise}
\begin{solution}
We expect that $\Pi_g(1,z) = \Omega(z)$.  This can be checked by
noting that for given $r$, the sum $\sum_i \pi_{i,r}(g) 1^i z^r$ is
$z^r \sum_i \pi_{i,r}(g) = z^r\omega_r(g)$.

So Theorem~\ref{thm:joint_size} states that $\Pi_g(1,z) = z
\mu(\Pi_{g-1}(1,z))$ which becomes $\Omega_g(z) = z \mu(\Omega_{g-1}(z))$.
\end{solution}

\begin{exercise}
\label{exercise:neg_binom}
Redo example~\ref{example:neg_binom} if $\hat{r}$ is a real number, rather than an integer.  It may be useful to use the $\Gamma$--function, which satisfies $\Gamma(x+1)=x\Gamma(x)$ for any $x$ and $\Gamma(n+1)=n!$ for integer $n$.
\end{exercise}
\begin{solution}
The key observation is that $\hat{r}j(\hat{r}j+1) \cdots (\hat{r}j + j
- 2)/(j-1)!$ becomes $\Gamma(\hat{r}j+j-1)/\Gamma(\hat{r}j-1)
(j-1)!$.  Thus we replace $\binom{\hat{r}j + j -2}{j-1}$ in the
expression to yield
\[
\frac{1}{j} \frac{\Gamma(\hat{r}j+j-2)}{\Gamma(\hat{r}j-1)(j-1)!}
q^{\hat{r}j} p^{j-1}
\]
\end{solution}

\begin{exercise}
\label{exercise:final_probs}
Except for the negative binomial case done in example~\ref{example:neg_binom}, derive the probabilities in Table~\ref{table:final_probs}.
\begin{myenumerate}
\item For the Poisson distribution, use Property~\ref{prop:coefficient}.
\item For the Uniform distribution, use Property~\ref{prop:coefficient}.
\item For the Binomial distribution, use the binomial theorem: $(a+b)^c= \sum_{i=0}^c \binom{c}{i} a^ib^{c-i}$.
\item For the Geometric distribution, follow example~\ref{example:neg_binom} (noting that $p$ and $q$ interchange roles).
\end{myenumerate}
\end{exercise}
\begin{solution}
For each we need to find the coefficient of $y^{j-1}$ in $[\mu(y)]^j$ and then divide it by $j$.  
\begin{myenumerate}
\item $[\mu(y)]^j = e^{j\lambda(y-1)}$  The coefficient of $y^{j-1}$ is 
\begin{align*}
\left. \frac{1}{(j-1)!} \left(\diff{}{y}\right)^{j-1} e^{j\lambda(y-1)}\right|_{y=0} &= \left. \frac{1}{j-1} (j\lambda)^{j-1} e^{j\lambda(y-1)} \right|_{y=0}\\
&= \frac{(j\lambda)^{j-1}}{(j-1)!} e^{-\lambda j}
\end{align*}
Taking $1/j$ times this yields $\frac{(\lambda j)^{j-1}}{j!} e^{-\lambda j}$.
\item Note that for this to make sense $\lambda$ must be a non-negative integer.  We have $[\mu(y)]^j = y^{\lambda j}$.  The first case we consider is $j=1$, \ $\lambda = 0$.  Then we have $[\mu(y)]^j = 1$.  Taking zero derivatives, setting $y=0$, and dividing by $(j-1)! =1$ and $j=1$ yields $1$.

Now consider $\lambda\geq 1$.  Then $\lambda j>j-1$.  So after taking $j-1$ derivatives, we still have a factor of $y$, which when we set $y=0$ yields $0$.

Now consider $\lambda = 0$, but $j>1$.  We have $[\mu(y)]^j = 1$.  Taking $j-1\geq 1$ derivatives yields $0$.

\item We have $[\mu(y)]^j = (q+py)^{jn}$.  By the binomial theorem, the coefficient of $y^{j-1}$ is $\binom{jn}{j-1} q^{jn-j+1}p^{j-1}$.  Taking $1/j$ times this yields the result
\item We have $[\mu(y)]^j = \left(p/(1-qy)\right)^j$.  Following the same steps as in the negative binomial distribution, but taking $\hat{r}=1$ and interchanging $p$ and $q$, we end up with $\frac{1}{j} \binom{2j-2}{j-1} p^jq^{j-1}$.
\end{myenumerate}
\end{solution}

\begin{exercise}
\label{exercise:discrete_two_variable}
To help model continuous-time epidemics, Section~\ref{sec:cts} will use a modified version of $\mu$, which in some contexts will be written as $\hat{\mu}(y,z)$.  To help motivate the use of two variables, we reconsider the discrete case.  We think of a recovery as an infected individual disappearing and giving birth to a recovered individual and a collection of infected individuals.  Look back at the discrete-time calculation of $\Omega_g$ and $\Pi_g$.  Define a two-variable version of $\mu$ as $\mu(y,z) = z\sum_i r_i y^i=z\mu(y)$.  
\begin{myenumerate}
\item What is the biological interpretation of $\mu(y,z)=z\mu(y)$? 
\item Rewrite the recursive relations for $\Omega_g$ 
using $\mu(y,z)$ rather than $\mu(y)$.
\item Rewrite the recursive relations for $\Pi_g$ 
using $\mu(y,z)$ rather than $\mu(y)$. 
\end{myenumerate}
The choice to use $\mu(y,z)$ versus $\mu(y)$ is purely a matter of convenience.
\end{exercise}
\begin{solution}\mbox{}
\begin{myenumerate}
\item After a generation, an individual contributes $\mu(y)$ new
  infections and $1$ new recovery.  This is captured by $\mu(y,z) = z \mu(y)$.
\item $\Omega_g(z) = \mu(\Omega_{g-1}(z),z)$.
\item $\Pi_g(y,z) = \mu(\Pi_{g-1}(y,z),z)$.
\end{myenumerate}
\end{solution}

\begin{exercise}
\label{exercise:inference}
Consider Example~\ref{example:inference}.  Assume that a third
outbreak is observed with $4$ infections. Calculate the probability of
$\Theta_1$ and $\Theta_2$ given the data starting
\begin{myenumerate}
\item with the assumption that $P(\Theta_1) = P(\Theta_2)=0.5$ and $X$
  consists of the three observations $j=7$, \ $j=8$, and $j=4$.
\item with the assumption that $P(\Theta_1) = 0.6546$ and $P(\Theta_2)
  = 0.3454 $ and $X$ consists only of the single observation $j=4$.
\item Compare the results and explain why they should have the
  relation they do.
\end{myenumerate}
\end{exercise}

\begin{solution}
\mbox{}
\begin{myenumerate}
\item There are now three observations $j=7$, \ $j=8$, and $j=4$.
  Adapting the result in the example, we
  have
\[
f(\Theta) = \left(\sum_{j=7,8,4} \log ((\hat{r}j+j-2)!) - \log(j!) - \log
  ((\hat{r}j-1)!) +\hat{r}j \log q + (j-1) \log p\right) + \log 0.5 \, .
\]
We find $f(\Theta_1) \approx  -11.3978$ and $f(\Theta_2) \approx -12.0858$.  So
$\hat{f}(\Theta_1) =0$ and $\hat{f}(\Theta_2) = -0.688$.  Then
$g(\Theta_1) = 1$ and $g(\Theta_2) = 0.5026$.  We finally have
\[
P(\Theta_1|X) = 0.6655 , \qquad P(\Theta_2|X) = 0.3345
\]
\item The difference appears at the beginning:
\[
f(\Theta_1) = \left(\log ((\hat{r}4+4-2)!) - \log(4!) - \log
  ((\hat{r}4-1)!) +\hat{r}4 \log q + (4-1) \log p\right) + \log 0.6546
\]
and 
\[
f(\Theta_2) = \left(\log ((\hat{r}4+4-2)!) - \log(4!) - \log
  ((\hat{r}4-1)!) +\hat{r}4 \log q + (4-1) \log p\right) + \log 0.3454
\, .
\]
Plugging in for $\hat{r}$, we get  $f(\Theta_1) = -3.326$
and $f(\Theta_2) = -4.014$.  Once we find $\hat{f}$ we find that it
takes the same value as in the previous part, and so the results follow.

\item If we update our beliefs with all of our observations, we should
  get the same final outcome.  This should depend on whether we do it
  all at once, or sequentially (or even what order we do it
  sequentially).  
\end{myenumerate}
\end{solution}

\begin{exercise}
Assume that we know \emph{a priori} that the offspring distribution
for a disease has a negative binomial distribution with $p=0.02$.
Assume that our \emph{a priori knowledge of} $\hat{r}$ is that it is
an integer uniformly distributed between $1$ and $80$ inclusive.  Given observed outbreaks
of sizes $1$, $4$, $5$, $6$, and $10$:
\begin{myenumerate}
\item For each $\hat{r}$, calculate $P(\hat{r}|X)$ where $X$ is the
  observed outbreak sizes.  Plot the result.
\item Find the probability that $\Ro = \mu'(1)$ is greater than $1$.
\end{myenumerate}
\end{exercise}
\begin{solution}
The PGF is $\left( \frac{0.98}{1-0.02x}\right)^{\hat{r}}$.  We start
off with $P(\Theta) = 1/80$ for all integers from $1$ to $80$.  Given
$\hat{r}$ and $p=0.02$, the probability of observing a particular size
$j$ is $\frac{1}{j} \binom{\hat{r} j + j-2}{j-1} 0.98^{\hat{r}j}
0.01^{j-1}$.

\begin{myenumerate}
\item So $P(X|\Theta)$ means the probability of observing sizes $1$, $4$,
  $5$, $6$, and $10$ in five outbreaks given the value $\hat{r}$.
  For simplicity, we note that $1\cdot 4\cdot 5 \cdot 6 \cdot 10
  =1200$ and $1+ 4+ 5 + 6 + 10 = 26$ This is
\[
P(1,4,5,6,10 | \hat{r}) = \frac{1}{1200} 0.98^{26 \hat{r}}0.02^{26-5}
\binom{\hat{r}-1}{0} \binom{4\hat{r} + 2}{3}\binom{5\hat{r}+3}{4}\binom{6\hat{r}+4}{5}\binom{10\hat{r}+8}{9}
\]
Following the steps we get
\raisebox{-0.5\height}{\includegraphics[width=0.5\textwidth]{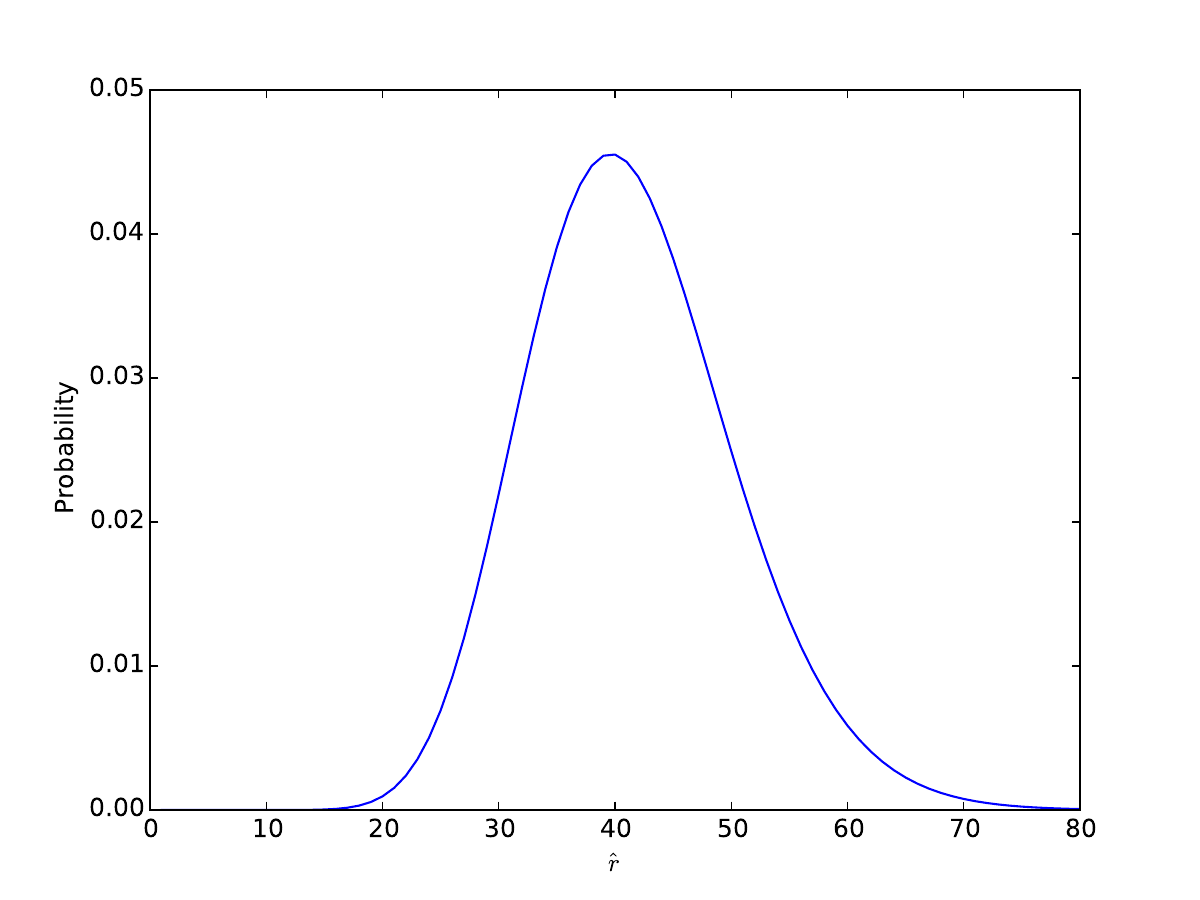}}
\item By summing the probabilities over all $\hat{r}$ for which $\Ro>1$, we find the
  probability that $\Ro>1$ is 0.181
\end{myenumerate}
\end{solution}
\section{Continuous-time spread of a simple disease}
\label{sec:cts}
We now develop PGF-based approaches adapting the results above to continuous-time processes.  In the continuous-time framework, generations will overlap, so we need a new approach if we want to answer questions about the probability of being in a particular state at time $t$ rather than at generation $g$.  Questions about the final state of the population can be answered using the same techniques as for the discrete case, but the techniques introduced here also apply and yield the same predictions.  Unlike Section~\ref{sec:discrete}, we do not do a detailed comparison with simulation.

In the continuous-time model, infected individuals have a constant rate of recovery $\gamma$ and a constant rate of transmission $\beta$.  Then $\gamma/(\beta + \gamma)$ is the probability that the first event is a recovery, while $\beta/(\beta + \gamma)$ is the probability it is a transmission.  If the event is a recovery, then the individual is removed from the infectious population.  If the event is a transmission, then the individual is still available to transmit again, with the same rate.  If the recipient of a transmission is susceptible, it becomes infectious.

Unlike the discrete-time case, we do not focus on the offspring distribution.  Rather, we focus on the resulting number of infected individuals after an event.  Early on we treat the process as if as if each infected individual were removed and replaced by either $2$ or $0$ new infections.  Although this is not the true process (she either recovers or she creates one additional infection and remains present), it is equivalent as far as the number of infections at any early time is concerned.  We focus on a PGF for the outcome of the next event. 
\begin{subequations}
\label{sys:muhat}
We define $\hat{\mu}(y) = \sum_i\hat{p}_i y^i$ and so
\begin{equation}
\hat{\mu}(y) = \frac{\beta}{\beta + \gamma}y^2 + \frac{\gamma}{\beta + \gamma}
\end{equation}
When we are calculating the number of completed  cases, it will be useful to have a two-variable version of $\hat{\mu}$:
\begin{equation}
\hat{\mu}(y,z) = \frac{\beta }{\beta + \gamma} y^2+
\frac{\gamma}{\beta + \gamma}z  \, .
\end{equation}
\end{subequations}

Most of the results in this section are the continuous-time analog of the discrete-time results above for the infinite population limit.  In the discrete-time approach we did not attempt to address outbreaks in finite populations.  However, we end the continuous-time section by deriving the equations for $\Xi(x,y,t)$, the PGF for the joint distribution of the number of susceptibles and active infections in a population of finite size $N$.

\ifsolns
\else
\subsection{Extinction probability}
\fi

\label{sec:cts_extinction}
For the extinction probability, we can apply the same methods derived in the discrete case to $\hat{\mu}(y)$.    Thus we can find the extinction probability iteratively starting from the initial guess $\alpha_0=0$ and setting $\alpha_g = \hat{\mu}(\alpha_{g-1})$.  

Exercises~\ref{exercise:cts_extinction1} and~\ref{exercise:continuous_vs_discrete} each show that
\begin{thm}
For the continuous-time Markovian model of disease spread in an infinite population, the probability of extinction given a single initial infection is
\begin{equation}
\alpha = \min(1,\gamma/\beta)
\end{equation}
\end{thm}

\ifsolns
\else
\subsubsection{Extinction probability as a function of time}
\fi
\label{sec:cts_extinction_time}
In the discrete-time case, we were interested in the probability of extinction after some number of generations.  When we are using a continuous-time model, we are generally interested in ``what is the probability of extinction by time $t$?''

To answer this, we set $\alpha(t)$ to be the probability of extinction within time $t$. We will calculate the derivative of $\alpha$ at time $t$ by using some mathematical sleight of hand to find $\alpha(t+\Delta t) - \alpha(t)$.  Then dividing this by $\Delta t$ and taking $\Delta t \to 0$ will give the result.  Our approach is closely related to \emph{backward Kolmogorov equations} (described later below).

We choose the time step $\Delta t$ to be  small enough that we
can assume that at most one event happens between time $0$ and $\Delta
t$.   The probabilities of having $0$, $1$,
or $2$ infections are $P(I(\Delta t)=0)=\gamma \Delta t +
\littleo(\Delta t)$, $P(I(\Delta t)=1) = 1-(\beta+\gamma)\Delta t +
\littleo(\Delta t)$ and $P(I(\Delta t) = 2)=\beta \Delta t + \littleo
(\Delta t)$ where the $\littleo$ notation means that the error goes
to zero fast enough that $\littleo(\Delta t)/\Delta t \to 0$ as
$\Delta t \to 0$.  The probability of having $3$ or more 
infections in the interval (that is, multiple transmission events) is $\littleo(\Delta
t)$ as well.

If there are two infected individuals at time $\Delta t$, then the probability of extinction by time $t+\Delta t$ is $\alpha(t)^2$.  Similarly, if there is one infected at time $\Delta t$, the probability of extinction by time $t+\Delta t$ is $\alpha(t)$; and if there are no infections at time $\Delta t$, then the probability of extinction by time $t+\Delta t$ is $1=\alpha(t)^0$.  So up to $\littleo(\Delta t)$ we have
\begin{align}
\alpha(t+\Delta t)&= \sum_{i=0}^\infty P(I(\Delta t)=i) \: \alpha(t)^i \nonumber\\
&= [\gamma \Delta t] \alpha(t)^0 + [1-(\beta+\gamma)\Delta t] \alpha(t) + [\beta \Delta t] \alpha(t)^2 + \littleo(\Delta t)\nonumber\\
& = \alpha(t)  + \Delta t (\beta+\gamma)\left[\hat{\mu}(\alpha(t)) - \alpha(t)\right] + \littleo(\Delta t)\label{eqn:alpha_dt}
\end{align}
Thus 
\[
\dot{\alpha} = \lim_{\Delta t \to 0}[\alpha(t+\Delta t) - \alpha(t)]/\Delta t = (\beta+\gamma) \left[\hat{\mu}(\alpha) - \alpha\right]
\]
and so 
\begin{thm}
\label{thm:cts_extinction}
Given an infinite population with constant transmission rate $\beta$
and recovery rate $\gamma$, then $\alpha(t)$, the probability of extinction by time $t$ assuming a single initial infection at time $0$ solves
\begin{equation}
\dot{\alpha} = (\beta+\gamma) \left[ \hat{\mu}(\alpha) - \alpha \right]
\label{eqn:cts_extinction}
\end{equation}
with  $\hat{\mu}(y) = (\beta y^2 + \gamma)/(\beta+\gamma)$ and the initial condition $\alpha(0)=0$.
\end{thm}
We could solve this analytically (Exercise~\ref{exercise:extinction}), but most results are easier to derive directly from the ODE formulation.

\ifsolns
\else
\subsection{Early-time outbreak dynamics}
\fi
\label{sec:cts_infection_PGF}
We now explore the number of infections at time $t$.  We define the PGF
\[
\Phi(y,t) = \sum_i\phi_i(t) y^i
\]
where $\phi_i(t)$ is the probability of $i$ actively infected individuals at time $t$.  We will derive equations for the evolution of $\Phi(y,t)$.  We assume that $\Phi(y,0) = y$ so a single infected individual exists at time $0$.

Our goal is to derive equations telling us how $\Phi$ changes in time.  We will use two approaches which were hinted at in exercise~\ref{exercise:discrete_backward_and_forward}, yielding two different partial differential equations.  Although their appearance is different, for the appropriate initial condition, their solutions are the same.  These equations are called the \emph{forward} and \emph{backward Kolmogorov equations}. 

We briefly describe the analogy between the forward and backward Kolmogorov equations and exercise~\ref{exercise:discrete_backward_and_forward}:
\begin{itemize}
\item Our first approach finds the \emph{forward Kolmogorov equations}.  This is akin to exercise~\ref{exercise:discrete_backward_and_forward} where we found $\Phi_g(y)$ by knowing the PGF $\Phi_{g-1}(y)$ for the number infected in generation $g-1$ and recognizing that since the PGF for the number of infections each of them causes is $\mu(y)$, we must have $\Phi_g(y) = \Phi_{g-1}(\mu(y))$.

\item Our second approach finds the \emph{backward Kolmogorov equations} which are more subtle and can be derived similarly to how we derived the ODE for extinction probability in Theorem~\ref{thm:cts_extinction}.  This is akin to exercise~\ref{exercise:discrete_backward_and_forward} where we found $\Phi_g(y)$ by knowing that the PGF for the number infected in generation $1$ is $\mu(y)$, and recognizing that after another $g-1$ generations each of those creates a number of infections whose PGF is $\Phi_{g-1}(y)$ and so $\Phi_g(y) = \mu(\Phi_{g-1}(y))$.
\end{itemize}

For both approaches, we make use of the observation that for $\Delta t \ll 1$, we can write the PGF for the number of infections resulting from a single infected individual at time $t=0$ to be 
\[
\Phi(y,\Delta t) = y +(y^2-y)\beta \Delta t+ (1-y)\gamma \Delta t  + \littleo(\Delta t)\,.
\]
This says that with probability approximately $\beta \Delta t$ a transmission happens and we replace $y$ by $y^2$ and with probability approximately $\gamma \Delta t$ a recovery happens and we replace $y$ by $1$.  With probability $\littleo(\Delta t)$ multiple events happen.  We can rewrite this as 
\[
\Phi(y, \Delta t)=y +(\beta+\gamma) [\hat{\mu}(y) - y] \Delta t + \littleo(\Delta t)\,.
\]
Note that $\Phi(y,0)=y$ and $\pd{}{t} \Phi(y,0) = (\beta+\gamma) [\hat{\mu}(y) - y]$.  

Both of our approaches rely on the observation that $\Phi(y, t_1 + t_2) = \Phi(\Phi(y,t_2),t_1)$ by Property~\ref{prop:composition}. This states that if we take the PGF at time $t_1$, and then substitute for each $y$ the PGF for the number of descendants of a single individual after $t_2$ units of time, the result is the PGF for the total number at time $t_1+t_2$.  

\ifsolns
\else
\paragraph{Forward equations}
\fi
  For this we use $\Phi(y, t_1 + t_2) = \Phi(\Phi(y,t_2),t_1)$ with $t_2$ playing the role of $\Delta t$ and $t_1$ playing the role of $t$.

So $\Phi(y,t+\Delta t) = \Phi(\Phi(y,\Delta t), t)$.  For small $\Delta t$ (and taking $\Phi_y(\Phi(y,0),t)$ to be the partial derivative of $\Phi$ with respect to its first argument), we have
\begin{align*}
\Phi(y, t+\Delta t) &= \Phi(\Phi(y,\Delta t), t)\\
&=\Phi(\Phi(y,0),t) + (\Delta t) \Phi_y(\Phi(y,0),t) \pd{}{t} \Phi(y,0) +\littleo(\Delta t)\\
&= \Phi(y,t) + (\Delta t) (\beta+\gamma) [\hat{\mu}(y) - y]\pd{}{y}\Phi(y,t) + \littleo(\Delta t) \, .
\end{align*}

Then
\begin{align*}
\dot{\Phi}(y,t) &= \lim_{\Delta t \to 0} \frac{\Phi(y,t+\Delta t) -\Phi(y,t)}{\Delta t}\\
&= \lim_{\Delta t \to 0} \frac{\Phi(y,t) + (\Delta t) (\beta+\gamma) [\hat{\mu}(y) - y]\pd{}{y}\Phi(y,t) + \littleo(\Delta t)- \Phi(y,t)}{\Delta t}\\
&=  (\beta+\gamma)[\hat{\mu}(y)-y]\pd{}{y}\Phi(y,t) \, .
\end{align*}
More generally, we can directly apply Property~\ref{prop:derivative_sum} to get this result.  Exercise~\ref{exercise:phi_forward} provides an alternate direct derivation of these equations.

\ifsolns
\else
\paragraph{Backward equations}
\fi
 In the backward direction we have $\Phi(y, t_1 + t_2) = \Phi(\Phi(y,t_2),t_1)$ with $t_2$ playing the role of $t$ and $t_1$ playing the role of $\Delta t$.

So $\Phi(y,t+\Delta t)= \Phi(y,\Delta t + t) = \Phi(\Phi(y,t), \Delta t)$.  Note that because $\Phi(y,0) = y$, we have $\Phi(\Phi(y,t),0) = \Phi(y,t)$.  Thus for small $\Delta t$, we expand $\Phi$ as a Taylor Series in its second argument $t$
\begin{align*}
\Phi(y,t+\Delta t) &= \Phi(\Phi(y,t),\Delta t)\\
&= \Phi(\Phi(y,t),0) + (\Delta t) \Phi_t(\Phi(y,t),0) + \littleo(\Delta t) \\ 
&= \Phi(y,t) + (\Delta t) \Phi_t(\Phi(y,t),0) + \littleo(\Delta t) \\
&= \Phi(y,t) + (\Delta t)(\beta+\gamma)[\hat{\mu}(\Phi(y,t))-\Phi(y,t)] + \littleo(\Delta t)\, .
\end{align*}
To avoid ambiguity, we use $\Phi_t$ to denote the partial derivative of $\Phi$ with respect to its second argument $t$.  So
\begin{align*}
\dot{\Phi}(y,t) &= \lim_{\Delta t \to 0} \frac{\Phi(y,t+\Delta t) - \Phi(y,t)}{\Delta t}\\
&= \lim_{\Delta t \to 0} \frac{\Phi(y,t) + (\Delta t)(\beta+\gamma)[\hat{\mu}(\Phi(y,t))-\Phi(y,t)] + \littleo(\Delta t) - \Phi(y,t)}{\Delta t}\\
&= (\beta+\gamma)[\hat{\mu}(\Phi(y,t))-\Phi(y,t)] \, .
\end{align*}
This result also follows directly from Property~\ref{prop:bKe2PGF}.

So we have
\begin{thm}
\label{thm:cts_current}
The PGF $\Phi(y,t)$ for the distribution of the number of current infections at time $t$ assuming a single introduced infection at time $0$ solves
\begin{equation}
\pd{}{t} \Phi(y,t) = (\beta+\gamma)\left[\hat{\mu}(y) - y \right] \pd{}{y}\Phi(y,t) \label{eqn:forward_PGF}
\end{equation}
as well as
\begin{equation}
\pd{}{t}\Phi(y,t) = (\beta+\gamma) \big[\hat{\mu}(\Phi(y,t)) - \Phi(y,t)\big] \, .
\label{eqn:backward_PGF}
\end{equation}
both with the initial condition $\Phi(y,0)=y$.
\end{thm}
It is perhaps remarkable that such seemingly different equations yield the same solution for the given initial condition.

\begin{example}
The expected number of infections in the infinite population limit is given by $[I] = \sum_i i p_i(t) = \pd{}{y} \Phi(1,t)$.  From this we have
\begin{align*}
\diff{}{t}[I] &= \left. \pd{}{t} \pd{}{y} \Phi(y,t)\right|_{y=1}\\  
& = \left. \pd{}{y} \left[ (\beta+\gamma) [ \hat{\mu}(y) - y] \pd{}{y} \Phi(y,t) \right] \right|_{y=1}\\
&= (\beta+\gamma) [ \hat{\mu}'(y) - 1]\pd{}{y} \Phi(y,t)  +  (\beta+\gamma) [ \hat{\mu}(y) - y] \left. \pds{}{y}\Phi(y,t)] \right|_{y=1}\\
&= (\beta+\gamma) [ \hat{\mu}'(1) - 1][I]  + (\beta+\gamma) [ \hat{\mu}(1) - 1] \left.\left[\pds{}{y}\Phi(y,t)\right]\right|_{y=1} \\
&= (\beta+\gamma) [(2\beta)/(\beta+\gamma) -1][I]\\
&= (\beta - \gamma)[I]
\end{align*}
We used $\hat{\mu}(1)=1$ to eliminate the $\pds{}{y}\Phi(y,t)$ term and replaced $\hat{\mu}'(1)$ with $2\beta/(\beta+\gamma)$.  Using this and $[I](0)=1$, we have 
\[
[I] = e^{(\beta-\gamma)t}\,.
\]
\end{example}

This example proves
\begin{cor}
In the  infinite population limit, if a disease starts with a single infection, then the expected number of active infections at time $t$ solves
\begin{equation}
[I] = e^{(\beta-\gamma)t}
\end{equation}
\end{cor}

\ifsolns
\else
\subsection{Cumulative and current outbreak size distribution}
\fi


\label{sec:cts_joint_size}
Let $\pi_{i,r}(t)$ be the probability of having $i$ currently infected
individuals and $r$ completed infections at time $t$.  We define
$\Pi(y,z,t) = \sum_{ir} \pi_{i,r}(t) y^iz^r$ to be the PGF at time
$t$.  We have $\Pi(y,z,0)=y$.  As before we assume the population is
large enough that the spread of the disease is not limited by the size
of the population.

We give an abbreviated derivation of the Kolmogorov equations for $\Pi$.  A full derivation is requested as an exercise.

\ifsolns
\else
\paragraph{Forward Kolmogorov formulation}
\fi
To derive the forward Kolmogorov equations for the PGF $\Pi(y,z,t)$, we use Property~\ref{prop:2Dderivative_sum}, noting that all transition rates are proportional to $i$.  The rate of transmission is $\beta i$ and the rate of recovery is $\gamma i$.  There are no interactions to consider.  So
\begin{align*}
\pd{}{t}\Pi(y,z,t) &= (\beta + \gamma)  \left(\frac{\beta y^2}{\beta+\gamma} + \frac{\gamma z}{\beta+\gamma} -y\right) \pd{}{y} \Pi(y,z,t)\\
&= (\beta+\gamma)\big[\hat{\mu}(y,z) - y \big] \pd{}{y}\Pi(y,z,t)
\end{align*}


\ifsolns
\else
\paragraph{Backward Kolmogorov formulation} 
\fi
To derive the backward Kolmogorov equations for the PGF $\Pi$, we use a modified version of Property~\ref{prop:bKe2PGF} to account for two types of individuals (Exercise~\ref{exercise:bkwds_2type}, with events proportional only to the infected individuals).  We find
\[
\dot{\Pi}(y,z,t) = (\beta+\gamma) [ \hat{\mu}(\Pi(y,z,t),z)-\Pi(y,z,t)]\, .
\]

Combining our backward and forward Kolmogorov equation results, we get
\begin{thm}
\label{thm:kolmogorov}
Assuming a single initial infection in an infinite population, the PGF $\Pi(y,z,t)$ for the joint distribution of the number of current and completed infections at time $t$ solves
\begin{equation}
\pd{}{t} \Pi(y,z,t) = (\beta+\gamma) \big[\hat{\mu}(y,z) - y \big] \pd{}{y} \Pi(y,z,t)
\label{eqn:forward_result}
\end{equation}
as well as
\begin{equation}
\label{eqn:backward_result}
\pd{}{t} \Pi(y,z,t) = (\beta+\gamma) \big[ \hat{\mu}(\Pi(y,z,t), z) - \Pi(y,z,t)\big]
\end{equation}
both with the initial condition $\Pi(y,z,0) = y$.
\end{thm}
It is again remarkable that these seemingly very different equations have the same solution.

\begin{example}
The expected number of completed infections at time $t$ is 
\[
[R]=\sum_{j,k}k p_{jk} = \left. \pd{}{z}\Pi(y,z,t)\right|_{y=z=1}
\]
 (although we use $R$, this approach is equally relevant for counting completed infections in the SIS model because of the infinite population assumption).  Its evolution is given by
\begin{align*} 
\diff{}{t} [R] &=\left. \pd{}{t} \pd{}{z}\Pi(y,z,t)\right|_{y=z=1}\\
&= \left. \pd{}{z}\left [ (\beta+\gamma)\big[\hat{\mu}(y,z) - y \big] \pd{}{y}\Pi(y,z,t)\right]\right|_{y,z=1}\\
&= \left.(\beta+\gamma) \left[ \pd{}{z}\hat{\mu}(y,z)\pd{}{y}\Pi(y,z,t) + [\hat{\mu}(y,z)-y] \pd{}{z}\pd{}{y} \Pi(y,z,t)\right]\right|_{y=z=1}\\
&= (\beta+\gamma)\left.\left[ \frac{\gamma}{\beta+\gamma} \pd{}{y}\Pi(y,z,t)  + 0 \pd{}{z}\pd{}{y} \Pi(y,z,t)\right]\right|_{y=z=1}\\
&= \gamma [I]
\end{align*}
where we use the fact that $\hat{\mu}(1,1) = 1$, \ $\pd{}{z}
\hat{\mu}(y,z)=\gamma/(\beta+\gamma)$, and $[I] = \left. \pd{}{y}
  \Pi(y,z,t)\right |_{y=z=1}$.  Our result says that the rate of change of the expected number of completed infections is $\gamma$ times the expected number of current infections.
\end{example}
This example proves
\begin{cor}
In the  infinite population limit
the expected number of recovered individuals as a function of time solves
\begin{equation}
\diff{}{t} [R] = \gamma [I]
\end{equation}
\end{cor}
We will see that this holds even in finite populations.

\ifsolns
\else
\subsection{Small outbreak final size distribution}
\fi
\label{sec:cts_final_size}
We define
\[
\Omega_\infty(z) = \left(\sum_{j<\infty} \omega_j z^j\right) + \omega_\infty z^\infty
\]
 to be  the PGF of the distribution of outbreak final sizes in an infinite population, with $\omega_\infty z^\infty$ representing epidemics and for $j<\infty$ $\omega_j$ representing the probability of an outbreak that infects exactly $j$ individuals.  We use the convention that $z^\infty = 0$ for $z<1$ and $1$ for $z=1$.  To calculate $\Omega_\infty$, we make observations that the outbreak size coming from a single infected individual is $1$ if the first thing that individual does is a recovery or it is the sum of the outbreak sizes of two infected individuals if the first thing the individual does is to transmit (yielding herself and her offspring).  

Thus we have
\begin{align*}
\Omega_\infty(z) &= \frac{\beta}{\beta+\gamma} [\Omega_\infty(z)]^2 + \frac{\gamma}{\beta+\gamma} z \\
&= \hat{\mu}(\Omega_\infty(z),z)
\end{align*}
As for the discrete-time case we may solve this iteratively, starting
with the guess $\Omega_\infty(z) = z$.  Once $n$ iterations have occurred, the first $n$ coefficients of $\Omega_\infty(z)$ remain constant.  Note that unlike the discrete case, here $\Omega_\infty(z) \neq z \hat{\mu}(\Omega_\infty(z))$.  This yields
\begin{thm}
\label{thm:cts_final_size}
The PGF $\Omega_\infty(z) = \sum_j \omega_j z^j + \omega_{\infty} z^\infty$ for the final size distribution assuming a single initial infection in an infinite population solves
\begin{equation}
\Omega_\infty(z) = \hat{\mu}(\Omega_\infty(z),z)
\label{eqn:cts_final_size}
\end{equation}
with $\Omega_\infty(1)=1$.  This function is discontinuous at $z=1$.  For the final size distribution conditional on the outbreak being finite, the PGF is continuous and equals  
\[
\begin{cases}
\Omega_\infty(z) / \alpha & 0 \leq z < 1\\
1 & z=1
\end{cases}
\]
\end{thm}

As in the discrete-time case, we can find the coefficients of $\Omega_\infty(z)$ analytically.  
\begin{thm}
\label{thm:cts_power_magic}
Consider continuous-time outbreaks with transmission rate $\beta$ and recovery rate $\gamma$ in an infinite population with a single initial infection.  The probability the outbreak causes exactly $j$ infections for $j<\infty$ [that is, the coefficient of $z^j$ in $\Omega_\infty(z)$] is
\[
\omega_j = \frac{1}{j}\frac{\beta^{j-1}\gamma^j}{(\beta+\gamma)^{2j-1}} \binom{2j-2}{j-1}
\]
\end{thm}

We prove this theorem in appendix~\ref{app:power_magic}.  The proof is based on observing that if there are $j$ total infected individuals, this requires $j-1$ transmissions and $j$ recoveries.  Of the sequences of $2j-1$ events that have the right number of recoveries and transmissions, a fraction $1/(2j-1)$ of these satisfy additional constraints required to be a valid sequence leading to $j$ infections (the sequence cannot lead to 0 infections prior to the last step).  Alternately, we can note that the offspring distribution is geometric and use Table~\ref{table:final_probs}.

\ifsolns
\else
\subsection{Full dynamics in finite populations}
\fi
\label{sec:full}
We now derive the PGFs for continuous time SIS and SIR outbreaks in a finite population.  

PGF-based techniques are easiest when we can treat events as independent.  In the continuous-time model, when we look at the system in a given state, each event is independent of the others.  Once the next event happens the possible events change, but conditional on the new state, they are still independent.  Thus we can use the forward Kolmogorov approach (the backward Kolmogorov approach will not work because descendants of any individual are not independent).  

We do not look at the discrete-time version because in a single time step, multiple events can occur, some of which affect one another.  So we would lose independence as we go from one time step to another.

For these reasons we focus on the forward Kolmogorov formulations for
the continuous-time models.  Much of our approach here was derived
previously in~\cite{bailey1953total,bartlett1949some}.  See also~\cite{allen2008introduction}

For a given population size $N$, we let $s$, $i$, and $r$ be the number of susceptible, infected and immune (removed) individuals.  For the SIS model $r=0$ and we have $s+i=N$ while for the SIR model we have $s+i+r=N$.  We let $p_{s,i}$ be the probability of $s$ susceptible and $i$ infected individuals.

\ifsolns
\else
\subsubsection{SIS}
\fi
\label{sec:cts_full_SIS}
We start with the SIS model.  We set $\xi_{s,i}(t)$ to be the probability of $s$ susceptible and $i$ actively infected individuals at time $t$.  We define the PGF for the joint distribution of susceptible and infected individuals
\[
\Xi(x,y,t) = \sum_i \xi_{s,i}(t) x^sy^i
\]
At rate $\frac{\beta}{N} si$, successful transmissions occur, moving the system from the state $(s,i)$ to $(s-1,i+1)$, which is equivalent to removing one susceptible individual and one infected individual, and replacing them with two infected individuals.  Following property~\ref{prop:2Dderivative_sum}, this is represented by
\[
\frac{\beta}{N} (y^2-xy) \pd{}{x}\pd{}{y} \Xi\, .
\]
At rate $\gamma i$, recoveries occur, moving the system from the state $(s,i)$ to $(s+1, i-1)$, which is equivalent to removing one infected individual and replacing it with a susceptible individual.  This is represented by 
\[
\gamma (x-y) \pd{}{y} \Xi\, .
\]
So the PGF solves
\[
\dot{\Xi} = \frac{\beta}{N}(y^2-xy) \pd{}{x}\pd{}{y} \Xi + \gamma (x-y) \pd{}{y} \Xi
\]
It is sometimes useful to rewrite this as
\[
\dot{\Xi} = (y-x) \left[\frac{\beta}{N} y \pd{}{x} - \gamma \right] \pd{}{y} \Xi
\]
We have 
\begin{thm}
\label{thm:SIS_full_dynamics}
For SIS dynamics in a finite population we have
\begin{equation}
\pd{}{t} \Xi = \frac{\beta}{N}(y^2-xy) \pd{}{x}\pd{}{y} \Xi + \gamma (x-y) \pd{}{y} \Xi
\label{eqn:SIS_full_dynamics}
\end{equation}
\end{thm}

We can use this to derive equations for the expected number of susceptible and infected individuals.
\begin{example}
\label{example:SIS_MA_eqns}
We use $[S]$ and $[I]$ to denote the expected number of susceptible and infected individuals at time $t$.  
We have 
\begin{align*}
[S] &= \sum_{s,i} s\xi_{si}(t) = \sum_{s,i} s \xi_{si} 1^{s-1} 1^i  = \pd{}{x} \Xi(1,1,t)\\
[I] &= \sum_{s,i} i \xi_{si}(t) = \sum_{s,i} i \xi_{si} 1^s 1^{i-1} = \pd{}{y} \Xi(1,1,t)
\end{align*}
We also define the expected value of the product $si$,
\[
[SI] = \sum_{s,i} si \xi_{si}(t) = \pd{}{x}\pd{}{y} \Xi(1,1,t) \, .
\]
Then we have
\begin{align*}
[\dot{S}] &= \pd{}{t} \pd{}{x} \Xi(1,1,t)\\
&= \left.\pd{}{x} \pd{}{t} \Xi(x,y,t) \right|_{x=y=1}\\
&= \left . \pd{}{x} \left((y-x) \left[\frac{\beta}{N} y \pd{}{x} - \gamma \right] \pd{}{y} \Xi\right)\right|_{x=y=1}\\
&= \left. (y-x) \pd{}{x} \left[ \left(\frac{\beta}{N} y \pd{}{x} - \gamma \right) \pd{}{y} \Xi\right] - \left[ \frac{\beta}{N} y \pd{}{x} - \gamma \right]\pd{}{y}\Xi \right|_{x=y=1}\\
&= - \frac{\beta}{N} [SI] + \gamma [I]
\end{align*}
\end{example}
In the final line, we eliminated the first term because $y-x$ is zero at $x=y=1$.  Similar steps show that 
\[
[\dot{I}] = \frac{\beta}{N}[SI] - \gamma [I]
\]
but the derivation is faster if we simply note $[S]+[I]=N$ is constant.  This proves
\begin{cor}
\label{cor:SIS_MA_eqns}
For SIS disease, the expected number infected and susceptible solves
\begin{align}
\diff{}{t} [S] &= -\frac{\beta}{N}[SI] + \gamma [I]\\
\diff{}{t} [I] &= \frac{\beta}{N}[SI] - \gamma [I]
\end{align}
where $[SI]$ is the expected value of the product $si$.
\end{cor}

\ifsolns
\else
\subsubsection{SIR}
\fi
\label{sec:cts_full_SIR}
Now we consider the SIR model.  A review of various techniques
(including PGF-based methods) to find
the final size distribution of outbreaks in finite-size populations can be found
in~\cite{house2013big}.  Here we focus on the application of PGFs to
find the full dynamics.  For a given $s$ and $i$, infection occurs at rate $\beta s i/N$.  It appears as a departure from the state $(s,i)$ and entry into  $(s-1,i+1)$.  Following property~\ref{prop:2Dderivative_sum}, this is captured by 
\[
\frac{\beta}{N} (y^2-xy)\pd{}{x}\pd{}{y} \Xi \, .
\]
Recovery is captured by 
\[
\gamma (1-y)  \pd{}{y} \Xi \, 
\]
[note the difference from the SIS case in the recovery term].
So we have
\begin{thm}
\label{thm:SIR_full_dynamics}
For SIR dynamics in a finite population we have
\begin{equation}
\label{eqn:SIR_full_dynamics}
\pd{}{t} \Xi  = \frac{\beta(y^2-xy)}{N}\pd{}{x}\pd{}{y}\Xi + \gamma(1-y) \pd{}{y} \Xi
\end{equation}
\end{thm}

We follow similar steps to example~\ref{example:SIS_MA_eqns} to derive equations for $[S]$ and $[I]$ in Exercise~\ref{exercise:SIR_mass_action}.  The result of this exercise should show
\begin{cor}
For SIR disease, the expected number of susceptible, infected, and recovered individuals solves
\begin{align}
\diff{}{t} [S] &= -\frac{\beta}{N}[SI] \\
\diff{}{t} [I] &= \frac{\beta}{N} [SI] - \gamma [I]\\
\diff{}{t} [R] &= \gamma [I]
\end{align}
where $[SI]$ is the expected value of the product $si$.
\end{cor}


\ifsolns
\else
\subsection{Exercises}
\fi
\begin{exercise}
\textbf{Extinction Probability}
\label{exercise:cts_extinction1}  

Let $\beta$ and $\gamma$ be given with $\hat{\mu}(y) = (\beta y^2 + \gamma)/(\beta + \gamma)$.
\begin{myenumerate}
\item Analytically find solutions to $y = \hat{\mu}(y)$.
\item Assume $\beta<\gamma$.  Find all solutions in $[0,1]$.
\item Assume $\beta > \gamma$.  Find all solutions in $[0,1]$.
\end{myenumerate}
\end{exercise}
\begin{solution}\mbox{}
\begin{myenumerate}
\item $y(\beta+\gamma) = \beta y^2 + \gamma$ so
\[
\beta y^2 - (\beta+ \gamma)y + \gamma = 0
\]
By the quadratic formula
\[
y = \frac{\beta+\gamma \pm \sqrt{(\beta+\gamma)^2 -
    4\beta\gamma}}{2\beta}
\]
The radical becomes $\sqrt{\beta^2 - 2\beta\gamma +\gamma^2} =
(\beta-\gamma)$.
So
\[
y = \frac{\beta+\gamma \pm (\beta-\gamma)}{2\beta} = \{ 1, \gamma/\beta\}
\].
\item If $\beta < \gamma$, then only $1$ is in the interval.
\item If $\gamma>\beta$ then both are in the interval.
\end{myenumerate}
\end{solution}

\begin{exercise}
\textbf{Consistency with discrete-time formulation.}

\label{exercise:continuous_vs_discrete}
Although we have argued that a transmission in the continuous-time disease transmission case can be treated as if a single infected individual has two infected offspring and then disappears, this is not what actually happens.  In this exercise we look at the true offspring distribution of an infected individual before recovery, and we show that the ultimate predictions of the two versions are equivalent.

Consider a disease in which individuals transmit at rate $\beta$ and recover at rate $\gamma$.  Let $p_i$ be the probability an infected individual will cause exactly $i$ new infections before recovering.
\begin{myenumerate}
\item Explain why $p_0 = \gamma/(\beta+\gamma)$.
\item Explain why $p_i =  \beta^i\gamma/(\beta+\gamma)^{i+1}$.  So $p_i$ form a geometric distribution.
\item Show that $\mu(y) = \sum_ip_i y^i$ can be expressed as $\mu(y) = \gamma/(\beta+\gamma-\beta y)$.  [This definition of $\mu$ without the hat corresponds to the discrete-time definition]
\item Show that the solutions to $y=\mu(y)$ are the same as the
  solutions to $y= \hat{\mu}(y) = (\beta y^2  + \gamma)/(\beta+\gamma)$.
  So the extinction probability can be calculated either way. (You do not have to find the solutions to do this, you can simply show that the two equations are equivalent).
\end{myenumerate}
\end{exercise}
\begin{solution}\mbox{}
\begin{myenumerate}
\item The event is either a recovery or a transmission, and the rates
  are $\gamma$ and $\beta$.  So the probability of recovery is the
  rate of recovery divided by the total rate.
\item While the individual is susceptible, the probability that the
  next event is a transmission is $\beta/(\beta+\gamma)$.  So the
  probability that the first $i$ events are transmissions is
  $\beta^i/(\beta+\gamma)^i$.  The probability that the next event is
  a recovery is $\gamma/(\beta+\gamma)$.  The product of these is the
  probability that the first $i$ events are transmissions and the next
  event is a recovery.  
\item Let $r = \beta/(\beta+\gamma)$.  Then $p_i = (1-r) r^i$.  Then $\sum_ip_i y^i$ is a geometric series: $(1-r) \sum_i (ry)^i = (1-r)/(1-ry)$.  Multiplying by $1=(\beta+\gamma)/(\beta+\gamma)$ gives the result. 
\item The solutions to $y = \mu(y)$ are the solutions to
  $y(\beta+\gamma-\beta y) = \gamma$.  The solutions to $y =
  \hat{\mu}(y) = (\beta y^2+\gamma)/(\beta+\gamma)$ are the solutions
  to $y(\beta+\gamma) = \beta y^2 + \gamma$.  By moving $\beta y^2$ to
  the other side of either equation we see that the equations are
  equivalent.
\end{myenumerate}
\end{solution}

\begin{exercise}
\textbf{Relation with $\Ro$}

Take $\mu(y) = \gamma/(\beta+\gamma-\beta y)$ as given in exercise~\ref{exercise:continuous_vs_discrete} and $\hat{\mu}=(\beta y^2 + \gamma)/(\beta+\gamma)$.
\begin{myenumerate}
\item Show that $\mu'(1) \neq \hat{\mu}'(1)$ in general.
\item Show that when $\Ro = \mu'(1) = 1$, then $\mu'(1) = \hat{\mu}'(1)=1$.  So both are still threshold parameters.
\end{myenumerate}
\end{exercise}
\begin{solution} \mbox{}
\begin{myenumerate}
\item $\mu'(y) = \gamma \beta/(\beta+\gamma-\beta y)^2$, so $\mu'(1) = \gamma\beta/\gamma^2 = \beta/\gamma$.  Similarly $\hat{\mu}'(y) = 2\gamma y/(\beta+\gamma)$, so $\hat{\mu}'(1) = 2\gamma/(\beta+\gamma)$.  These are generally unequal.
\item When $\Ro = 1$, we have $\beta = \gamma$.  So $\mu'(1) = \gamma/\gamma=1$ and $\hat{\mu}'(1) = 2\gamma/(\gamma+\gamma)=1$.  So the two are equal.
\end{myenumerate}
\end{solution}

\begin{exercise} 
\label{exercise:extinction}\textbf{Revisiting eventual extinction
  probability}.  

We revisit the results of exercise~\ref{exercise:cts_extinction1} using Eq.~\eqref{eqn:cts_extinction} (without solving it).
\begin{myenumerate}
\item By substituting for $\hat{\mu}(\alpha)$, show that $\dot{\alpha} = (1-\alpha)(\gamma-\beta\alpha)$.
\end{myenumerate}
We have $\alpha(0)=0$.  Taking this initial condition and expression for $\dot{\alpha}$, show that
\begin{continuemyenumerate}
\item $\alpha \to 1$ as $t \to \infty$ if $\beta<\gamma$ (i.e., $\Ro<1$) and
\item $\alpha \to \gamma/\beta$ as $t \to \infty$ if $\beta>\gamma$ (i.e., $\Ro>1$).
\item Set up (but do not solve) a partial fraction integration that would give $\alpha(t)$ analytically.
\end{continuemyenumerate}
\end{exercise}
\begin{solution}\mbox{}
\begin{myenumerate}
\item We have 
\begin{align*} 
\dot{\alpha} &= (\beta+\gamma)\left(\frac{\beta \alpha^2+ \gamma}{\beta+\gamma}-\alpha\right)\\
&= \beta \alpha^2 + \gamma  - \alpha\\
&= (1-\alpha)(\gamma-\beta \alpha)
\end{align*}
\end{myenumerate}
Starting from $\alpha(0)=0$, we see that $\dot{\alpha}(0) =\gamma >0$.  So $\alpha$ will increase and approach the smallest positive equilibrium value. The two equilibria are at $\alpha=1$ and $\alpha = \gamma/\beta$.
\begin{continuemyenumerate}
\item If $\beta<\gamma$, then the smaller equilibrium is $\alpha=1$.
\item If $\beta>\gamma$, then the smaller equilibrium is $\alpha = \gamma/\beta$.
\item We have 
\[
\frac{\diff{\alpha}{t}}{(1-\alpha)(\gamma-\beta\alpha)} = 1
\]
So using separation of variables, we have
\[
\int \frac{1}{(1-\alpha)(\gamma-\beta\alpha)} \, \mathrm{d}\alpha = \int 1 \mathrm{d}t
\]
Using partial fractions this becomes
\[
\int \frac{A}{1-\alpha} + \frac{B}{\gamma-\beta \alpha} \, \mathrm{d}\alpha = \int 1 \, \mathrm{d}t
\]
where $A$ and $B$ are chosen such that the sum in the integrand has a
numerator of $1$.  Once $A$ and $B$ are found, both sides can be integrated analytically.
\end{continuemyenumerate}
\end{solution}

\begin{exercise}
This exercise is intended to help with understanding the backward Kolmogorov equations. 

Let $\phi_i(t)$ denote the probability of having $i$ active infections at time $t$ given that at time $0$ there was a single infection [$\phi_1(0)=1$].  We have $\phi_0(t)=\alpha(t)$.  We extend the derivation of Eq.~\eqref{eqn:alpha_dt} to $\phi_1$.  Assume $\phi_0(t_0)$ and $\phi_1(t_0)$ are known.
\begin{myenumerate}
\item Following the derivation of Eq.~\eqref{eqn:alpha_dt}, approximate $\phi_0(\Delta t)$, $\phi_1(\Delta t)$, and $\phi_2(\Delta t)$ for small $\Delta t$.
\item From biological grounds explain why if there are $0$ infections at time $\Delta t$ then there are also $0$ infections at time $t_0+\Delta t$.
\item If there is $1$ infection at time $\Delta t$, what is the probability of $1$ infection at time $t_0+\Delta t$?
\item If there are $2$ infections at time $\Delta t$, what is the probability of $1$ infection at time $t_0+\Delta t$?
\item Write $\phi_1(t_0+\Delta t)$ in terms of $\phi_0(t_0)$, $\phi_1(t_0)$, $\phi_1(\Delta t)$, and $\phi_2(\Delta t)$.
\item Using the definition of the derivative, find an expression for $\dot{\phi}_1$ in terms of $\phi_1(t)$ and $\phi_2(t)$.
\end{myenumerate}
\end{exercise}
\begin{solution}\mbox{}
\begin{myenumerate}
\item We have 
\begin{align*}
\phi_0(\Delta t) &= \gamma \Delta t + \littleo(\Delta t)\\
\phi_1(\Delta t) &= 1- (\beta+\gamma) \Delta t + \littleo(\Delta t)\\
\phi_2(\Delta t) &= \beta \Delta t + \littleo(\Delta t)
\end{align*}
\item If the disease is extinct at time $\Delta t$, then it remains extinct.
\item This equals the probability of having $1$ infection at time $t_0$ if there is one infection at time $0$.  So it is is $\phi_1(t_0)$.
\item One of the infections has to have its descendants go extinct within $t_0$ units of time, and the other must have 1 descendant after $t_0$ units of time.  So this is $\phi_0(t_0)\phi_1(t_0) + \phi_1(t_0)\phi_0(t_0) = 2\phi_0(t_0) \phi_1(t_0)$
\item So  [using the additional fact that the probability of 3 or more infections at time $\Delta t$ is $\littleo(\Delta t)$]
\[
\phi_1(t_0+\Delta t) = \phi_1(\Delta t)\phi_1(t_0) + 2\phi_2(\Delta t)\phi_0(t_0)\phi_1(t_0) + \littleo(\Delta t)
\]
\item So 
\begin{align*}
\dot{\phi}_1(t) &= \lim_{\Delta t \to 0} \frac{\phi_1(t+\Delta t) - \phi_1(t)}{\Delta t}\\
&= \lim_{\Delta t \to 0} \frac{\phi_1(\Delta t)\phi_1(t) + 2\phi_2(\Delta t)\phi_0(t)\phi_1(t) + \littleo(\Delta t) - \phi_1(t)}{\Delta t}\\
&= \lim_{\Delta t \to 0} \frac{[1-(\beta+\gamma)\Delta t] \phi_1(t) + 2\beta \Delta t\phi_0(t)\phi_1(t) + \littleo(\Delta t) - \phi_1(t)}{\Delta t}\\
&= \lim_{\Delta t \to 0} \frac{-(\beta+\gamma)\Delta t \phi_1(t) + 2\beta \Delta t\phi_0(t)\phi_1(t) + \littleo(\Delta t)}{\Delta t}]\\
&= -(\beta+\gamma)\phi_1(t) + 2\beta \phi_0(t)\phi_1(t)
\end{align*}
\end{myenumerate}
\end{solution}

\begin{exercise}
\label{exercise:phi_forward}
In this exercise we derive the PGF version of the forward Kolmogorov equations by directly calculating the rate of change of the probabilities of the states.  Define $\phi_j(t)$ to be the probability that there are $j$ active infections at time $t$.  

We have the \emph{forward Kolmogorov equations}:
\[
\dot{\phi}_j = \beta (j-1) \phi_{j-1} + \gamma(j+1)\phi_{j+1} - (\beta+\gamma)j\phi_j \, .
\]
\begin{myenumerate}
\item Explain each term on the right hand side of the equation for $\dot{\phi}_j$.
\item By expanding $\dot{\Phi}(y,t) = \pd{}{t} \sum_j \phi_j y^j$, arrive at Equation~\eqref{eqn:forward_PGF}.
\end{myenumerate}
\end{exercise}

\begin{solution}\mbox{}
\begin{myenumerate}
\item If there are $j$ infected individuals, then a new transmission will result in $j+1$ total infections and occurs at total rate $\beta j$.  This appears as a loss to $\phi_j$ at rate $j\beta \phi_j$, but a gain at rate $(j-1)\beta\phi_{j-1}$ from the smaller state.  
Similarly a recovery will result in $j-1$ total infections and occurs at total rate $\gamma j$.  No other possibilities are considered in our model.   This appears as a loss at rate $j \gamma \phi_j$ and a gain at rate $(j+1)\gamma \phi_{j+1}$.
\item
We convert this into an equation for the PGF:
\begin{align*}
\pd{}{t} \Phi(y,t) &=\pd{}{t} \sum_j \phi_j(t) y^j \\
&= \sum_j \dot{\phi}_j(t) y^j\\
&= \sum_j \beta(j-1)\phi_{j-1}y^j + \gamma(j+1)\phi_{j+1}y^j-(\beta+\gamma)j \phi_j y^j\\
&=  \beta \sum_j (y^2-y)\pd{}{y} \phi_{j-1}y^{j-1}  + \gamma \sum_j \pd{}{y} \phi_{j+1}y^{j+1} - (\beta+\gamma) \sum_j y \pd{}{y} \phi_j y^j\\
&=[\beta(y^2-y) +\gamma - (\beta+\gamma)y\pd{}{y}\Phi(y,t)\\
&= (\beta+\gamma)\left[\hat{\mu}(y) - y \right] \pd{}{y}\Phi(y,t) 
\end{align*}
\end{myenumerate}
\end{solution}

\begin{exercise}
In this exercise we follow~\cite{allen2017primer,bailey1964elements} and derive the PGF version of the backward Kolmogorov equations by directly calculating the rate of change of the probabilities of the states.  Define $\phi_{ki}(t)$ to be the probability of $i$ infections at time $t$ given that there were $k$ infections at time $0$.  Although we assume that at time $0$ there is a single infection, we will need to derive the equations for arbitrary $k$.  
\begin{myenumerate}
\item Explain why 
\[
\phi_{ki}(t+\Delta t) = \phi_{ki}(t) - k(\beta+\gamma)\phi_{ki}(t) \Delta t + k(\beta \phi_{(k+1)i}(t) + \gamma \phi_{(k-1)i}(t)) + \littleo(\Delta t)
\]
 for small $\Delta t$.
\item By using the definition of the derivative
$\dot{\phi}_{ki} = \lim_{\Delta t \to 0} \frac{\phi_{ki}(t+\Delta t)-\phi_{ki}(t)}{\Delta t}$, find $\dot{\phi}_{ki}$
\end{myenumerate}
Define $\Phi(y,t|k) = \sum_i \phi_{ki}y^i$ to be the PGF for the number of active infections assuming that there are $k$ initial infections.
\begin{continuemyenumerate}
\item Show that
\[
\dot{\Phi}(y,t|1) = -(\beta+\gamma) \Phi(y,t|1) + \beta \Phi(y,t|2) + \gamma\Phi(y,t|0)
\]
\item Explain why $\Phi(y,t|k) = \Phi(y,t|1)^k$.
\item Complete the derivation of Equation~\eqref{eqn:backward_PGF}.
\end{continuemyenumerate}
\end{exercise}
\begin{solution}\mbox{}
\begin{myenumerate}
\item If $\Delta t$ is very small then to leading order either $0$ or $1$ event occurs in the first $\Delta t$ units of time  
\begin{itemize}
\item The probability of $0$ events if we start with $k$ infections is $1-k(\beta+\gamma)\Delta t + \littleo(\Delta t) $.  In this case the probability of $i$ infections at time $t+\Delta t$ is the same as the probability of $i$ infections at time $t$ given the initial condition.
\item The probability of $1$ event occurring and it being an infection is $k\beta \Delta t + \littleo(\Delta t)$.  In this the probability of $i$ infections at time $t+\Delta t$ is the same as the probability of $i$ infections at time $t$ if we start with $k+1$ infections, $\phi_{(k+1)i}$.
\item The probability of 1 event occurring and it being a recovery is $k \gamma \Delta t + \littleo(\Delta t)$.  Following the previous case, this corresponds to starting with $k-1$ infections, $\phi_{(k-1)i}$.
\end{itemize}
Adding these together gives the result.
\item \[
\dot{\phi}_{ki} = -k(\beta+\gamma)\phi_{ki} + k\beta \phi_{(k+1)i} + k\gamma \phi_{(k-1)i}
\]
\item 
\begin{align*}
\dot{\Phi}(y, t | 1) &= \sum_i\dot{\phi}_{1i}y^i\\
&= \sum_i -(\beta+\gamma)\phi_{1i}y^i + \beta \phi_{2i}y^i+\gamma\phi_{0i}y^i\\
&= - (\beta+\gamma) \Phi(y,t|1) + \beta \Phi(y,t|2) + \gamma \Phi(y,t|0)
\end{align*}
\item This follows from $k$ applications of Property~\ref{property:sum2product}.
\item So we use $\Phi(y,t|2)=\Phi(y,t|1)^2$ and $\Phi(y,t|0)=1$ to give
\begin{align*}
\dot{\Phi}(y,t|1) &= -(\beta+\gamma)\Phi(y,t|1)+ \beta \Phi(y,t|1)^2 + \gamma\\
&= (\beta+\gamma) \left[ \frac{\beta \Phi(y,t|1)^2 + \gamma}{\beta+\gamma} - \Phi(y,t|1) \right]\\
&= (\beta+\gamma) [\hat{\mu(\Phi(y,t|1))} - \Phi(y,t|1)]
\end{align*}
Replacing $\Phi(y,t|1)$ with $\Phi(y,t)$ completes the derivation.
\end{myenumerate}
\end{solution}

\begin{exercise}
Define $\Phi(y,t|k)$ to be the PGF for the probability of having $i$ infections at time $t$ given $k$ infections at time $0$.

\begin{myenumerate}
\item Explain why $\Phi(y,t|k) = [\Phi(y,t)]^k$. 
\item Show that if we substitute $\Phi(y,t|k)=[\Phi(y,t)]^k$ in place of $\Phi(y,t)$ in Eq.~\eqref{eqn:forward_PGF} the equation remains true with the initial condition $y^k$.
\item Show that if we substitute $\Phi(y,t|k)=[\Phi(y,t)]^k$ in place of $\Phi(y,t)$ in equation~\eqref{eqn:backward_PGF} we do not get a true equation.
\end{myenumerate} 
So Eq.~\eqref{eqn:forward_PGF} applies regardless of the initial condition, but Eq.~\eqref{eqn:backward_PGF} is only true for the specific initial condition of one infection.
\end{exercise}
\begin{solution}\mbox{}
\begin{myenumerate}
\item This follows from $k$ applications of Property~\ref{property:sum2product}.
\item On the left hand side, the substitution yields 
\[
\pd{}{t} \Phi(y,t)^k = k \Phi(y,t)^{k-1} \pd{}{t} \Phi(y,t)
\]
and on the right hand side we get
\[ 
  (\beta+\gamma) [\hat{\mu}(y) - y] \pd{}{y} \Phi(y,t)^k = k \Phi(y,t)^{k-1} (\beta+\gamma) [\hat{\mu}(y) - y] \pd{}{y} \Phi(y,t)
\]
The $k \Phi(y,t)^{k-1}$ term on each side cancel, and we have a correct equation.
\item On the left hand side the substitution yields
\[
\pd{}{t} \Phi(y,t)^k = k \Phi(y,t)^{k-1} \pd{}{t} \Phi(y,t)
\]
as before, but on the right hand side we get
\[
(\beta+\gamma) [\hat{\mu}(\Phi(y,t)^k) - \Phi(y,t)^k]
= \beta\Phi(y,t)^{2k} + \gamma - (\beta+\gamma)\Phi(y,t)^k
\]
There is no common term we can cancel to get back to the original equation.
\end{myenumerate}
\end{solution}

\begin{exercise}
Let $\Phi(y,t|k)$ be the PGF for the number of infections assuming there are initially $k$ infections.  Derive the backward Kolmogorov equation for $\Phi(y,t|k)$.  Note that some of the $\Phi$s in the derivation above would correspond to $\Phi(y,t|1)$ and some of them to $\Phi(y,t|k)$.
\end{exercise}
\begin{solution}
We have 
\begin{align*}
\dot{\Phi}(y, t | k ) &= \sum_i \dot{\phi}_{ki} y^i\\
&= \sum_i\left[ -k (\beta+\gamma) \phi_{ki} + k\beta \phi_{(k+1)i} + k\gamma
  \phi_{(k-1)i} \right]y^i\\
&= -k(\beta+\gamma)  \Phi(y, t |k)  + k \beta \Phi(y, t| k+1) + k
  \gamma \Phi(y, t | k-1)
\end{align*}
\end{solution}

\begin{exercise}
\textbf{Comparison of the formulations}
\begin{myenumerate}
\item Using Eq.~\eqref{eqn:forward_PGF} derive an equation for $\dot{\alpha}$ where $\alpha(t) = \Phi(0,t)$.  What, if any, additional information would you need to solve this numerically?
\item Using Eq.~\eqref{eqn:backward_PGF}, derive Equation~\eqref{eqn:cts_extinction} for $\dot{\alpha}$ where $\alpha(t) = \Phi(0,t)$.  What, if any, additional information would you need to solve this numerically?
\end{myenumerate}
\end{exercise}
\begin{solution}\mbox{}
\begin{myenumerate}
\item We know that $\alpha(0)=0$. Substituting into Equation~\eqref{eqn:forward_PGF}
\begin{align*}
\dot{\alpha}(t) &= (\beta+\gamma) [\hat{\mu}(0)-0] \pd{}{y} \Phi(0,t)\\
&= \gamma \pd{}{y} \Phi(0,t)\\
&= \gamma \phi_1(t)
\end{align*}
We would also need to know the probability of having one infection at any given time.
\item Again, we know that $\alpha(0)=0$.  Substituting into Equation~\eqref{eqn:backward_PGF} yields
\[
\dot{\alpha} = (\beta+\gamma) [\hat{\mu}(\alpha) - \alpha]
\]
We would not need any additional information to solve this
\end{myenumerate}
\end{solution}

\begin{exercise}
\textbf{Full solution}
\begin{myenumerate}
\item Show that Eq.~\eqref{eqn:backward_PGF} can be written 
\[
\pd{}{t}\Phi(y,t) = (\gamma - \beta \Phi(y,t)) (1-\Phi(y,t))
\]
\item Using partial fractions, set up an integral which you could use to solve for $\Phi(y,t)$ analytically (you do not need to do all the algebra to solve it).
\end{myenumerate}
\end{exercise}
\begin{solution}
\mbox{}
\begin{myenumerate}
\item If we substitute for $\hat{\mu}$, we get
\begin{align*}
\pd{}{t} \Phi(y,t) &= (\beta+\gamma) \left( \frac{\beta [\Phi(y,t)]^2 +
    \gamma}{\beta+\gamma}  - \Phi(y,t)\right)\\
&= \beta[\Phi(y,t)]^2 + \gamma - (\beta+\gamma)\Phi(y,t)
\end{align*}
and
\begin{align*}
(\gamma-\beta \Phi(y,t) ) (1-\Phi(y,t)) &= \gamma - \gamma\Phi(y,t) -
                                          \beta\Phi(y,t) + \beta[\Phi(y,t)]^2\\
&=  \beta [\Phi(y,t)]^2 - (\beta+\gamma)\Phi(y,t) + \gamma
\end{align*}
So these are equal.
\item We have
\[
\int \frac{1}{(\gamma - \beta \Phi(y,t)) (1-\Phi(y,t))}
\mathrm{d}\Phi(y,t) = \int \mathrm{d}t
\]
We write the first integrand as $ \frac{1}{(\gamma - \beta \Phi(y,t)) (1-\Phi(y,t))}=\frac{A}{\gamma-\beta\Phi(y,t)} +
\frac{B}{1-\Phi(y,t)}$ and must solve for $A$ and $B$.  Then this is integrable.
\end{myenumerate}
\end{solution}


\begin{exercise}
Argue from their definitions that $\Phi(y,t) = \Pi(y,z,t)|_{z=1}$.
\end{exercise}
\begin{solution}
We have $\Phi(y,t) = \sum_i \phi_i(t) y^i$ where $\phi_i(t)$ is the
probability of $i$ infections at time $t$.  Similarly
\begin{align*}
\Pi(y,z,t)|_{z=1} &= \left.\sum_{i,r} \pi_{i,r} y^i z^r \right|_{z=1}\\
&= \sum_{i}\sum_r  \pi_{i,r} y^i1^r\\
&= \sum_i \left(y^i \sum_r \pi_{i,r}\right)
\end{align*}
and because $\sum_r \pi_{i,r} = \phi_i$, the result follows.
\end{solution}

\begin{exercise}
Derive Theorem~\ref{thm:cts_current} from Theorem~\ref{thm:kolmogorov}.
\end{exercise}
\begin{solution}
Note that $\Phi(y,t) = \Pi(y,1,t)$.  Setting $z=1$ into the equation for $\Pi$ yields the equation for $\Phi$.
\end{solution}

\begin{exercise}
Derive Theorem~\ref{thm:cts_final_size} from Theorem~\ref{thm:kolmogorov}.
\end{exercise}
\begin{solution}
Note that for $z<1$, \ $\Omega_\infty(z) = \lim_{t \to \infty}
\Pi(0,z,t)$ and for $z=1$ it is $1$.  

As $t \to \infty$, we must have $\pd{}{t} Pi(0,z,t) = 0$ since the
system approaches a disease-free state and thus all coefficients
converge to a constant.  For this to hold,
Equation~\eqref{eqn:backward_result} yields
\[
\lim_{t\to\infty} \hat{\mu}(\Pi(0,z,t), z) = \lim_{t\to\infty}\Pi(0, z, t)
\]
Substituting with $\Omega_\infty$ completes the result.

\end{solution}


\begin{exercise}
\label{exercise:cts_final}
\textbf{Equivalence of continuous and discrete final size
  distributions.}

  Show by direct substitution that if
  $\Omega_\infty(z) = \hat{\mu}(\Omega_\infty(z),z)$
 then
  $\Omega_\infty(z) = z \mu(\Omega_\infty(z))$ where
  $\mu(y) = \gamma/(\beta+\gamma-\beta y)$ is the PGF for the offspring distribution found in Exercise~\ref{exercise:continuous_vs_discrete}.  
\end{exercise}
\begin{solution}
We take
\begin{align*}
\Omega_\infty(z) &= \hat{\mu}(\Omega_\infty(z), z)\\
&= \frac{\beta [\Omega_\infty(z)]^2 + \gamma z}{\beta+\gamma}
\end{align*}
So
\begin{align*}
(\beta+\gamma) \Omega_\infty(z) &= \beta [\Omega_\infty(z)]^2 + \gamma
                                  z\\
\Omega_\infty(z)[\beta+\gamma - \beta \Omega_\infty(z)]  &= \gamma z\\
\Omega_\infty(z) & = \frac{\gamma z}{\beta+\gamma-\beta\Omega_\infty(z)}
\end{align*}
\end{solution}


\begin{exercise} 
We revisit the derivations of the usual mass action SIR ODEs. Following Example~\ref{example:SIS_MA_eqns},
\label{exercise:SIR_mass_action}
\begin{myenumerate}
\item Derive $[\dot{S}]$ in terms of $[SI]$.
\item Derive $[\dot{I}]$ in terms of $[SI]$ and $[I]$.
\item Using $[S]+[I]+[R]=N$, derive $[\dot{R}]$.
\end{myenumerate}
\end{exercise}
\begin{solution}
\mbox{}
\begin{myenumerate}
\item We have 
\begin{align*}
[\dot{S}] &= \left . \pd{}{t} \pd{}{x} \Xi(x,y,t)\right|_{x=y=1}\\
& = \left . \pd{}{x} \pd{}{t} \Xi(x,y,t)\right|_{x=y=1}\\
&= \left . \pd{}{x}\left( \frac{\beta(y^2-xy)}{N}\pd{}{x}\pd{}{y}\Xi(x,y,t) +
  \gamma(1-y) \pd{}{y} \Xi(x,y,t) \right)\right |_{x=y=1}\\
&= \left. -\frac{\beta y}{N} \pd{}{x}\pd{}{y} \Xi(x,y,t)  +
  \frac{\beta(y^2-xy)}{N}\pds{}{x} \pd{}{y} \Xi(x,y,t) + \gamma (1-y)
  \pd{}{x}\pd{}{y} \Xi(x,y,t) \right|_{x=y=1}\\
&= - \frac{\beta}{N} \pd{}{x}\pd{}{y} \Xi(1,1,t)\\
&= - \frac{\beta}{N} [SI]
\end{align*}
\item We have
\begin{align*}
[\dot{I}] &= \left . \pd{}{t} \pd{}{y} \Xi(x,y,t) \right|_{x=y=1}\\
&= \left . \pd{}{y} \left( \frac{\beta(y^2-xy)}{N}\pd{}{x}\pd{}{y}\Xi(x,y,t) +
  \gamma(1-y) \pd{}{y} \Xi(x,y,t) \right)\right |_{x=y=1}\\
&= \left. \frac{\beta(2 y - x)}{N} \pd{}{x}\pd{}{y} \Xi(x,y,t) +
  \frac{\beta(y^2-xy)}{N}\pd{}{x}\pds{}{y} \Xi(x,y,t) - \gamma
  \pd{}{y} \Xi(x,y,t) + \gamma(1-y)\pds{}{y} \Xi(x,y,t)
  \right|_{x=y=1}\\
&=\frac{\beta}{N} \pd{}{x}\pd{}{y} \Xi(1,1,t)  - \gamma
  \pd{}{y}\Xi(1,1,t)\\
&= \frac{\beta}{N} [SI] - \gamma [I]
\end{align*}
\item As $[\dot{S}] + [\dot{I}] + [\dot{R}] = 0$, we conclude
\[
[\dot{R}] = -[\dot{S}] - [\dot{I}] = \gamma [I]
\]
\end{myenumerate}
\end{solution}


\section{Large-time dynamics}
\label{sec:large-time}

We now look at how PGFs can be used to develop simple models of SIR
disease spread in the large population limit when the disease infects
a nonzero fraction of the population.  In this limit, the early-time
approaches derived before break down because depletion of the susceptible population is important.  The later-time models of Section~\ref{sec:full} are impractical because of the $N \to \infty$ limit and are more restricted due to the continuous-time assumption.

\ifsolns
\else
\subsection{SIR disease and directed graphs.}
\fi
\label{sec:EPN}

\begin{figure}
\begin{center}
\framebox{\begin{tikzpicture}[scale=0.7]
\node[SusCirc] (node0) at (4.226119,3.383105) {$0$};
\node[SusCirc] (node1) at (1.400000,1.502243) {$1$};
\node[SusCirc] (node2) at (5.066435,1.502939) {$2$};
\node[SusCirc] (node3) at (1.485215,3.000000) {$3$};
\node[SusCirc] (node4) at (4.666105,4.506938) {$4$};
\node[SusCirc] (node5) at (6.950803,2.727808) {$5$};
\node[SusCirc] (node6) at (2.411113,4.594994) {$6$};
\node[SusCirc] (node7) at (2.717351,3.637824) {$7$};
\node[SusCirc] (node8) at (3.312308,2.302973) {$8$};
\node[SusCirc] (node9) at (7.000000,4.214504) {$9$};
\node[SusCirc] (node10) at (3.200000,0.502243) {$10$};
\node[SusCirc] (node11) at (6.400000,0.502243) {$11$};
\draw [-{Latex[length=3mm,width=1mm,angle'=30]}] (node0) -- (node2);
\draw [-{Latex[length=3mm,width=1mm,angle'=30]}] (node0) -- (node7);
\draw [-{Latex[length=3mm,width=1mm,angle'=30]}] (node1) -- (node10);
\draw [-{Latex[length=3mm,width=1mm,angle'=30]}] (node2) -- (node10);
\draw [-{Latex[length=3mm,width=1mm,angle'=30]}] (node3) -- (node8);
\draw [-{Latex[length=3mm,width=1mm,angle'=30]}] (node3) -- (node7);
\draw [-{Latex[length=3mm,width=1mm,angle'=30]}] (node4) -- (node5);
\draw [-{Latex[length=3mm,width=1mm,angle'=30]}] (node4) -- (node7);
\draw [-{Latex[length=3mm,width=1mm,angle'=30]}] (node5) -- (node4);
\draw [-{Latex[length=3mm,width=1mm,angle'=30]}] (node5) -- (node8);
\draw [-{Latex[length=3mm,width=1mm,angle'=30]}] (node6) -- (node4);
\draw [-{Latex[length=3mm,width=1mm,angle'=30]}] (node7) -- (node8);
\draw [-{Latex[length=3mm,width=1mm,angle'=30]}] (node7) -- (node4);
\draw [-{Latex[length=3mm,width=1mm,angle'=30]}] (node8) -- (node5);
\draw [-{Latex[length=3mm,width=1mm,angle'=30]}] (node9) -- (node2);
\draw [-{Latex[length=3mm,width=1mm,angle'=30]}] (node10) -- (node2);
\draw [-{Latex[length=3mm,width=1mm,angle'=30]}] (node11) -- (node5);
\draw [-{Latex[length=3mm,width=1mm,angle'=30]}] (node11) -- (node10);

\end{tikzpicture}}\qquad
\framebox{\begin{tikzpicture}[scale=0.7]
\node[SusCirc] (node0) at (4.226119,3.383105) {$0$};
\node[SusCirc] (node1) at (1.400000,1.502243) {$1$};
\node[SusCirc] (node2) at (5.066435,1.502939) {$2$};
\node[SusCirc] (node3) at (1.485215,3.000000) {$3$};
\node[InfCirc] (node4) at (4.666105,4.506938) {$4$};
\node[InfCirc] (node5) at (6.950803,2.727808) {$5$};
\node[InfCirc] (node6) at (2.411113,4.594994) {$6$};
\node[InfCirc] (node7) at (2.717351,3.637824) {$7$};
\node[InfCirc] (node8) at (3.312308,2.302973) {$8$};
\node[SusCirc] (node9) at (7.000000,4.214504) {$9$};
\node[SusCirc] (node10) at (3.200000,0.502243) {$10$};
\node[SusCirc] (node11) at (6.400000,0.502243) {$11$};
\draw [-{Latex[length=3mm,width=1mm,angle'=30]}] (node0) -- (node2);
\draw [-{Latex[length=3mm,width=1mm,angle'=30]}] (node0) -- (node7);
\draw [-{Latex[length=3mm,width=1mm,angle'=30]}] (node1) -- (node10);
\draw [-{Latex[length=3mm,width=1mm,angle'=30]}] (node2) -- (node10);
\draw [-{Latex[length=3mm,width=1mm,angle'=30]}] (node3) -- (node8);
\draw [-{Latex[length=3mm,width=1mm,angle'=30]}] (node3) -- (node7);
\draw [-{Latex[length=3mm,width=1mm,angle'=30]}] (node5) -- (node4);
\draw [-{Latex[length=3mm,width=1mm,angle'=30]}] (node7) -- (node4);
\draw [-{Latex[length=3mm,width=1mm,angle'=30]}] (node8) -- (node5);
\draw [-{Latex[length=3mm,width=1mm,angle'=30]}] (node9) -- (node2);
\draw [-{Latex[length=3mm,width=1mm,angle'=30]}] (node10) -- (node2);
\draw [-{Latex[length=3mm,width=1mm,angle'=30]}] (node11) -- (node5);
\draw [-{Latex[length=3mm,width=1mm,angle'=30]}] (node11) -- (node10);
\draw [-{Latex[length=3mm,width=1mm,angle'=30]}, line width=0.5mm, color = colorI] (node4) -- (node5);
\draw [-{Latex[length=3mm,width=1mm,angle'=30]}, line width=0.5mm, color = colorI] (node4) -- (node7);
\draw [-{Latex[length=3mm,width=1mm,angle'=30]}, line width=0.5mm, color = colorI] (node5) -- (node8);
\draw [-{Latex[length=3mm,width=1mm,angle'=30]}, line width=0.5mm, color = colorI] (node6) -- (node4);
\draw [-{Latex[length=3mm,width=1mm,angle'=30]}, line width=0.5mm, color = colorI] (node7) -- (node8);
\end{tikzpicture}}\qquad
\framebox{\begin{tikzpicture}[scale=0.7]
\node[RecCirc] (node0) at (4.226119,3.383105) {$0$};
\node[SusCirc] (node1) at (1.400000,1.502243) {$1$};
\node[SusCirc] (node2) at (5.066435,1.502939) {$2$};
\node[RecCirc] (node3) at (1.485215,3.000000) {$3$};
\node[RecCirc] (node4) at (4.666105,4.506938) {$4$};
\node[RecCirc] (node5) at (6.950803,2.727808) {$5$};
\node[RecCirc] (node6) at (2.411113,4.594994) {$6$};
\node[RecCirc] (node7) at (2.717351,3.637824) {$7$};
\node[RecCirc] (node8) at (3.312308,2.302973) {$8$};
\node[SusCirc] (node9) at (7.000000,4.214504) {$9$};
\node[SusCirc] (node10) at (3.200000,0.502243) {$10$};
\node[RecCirc] (node11) at (6.400000,0.502243) {$11$};
\draw [-{Latex[length=3mm,width=1mm,angle'=30]}] (node0) -- (node2);
\draw [-{Latex[length=3mm,width=1mm,angle'=30]}] (node1) -- (node10);
\draw [-{Latex[length=3mm,width=1mm,angle'=30]}] (node2) -- (node10);
\draw [-{Latex[length=3mm,width=1mm,angle'=30]}] (node3) -- (node8);
\draw [-{Latex[length=3mm,width=1mm,angle'=30]}] (node4) -- (node5);
\draw [-{Latex[length=3mm,width=1mm,angle'=30]}] (node4) -- (node7);
\draw [-{Latex[length=3mm,width=1mm,angle'=30]}] (node5) -- (node8);
\draw [-{Latex[length=3mm,width=1mm,angle'=30]}] (node7) -- (node8);
\draw [-{Latex[length=3mm,width=1mm,angle'=30]}] (node9) -- (node2);
\draw [-{Latex[length=3mm,width=1mm,angle'=30]}] (node10) -- (node2);
\draw [-{Latex[length=3mm,width=1mm,angle'=30]}] (node11) -- (node10);
\draw [-{Latex[length=3mm,width=1mm,angle'=30]}, line width=0.5mm, color=colorR] (node11) -- (node5);
\draw [-{Latex[length=3mm,width=1mm,angle'=30]}, line width=0.5mm, color=colorR] (node5) -- (node4);
\draw [-{Latex[length=3mm,width=1mm,angle'=30]}, line width=0.5mm, color=colorR] (node8) -- (node5);
\draw [-{Latex[length=3mm,width=1mm,angle'=30]}, line width=0.5mm, color=colorR] (node0) -- (node7);
\draw [-{Latex[length=3mm,width=1mm,angle'=30]}, line width=0.5mm, color=colorR] (node3) -- (node7);
\draw [-{Latex[length=3mm,width=1mm,angle'=30]}, line width=0.5mm, color=colorR] (node6) -- (node4);
\draw [-{Latex[length=3mm,width=1mm,angle'=30]}, line width=0.5mm, color=colorR] (node7) -- (node4);
\end{tikzpicture}}
\end{center}
\caption{(\textbf{Left}) A twelve-individual population, after the \emph{a priori}
  assignment of who would transmit to whom if ever infected by the SIR
  disease  (the delay
  until transmission is not shown).  Half of
  the nodes have zero potential infectors and half have $3$.  Half of
  the nodes have $1$ potential offspring and half have $2$.  So the
  offspring distribution has PGF $(x+x^2)/2$ while the ancestor distribution
  has PGF
  $\chi(x) = (1+x^3)/2$.  (\textbf{Middle}) If node $6$ is initially
  infected, the infection will reach node $4$ who will transmit to $5$
and $7$, and eventually infection will also reach $8$ and $2$ before
further transmissions fail because nodes are already infected.  If
however, it were to start at $9$, then it would reach $10$, from which
it would spread only to $2$.  (\textbf{Right}) By tracing backwards from
an individual, we can determine which initial infections would lead to
infection of that individual.  For example individual $4$ will 
become infected if and only if it is initially infected or 
  $0$, $3$, $5$, $6$, $7$, $8$, or $11$ is an initial infection.}
\label{fig:EPN}
\end{figure}
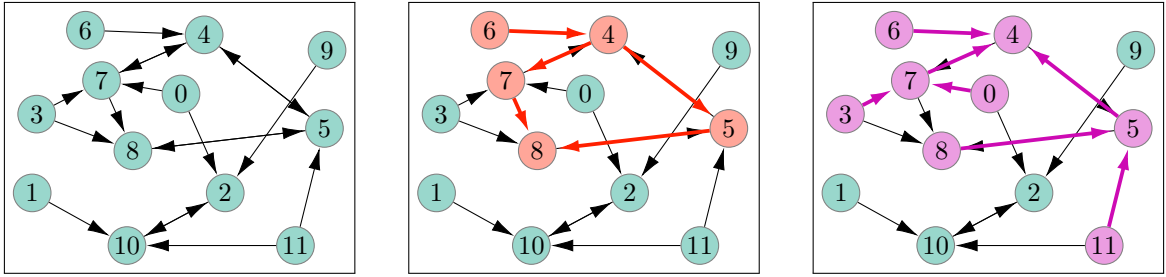

In Section~\ref{sec:discrete_general} we argued that for early times
the continuous-time predictions are equivalent to discrete-time
predictions because we can classify infections by the length of the
transmission chain to them from the index case.  For SIR disease this
argument extends beyond early times.

To see this, we assume that prior to the disease introduction, we know
for each individual what would happen if he ever becomes infected as
in Figure~\ref{fig:EPN}.  In particular, we know how long his
infection would last, to whom he would transmit, and how long the
delays from his infection to onwards transmission would be.  The process of choosing these in advance, selecting the initial infection(s), and tracing infection from there is equivalent to choosing the initial infection(s) and then choosing the transmissions while the infection process is traced out.

By assigning who transmits to whom (and how long the delays are), we
have defined a weighted directed graph whose edges represent the
potential transmissions and weights represent the
delays~\cite{kenah:EPN,EoNbook}.  A node $v$ will become infected if and only
if there is at least one directed path from an initially infected node
$u$ to $v$.  The time of $v$'s infection is given by the least sum of
all paths from initially infected nodes to $v$.  We note that the
transmission process could be quite complex: the duration of a node's
infection and the delays from time of infection to time of onwards
transmissions can have effectively arbitrary distributions, and we
could still build a similar directed graph.

This directed graph is a useful structure to study because it encodes
the outbreak in a single static object, as opposed to a dynamic
process.  There is significant study of the structure of such directed
graphs~\cite{broder,dorogovtsev}.   Much of it focuses on the size of
out-compoents of a node (that is, for a given node, what fraction of
the population can be reached following the edges forwards) or the
in-components (that is, from what fraction of the population is it
possible to reach a given node by following edges forwards).

\ifsolns
\else
\subsection{Final size relations for SIR epidemics}
\fi
\label{sec:final_size}
We now derive final size relations for SIR epidemics in the large
population limit.  We begin with the assumption that a single node is
initially infected and that an epidemic happens.  

We use the mapping of the SIR epidemic to a directed graph $G$.  Assume that a single node $u$ is chosen to be infected.  Consider a node $v$.  
%
The probability $v$ is infected is the probability that $u$ is in her in-component, and so it equals the proportion of $G$ that is in the in-component of $v$.  In the limit as $G$ becomes infinite, there are a few possibilities.  We are interested in what happens when an epidemic occurs, so we can assume that $u$ has a large out-component (in the sense that the out-component takes up a non-zero fraction of $G$ in the $N\to\infty$ limit)~\cite{broder}:
\begin{itemize}
\item If $v$ has a small in-component, then almost surely $u$ is not in the in-component and so almost-surely $v$ is not infected.
\item If $v$ has a large in-component, then almost surely it contains a node $w$ that lies in the out-component of $u$.  The existence of $w$ then implies the existence of a path from $u$ to $w$ to $v$, so $v$ is in $u$'s out-component and $v$ becomes infected.
\end{itemize}
Thus, if $u$ causes an epidemic in the large $N$ limit, then the probability that $v$ becomes infected equals the probability that $v$ has a large in-component.  So the size of an epidemic (if it happens) is simply the probability a random individual has a large in-component.

We approach the question of whether $v$ has a large in-component in the same way we approached the question of whether $u$ causes a large chain of infections (i.e., whether $u$ has a large out-component).  We define the PGF of the \emph{ancestor distribution} to be the function $\chi(x)$ defined by 
\[
\chi(x) = \sum_i p_i x^i
\]
where $p_i$ is the probability that a random node in the directed
graph has in-degree $i$.  That is, there are exactly $i$ nodes that
would directly transmit to the randomly chosen node if they were ever
infected.  So the probability an individual is not infected
$S(\infty)/N$ solves $x
= \chi(x)$, choosing the smaller solution when two solutions exist.
Since the proportion infected is $r(\infty) = R(\infty)/N = 1
-S(\infty)/N$, we can conclude
\begin{thm}
\label{thm:chi_final}
  Assume that an outbreak begins with a single infected individual and an epidemic results.  In the large $N$ limit, the expected cumulative proportion infected $r(\infty) = R(\infty)/N$ solves
\[
r(\infty) = 1-\chi(1-r(\infty))
\]
where $\chi(x)$ is the PGF of the ancestor distribution.  If there are multiple solutions we choose the larger solution for $r(\infty)$ in $[0,1]$.
\end{thm}

Under common assumptions, the population is large, the average number
of transmissions an individual causes is $\Ro$, and the recipient is
selected uniformly at random.  Under these assumptions the ancestor
distribution is Poisson with mean $\Ro$.  
So $\chi(x) = e^{-\Ro(1-x)}$.   xThen
\begin{equation}
r(\infty) = 1-e^{-\Ro r(\infty)} \, .
\end{equation}
Deriving this result does not depend on the duration of infections, or even
on the distribution of factors affecting infectiousness.  The
assumptions required are that an epidemic starts from a single infected
individual, that each transmission reaches a randomly chosen member of
the population, that all individuals have equal susceptibility, and
the average individual will transmit to $\Ro$ others.  This result is general across a wide range of assumptions about the infectious process.  

Restating this we have:
\begin{cor}
Assume that an SIR disease is spreading in a well-mixed population with homogeneous susceptibility.  Assuming that the initial fraction infected is infinitesimal and an epidemic occurs, the final size satisfies
\begin{equation}
r(\infty) = 1- e^{-\Ro r(\infty)}
\end{equation}
where $\Ro$ is the reproductive number of the disease.
\end{cor}
This explains many of the results of~\cite{ma, miller:final}, and our observation in example~\ref{example:cdf} that the epidemic size depends on $\Ro$ and not on any other property of the offspring distribution.  A closely-related derivation is provided by~\cite[Section 1.3]{diekmann:textbook}.

\ifsolns
\else
\subsection{Discrete-time SIR dynamics}
\fi
\label{sec:large_time_discrete_SIR}
We now take a discrete-time approach, similar
to~\cite{valdez2012temporal,miller:ebcm_overview} and~\cite[chapter
6]{EoNbook}. We will assume that at generation $g=0$ the disease is
introduced by infecting a proportion $\rho$ uniformly at random
leaving the remainder susceptible.  We assume that the population is
very large and that the number of infections is large enough that the
dynamics can be treated as deterministic.   Our results can be adapted to other initial conditions (for
example, to account for nonzero $R$ in the initial condition). 

We assume that $\chi(x)$ is known and that there is no correlation between how susceptible an individual is and how infectious that individual is.  Thus at generation $g$, the expected number of transmissions occurring is $\Ro I(g)$, and how the recipients are chosen depends on $\chi$.

Let $v$ be a randomly chosen member of the population.  The probability that $v$'s randomly chosen ancestor has not yet been infected by generation $g-1$ is $S(g-1)/N$.  The probability $v$ is susceptible at generation $g$ is the probability $v$ was initially susceptible, $1-\rho$, times the probability $v$ has not received any transmissions, $\chi(S(g-1)/N)$ (see Exercise~\ref{exercise:chi}).

So for $g>0$ we arrive at
\begin{align*}
S(g) &= (1-\rho)N\chi(S(g-1)/N)\\
I(g) &= N-R(g)-S(g)\\
R(g) &= R(g-1)+I(g-1)
\end{align*}
with 
\[
S(0)=1-\rho, \quad I(0) = \rho, \quad R(0)=0 \, .
\]
So we have
\begin{thm}
\label{thm:discrete_large_dynamics}
Assume that $\chi(x)$ is the PGF of the ancestor distribution and
assume there is no correlation between infectiousness and
susceptibility of a given individual.  Further assume that at
generation $0$ a fraction $\rho$ is randomly infected in the
generation-based discrete-time model. Then in the large population limit
\begin{subequations}
\begin{align}
S(g) &= (1-\rho)N\chi(S(g-1)/N)\\
I(g) &= N-R(g)-S(g) \\
R(g) &= R(g-1)+I(g-1) \, .
\end{align}
With initial conditions
\begin{equation}
S(0)=(1-\rho)N, \quad I(0) = \rho N, \quad R(0)=0 \, .
\end{equation}
\end{subequations}
\end{thm}

We can interpret this in the context of survival functions.  The
function  $(1-\rho)\chi(S(g-1)/N)$ gives the probability that a node has lasted $g$ generations without being infected.

\ifsolns
\else
\subsection{Continuous-time SIR epidemic dynamics}
\fi
\label{sec:large_time_cts_SIR}
We now move to continuous-time SIR epidemics.  We allow for
heterogeneity, assuming that each susceptible individual $u$ receives transmissions at some rate $\kappa_u \beta I(t) /N\ave{K}$, and that the PGF of $\kappa$ is $\psi(x) = \sum_\kappa P(\kappa) x^\kappa$.  We assume $\kappa$ takes only non-negative integer values.

For an initially susceptible individual $u$ with a given $\kappa_u$, the probability of not yet receiving a transmission by time $t$ solves
$\dot{s}_u = - \kappa_u\beta I(t) s_u/N\ave{K}$, which has solution
\[
s_u = e^{-\kappa_u\beta \frac{\int_0^t I(\tau) \mathrm{d}\tau}{N\ave{K}}} \, .
\]
So we can write 
\[
s_u = \theta^{\kappa_u}
\]
where $\theta = e^{-\beta \frac{\int_0^t I(\tau) \mathrm{d}\tau}{N\ave{K}}}
$ and 
\[
\dot{\theta} = -\beta \theta I/N\ave{K} \, .
\]
Considering a random individual of unknown $\kappa$, the probability she was initially susceptible is $1-\rho$ and the probability she has not received any transmissions is $\psi(\theta)$.  So
\[
S(t) = (1-\rho) N\psi(\theta)
\]

Taking $\dot{R} = \gamma I$, we have 
\begin{align*}
\dot{R} &= \gamma I\\
&= -\frac{\gamma N \ave{K}}{\beta} \frac{\dot{\theta}}{\theta} \, .
\end{align*}
Integrating both sides, taking $\theta(0)=1$ and $R(0)=0$, we have
\[
R = -\frac{\gamma N \ave{K}}{\beta} \ln \theta
\]
Taking $I = N-S-R$ we get
\[
I = N\left(1- (1-\rho)\psi(\theta) + \frac{\gamma\ave{K}}{\beta}\ln
  \theta\right)
\]
and so $\dot{\theta}$ becomes
\[
\dot{\theta} = -\beta \theta \left(1- (1-\rho)\psi(\theta) + \frac{\gamma\ave{K}}{\beta}\ln
  \theta\right)/\ave{K}
\]
\begin{thm}
\label{thm:cts_time_EBCM_like}
Assuming that at time $t=0$ a fraction $\rho$ of the population is
randomly infected and that the susceptible individuals each have a
$\kappa$ such that they become infected as a Poisson process with rate
$\kappa \beta I /N \ave{K}$, in the large population limit we have 
\begin{subequations}
\label{sys:cts_time_EBCM_like}
\begin{align}
S &= N(1-\rho)\psi(\theta)\\
I &= N \left( 1- (1-\rho)\psi(\theta) + \frac{\gamma\ave{K}}{\beta}\ln \theta\right)\\
R &= -\frac{\gamma N \ave{K}}{\beta} \ln \theta
\end{align}
where $\psi(x) = \sum_k P(k) x^k$ and the system is governed by a single ODE
\begin{equation}
\dot{\theta} = \frac{-\beta \theta \left(1- (1-\rho)\psi(\theta) + \frac{\gamma\ave{K}}{\beta}\ln
  \theta\right)}{\ave{K}}
\end{equation}
and initial condition 
\begin{equation}
\theta(0)=1 \, .
\end{equation}
\end{subequations}
\end{thm}
As in the discrete-time case, this can be interpreted as a survival function formulation of the SIR model.  Most, if not all, mass-action formulations of the SIR model can be re-expressed in a survival function formulation.  Some examples are shown in the Exercises.

Some very similar systems of equations are developed in~\cite[chapter
6]{EoNbook} and~\cite{miller:ebcm_overview,
  valdez2012temporal,volz:cts_time, miller:volz} where the focus is on networks for which the value of $\kappa$ not only affects the probability of becoming infected, but also of transmitting further.  These references focus on the assumption that an individual's infector remains a contact after transmission, but they contain techniques for studying partnerships with varying duration.

\ifsolns
\else
\subsection{Exercises}
\fi
\begin{exercise}
\textbf{Ancestor distribution for homogeneous well-mixed population.}

Consider an SIR disease in a well-mixed population having $N$
individuals and a given $\Ro$.  Let $v$ be a randomly chosen
individual from the directed graph created by placing  edges from
each node to all those nodes they would transmit to if infected.
\begin{myenumerate}
\item Show that if the average number of offspring is $\Ro$, then so is the average number of infectors.
\item If there are exactly $\Ro N$ edges in the directed graph and
  each recipient is chosen uniformly at
  random from the population (independent of any previous choice),
  argue that the number of transmissions $v$ receives has a binomial
  distribution with $\Ro N$ trials and probability $\Ro/N$.
  (technically we must allow edges from $v$ to $v$)
\item Argue that if $\Ro$ remains fixed as $N \to \infty$, then the number of transmissions $v$ receives is Poisson distributed with mean $\Ro$.
\end{myenumerate}
\end{exercise}
\begin{solution}
\mbox{}
\begin{myenumerate}
\item From Graph Theory, the average in-degree must equal the average
  out-degree.  

Alternately we can note that the average number of
  edges coming out of a node is the total number of edges divided by
  $N$, so the total number of edges is expected to be $\Ro N$.  Then
  the average number of edges in is also the total number divided by
  $N$, that is, $\Ro$.
\item Individual $v$ receives a given transmission with probability
  $1/N$.  There are $\Ro N$ edges, each of which could reach $v$ with
  probability $1/N$.  So this defines a binomial distribution.
\item As $N$ increases, the probability a particular edge goes to $v$
  is $1/N$, while the number of edges is $\Ro N$.  This defines a
  Poisson distribution with mean $\Ro$.
\end{myenumerate}
\end{solution}

\begin{exercise}
\label{exercise:chi}
Explain why for large $N$ the probability $v$ is still
susceptible at generation $g$ if she was initially susceptible is $\chi(S(g-1)/N)$.
\end{exercise}
\begin{solution}
Consider the directed graph created by placing an edge from a node to each
node it would infect if given the opportunity.  $v$ is susceptible at
generation $g$ iff $v$ was initially susceptible, and no ancestor of
$v$ was infected by generation $g-1$.

The probability a randomly chosen ancestor was susceptible at
generation $g-1$ is $S(g-1)/N$.  If $p_i$ is the probability of having
$i$ ancestors, then the probability of being susceptible at generation
$g$ is
\[
\sum_i p_i \left(\frac{S(g-1)}{N}\right)^i
\]
but this is just $\chi(S(g-1)/N)$.
\end{solution}

\begin{exercise}
Use Theorem~\ref{thm:discrete_large_dynamics} to derive a result like Theorem~\ref{thm:chi_final}, but with nonzero $\rho$.
\end{exercise}
\begin{solution}
We have $\lim_{g\to\infty} S(g) = \lim_{g\to\infty}(1-\rho) N
\chi(S(g-1)/N)$ or $S(\infty) = (1-\rho) N \chi(S(\infty)/N)$.
Substituting $r(\infty)  = (N-S(\infty)/N)$ yields
\[
r(\infty) =  1- (1-\rho) \chi(1-r(\infty))
\]
\end{solution}

\begin{exercise}
\label{exercise:final_size}
\textbf{Final size relations}

Consider the continuous time SIR dynamics as given in System~\eqref{sys:cts_time_EBCM_like}
\begin{myenumerate}
\item Assume $\kappa = 1$ for all individuals, and write down the corresponding equations for $S$, $I$, $R$, and $\theta$.
\item At large time $I\to 0$, so $S(\infty) = N-R(\infty)$.  But also $S(\infty)= S(0) \psi(\theta(\infty))$.  By writing $\theta(\infty)$ in terms of $R(\infty)$, derive a recurrence relation for $r(\infty)=R(\infty)/N$ in terms of $r(\infty)$ and $\Ro=\beta/\gamma$.
\item Comment on the relation between your result and Theorem~\ref{thm:chi_final}
\end{myenumerate}
\end{exercise}
\begin{solution}
\mbox{}
\begin{myenumerate}
\item $\psi(x)$ is simply $x$ and $\ave{K}=1$.  We have
\begin{align*}
S &= N(1-\rho) \theta\\
I & = N\left( 1- (1-\rho)\theta + \frac{\gamma }{\beta} \ln
    \theta \right)\\
R &= - \frac{\gamma N }{\beta} \ln \theta\\
\dot{\theta} &= - \frac{\beta I }{N} \theta
\end{align*}
\item Exponentiating the equation for $R$ yields 
\begin{align*}
\theta(\infty) &=
  \exp (-\beta R(\infty)/\gamma N )\\
& = \exp(-\Ro
  r(\infty))
\end{align*}
So we have $\theta(\infty)$ in terms of $r(\infty)$.  Now
\begin{align*}
r(\infty) &= 1- S(\infty)/N \\
&= 1- N(1-\rho)\theta(\infty)/N \\
&=  1- (1-\rho)\theta(\infty)\\
&= 1- (1-\rho) e^{-\Ro r(\infty)}
\end{align*}
\item This generalizes the result of Theorem~\ref{thm:chi_final} for
arbitrary $\rho$.
\end{myenumerate}
\end{solution}

\begin{exercise}
\textbf{Other relations}
\begin{myenumerate}
\item Using the equations from Exercise~\ref{exercise:final_size},
  derive the peak prevalence relation, an expression for the maximum value
  of $I$.  [at the maximum $\dot{I}=0$, so we start by finding
  $\theta$ so that $\dot{S}+\dot{R}=0$.]
\item Similarly, find the peak incidence relation, an expression for
  the maximum rate at which infections occur, $-\dot{S}$.
\end{myenumerate}
\end{exercise}
\begin{solution}
\mbox{}
\begin{myenumerate}
\item If $\dot{S}+\dot{R}=0$, then
\[
N(1-\rho)\dot{\theta} - \frac{\gamma N}{\beta} \frac{1}{\theta}
  \dot{\theta} = 0
\]
Factoring out $N \dot{\theta}$ and taking $\Ro = \beta/\gamma$, we are left
\[
(1-\rho) - \frac{1}{\Ro \theta} = 0
\]
So
\[
\theta = \frac{1}{\Ro(1-\rho)}
\]

Then peak prevalence is
\[
N\left( 1- \frac{1}{\Ro} + \frac{1}{\Ro} \ln
  \frac{1}{\Ro(1-\rho)}\right) = N\left( 1- \frac{1}{\Ro} -
  \frac{1}{\Ro} \ln \Ro(1-\rho) \right)
\]
\item At peak incidence $-\dot{S}$ takes a maximum.  That is,
  $-\dot{\theta}$ is a maximum.  So we need $\diff{}{t} I \theta=0$.
  That is,
\begin{align*}
0 &= \diff{}{t} \left( \theta - (1-\rho)\theta^2 + \frac{\theta}{\Ro}
    \ln \theta\right)\\
&= \dot{\theta} - (1-\rho)2\theta\dot{\theta} +
  \frac{\dot{\theta}}{\Ro}\ln \theta + \frac{1}{\Ro} \dot{\theta}
\end{align*}
Factoring out $\dot{\theta}$ yields
\[
0 = 1 - 2(1-\rho)\theta + \frac{\ln \theta}{\Ro}  + \frac{1}{\Ro}
\]
So
\[
\theta = \frac{1 + (\ln \theta + 1)/\Ro}{2(1-\rho)}
\]
This needs to be solved numerically.  Then plugging this result for
$\theta$ in $-\dot{S} = -(1-\rho)\beta
N\left(1-(1-\rho)\theta+\frac{1}{\Ro} \ln \theta \right)\theta$ yields the peak incidence.
\end{myenumerate}
\end{solution}
\begin{exercise}\textbf{Alternate derivation of $s_u$}. 

 If the rate of transmissions to $u$ is $\beta I \kappa_u /N\ave{K}$, then the expected number of transmissions $u$ has received is $\beta \kappa_u\int_0^t I(\tau) \, \mathrm{d}\tau / N\ave{K}$ and this is Poisson distributed.
\begin{myenumerate}
\item Let $f_u(x)$ be the PGF for the number of transmissions $u$ has received.  Find an expression for $f_u(x)$ in terms of the integral $\int_0^t I(\tau) \mathrm{d}\tau$.
\item Explain why $f_u(0)$ is the probability $u$ is still susceptible.
\item Find $f_u(0)$.
\end{myenumerate}
\end{exercise}
\begin{solution}
\mbox{}
\begin{myenumerate}
\item $f$ is the PGF for the Poisson distribution 
\[
f_u(x) = \exp \left( \beta \kappa_u\frac{\int_0^t I(\tau) \,
  \mathrm{d}\tau}{N\ave{K}}  (x-1)\right)
\]
\item The probability of still be susceptible is the probability of
  having received $0$ transmissions, which in turn is the coefficient
  of $x^0$ in the series expansion of $f_u(x)$.  This is found by
  setting $x=0$.
\item $f_u(0) = exp \left( -\beta \kappa_u\frac{\int_0^t I(\tau) \,
  \mathrm{d}\tau}{N\ave{K}} \right)$
\end{myenumerate}
\end{solution}

\begin{exercise}\textbf{Alternate derivation of
    Theorem~\ref{thm:cts_time_EBCM_like} in the homogeneous case.}

\label{exercise:integrating_factor}
The usual homogeneous SIR equations are
\begin{align*}
\dot{S} &= -\beta I S/N\\
\dot{I} &= \beta I S/N - \gamma I\\
\dot{R} &= \gamma I
\end{align*}
We will derive system~\eqref{sys:cts_time_EBCM_like} for fixed
$\kappa=1$ from this system through the use of an integrating factor.
Set $\theta= e^{-\beta \int_0^t I(\tau) \mathrm{d}\tau/N}$.
\begin{myenumerate}
\item Show that $\dot{\theta} = - \beta I \theta/N$ and so $\dot{\theta}/\theta = -\beta \dot{R}/N\gamma$. 
\item Using the equation for $\dot{S}$ add $\beta I S/N$ to both sides and then divide by (the factor $1/\theta$ is an \emph{integrating factor}).  Show that the expression on the left hand side is $\diff{}{t} S/\theta$ and so
\[
\diff{}{t} S/\theta= 0 \, .
\]
\item Solve for $R$ in terms of $\theta$.
\item Solve for $S$ in terms of $\theta$.
\item Solve for $I$ in terms of $\theta$ using $S+I+R=N$.
\end{myenumerate}
This equivalence was found in~\cite{miller:final} and~\cite{harko2014exact}.
\end{exercise}
\begin{solution}
\mbox{}
\begin{myenumerate}
\item 
By direct observation 
\[
\dot{\theta} = -\frac{\beta I(t)}{N } e^{-\beta \int_0^t I(\tau)
  \mathrm{d}\tau/N} = -\beta I\theta/N
\]
and dividing by $\theta$ and susbtituting $\dot{R} = \gamma I$ gives
\[
\dot{\theta}/\theta = -\beta \dot{R}/N\gamma
\]

\item The operations yield
\[
\dot{S}/\theta + \beta I S/N\theta = 0
\]

So
\begin{align*}
\diff{}{t}  S/\theta &= \frac{\dot{S}}{\theta} - S
                       \dot{\theta}/\theta^2\\
&= -\beta I S/N\theta - S \beta I \theta/N\theta^2\\
&= 0
\end{align*}
\item Since $\dot{R} =\frac{ N\gamma}{\beta}\dot{\theta}/\theta$
we have $R = R(0) + \frac{N\gamma}{\beta} \ln \theta $
\item Since $\diff{}{t} S/\theta=0$, we have $S= S(0) \theta$.
\item So $I = N - S(0)\theta -  R(0) - \frac{N\gamma}{\beta} \ln \theta$.
\end{myenumerate}
\end{solution}

\begin{exercise}
\textbf{Alternate derivation of Theorem~\ref{thm:cts_time_EBCM_like}}.

Consider now a population having many subgroups of susceptibles
denoted by $\kappa$ with the group $\kappa$ receiving transmissions at
rate $\beta \kappa I/N$ per individual.  Once infected, each individual transmits with rate $\beta \ave{K}$ and recovers
with rate $\gamma$.  These assumptions lead to
\begin{align*}
\dot{S}_\kappa &= -\beta\kappa\frac{I}{N\ave{K}}  S_\kappa \\
\dot{I} &= -\gamma I + \beta\frac{I}{N\ave{K}}\sum_\kappa  \kappa S_\kappa\\
\dot{R} &= \gamma I
\end{align*}
Following Exercise~\ref{exercise:integrating_factor}, set $\theta =
e^{-\beta \int_0^t I(\tau) \ \mathrm{d}\tau/N}$ and
derive system~\eqref{sys:cts_time_EBCM_like} from these equations by use of an integrating factor.
\end{exercise}
\begin{solution}\mbox{}
We write
\[
\dot{S}_\kappa + \beta \kappa \frac{I}{N\ave{K}} S_\kappa = 0
\]
and multiply by $1/\theta^\kappa$.  Then 
\[
\diff{}{t} \frac{S_\kappa}{\theta^\kappa} = 0
\]
and
\[
S_\kappa(t) = S_\kappa(0) \theta^\kappa
\]

We assume that $\psi(x) = \sum_\kappa P(\kappa) x^\kappa$. where
$P(\kappa) = N_\kappa/N$.  Starting with a fraction $1-\rho$ randomly
infected yields our expression for $S = \sum_\kappa S_\kappa(t)$.

Since $\dot{R} = \gamma I$ and $\dot{\theta} = -\beta I \theta/N$, we
can substitute to find $\dot{R} = -\gamma N \dot{\theta}/\theta
\beta$.  This can be integrated to find $R$ in terms of $\theta$.

Then simply set $I = N - S - R$ to complete the calculation.
\end{solution}

\section{Multitype populations}
\label{sec:multitype}

We now briefly discuss how PGFs can be applied to multitype populations.  This section is intended primarily as a pointer to the reader to show that it is possible to apply these methods to such populations.  We do not perform a detailed analysis.  

Many populations can be divided into subgroups.  These may be patches in a metapopulation model, genders in a heterosexual sexually transmitted infection model, age groups in an age-structured population, or any of a number of other groupings.  Applications of PGFs to such models have been studied in multiple contexts~\cite{kucharski2015characterizing,reluga2007reservoir}.  

\ifsolns
\else
\subsection{Discrete-time epidemic probability}
\fi
We begin by considering the probability of an epidemic in a discrete-time model.  To set the stage, assume there are $M$ groups and let $p_{i_1,i_2,\cdots, i_M|k}$ be the probability that an individual of group $k$ will cause $i_\ell$ infections in group $\ell$.  Define $\alpha_{g|k}$ to be the probability that a chain of infections starting from an individual of group $k$ becomes extinct within $g$ generations.  

It is straightforward to show that if we define
\[
\psi_k(x_1, x_2, \ldots, x_M) = \sum_{i_1, i_2, \ldots, i_M}
p_{i_1,i_2,\ldots, i_M} x_1^{i_1} x_2^{i_2} \cdots x_M^{i_M}
\] 
then 
\begin{align*}
\alpha_{g|k} &= \sum_{i_1,i_2,\ldots,i_M} p_{i_1,i_2,\cdots, i_M|k} \alpha_{g-1|1}^{i_1}\alpha_{g-1|2}^{i_2}\cdots \alpha_{g-1|M}^{i_M}\\ 
&=\psi_k(\alpha_{g-1|1},\alpha_{g-1|2},\cdots, \alpha_{g-1|M})
\end{align*}
After converting this into vectors we get $\vec{\alpha}_1 =
\vec{\psi}(\vec{0}$.  Iterating $g$ times we have
\begin{equation}
\vec{\alpha}_g= \vec{\psi}^{[g]}(\vec{0})
\label{eqn:multitype_alpha_g}
\end{equation}
Setting $\vec{\alpha}$ to be the limit as $g$ goes to infinity, we
find the extinction probabilities.  Specifically, the $k$-th component
of $\vec{\alpha}$ is the probability of extinction given that the
first individual is of type $k$.  Thus we have:
\begin{thm}
Let 
\begin{itemize}
\item $\vec{\alpha}_{g}=(\alpha_{g|0}, \alpha_{g|1}, \ldots, \alpha_{g|M})$ where
$\alpha_{g|k}$ is the probability a chain of infections starting with
a type $k$ individual will end within $g$ generations
\item and $\vec{\psi} = (\psi_1, \psi_2, \ldots, \psi_M)$ where
  $\psi_k(\vec{x}) = \sum_{i_1, i_2, \ldots, i_M} p_{i_1, i_2, \ldots,
    i_M|k} x_1^{i_1}x_2^{i_2} \cdots x_M^{i_M}$.
\end{itemize}
Then $\vec{\alpha}_g = \vec{\psi}^{[g]}(\vec{0})$.

The vector of eventual extinction probabilities in the infinite
population limit is given by $\vec{\alpha}_\infty = \lim_{g\to\infty}
\vec{\alpha}_g$ and is a solution to $\vec{\alpha}_\infty = \vec{\psi}(\vec{\alpha}_\infty)$.
\end{thm}

We could have derived this directly by showing that the extinction probabilities solve $\vec{\alpha} = \vec{\psi}(\vec{\alpha})$.  In this case it might not be obvious how to solve this multidimensional system of nonlinear equations or how to be certain that the solution found is the appropriate one.  However, by interpreting the iteration in Eqn.~\eqref{eqn:multitype_alpha_g} in terms of the extinction probability after $g$ generations, it is clear that simply iterating starting from $\vec{\alpha}_0 = \vec{0}$ will converge to the appropriate values.  Additionally the values calculated in each iteration have a meaningful interpretation.

\begin{example}
\label{example:ed}
Consider a population made up of many large communities.  We assume an unfamiliar disease is spreading through the population.  When the disease begins to spread in a community, the community learns to recognize the disease symptoms and infectiousness declines.  We assume that we can divide the population into 3 types: primary cases $T_0$, secondary cases $T_1$, and tertiary cases $T_2$.  The infectiousness of primary cases is higher than that of secondary cases which is higher than that of tertiary cases.  Within a community a primary case can cause secondary cases, while secondary and tertiary cases can cause tertiary cases.  All cases can cause new primary cases in other communities.  We ignore multiple introductions to the same community.

We define $n_{ij}$ to be the number of infections of type $T_i$ caused by a type $T_j$ individual, and we assume that we know the joint distribution $p_{n_{00}n_{10}}$, $p_{n_{01}n_{21}}$, and $p_{n_{02}n_{22}}$.  We define
\begin{align*}
\psi_1(x,y,z) &= \sum_{n_{00},n_{10}} p_{n_{00}n_{10}} x^{n_{00}}y^{n_{10}}\\
\psi_2(x,y,z) &= \sum_{n_{01},n_{21}} p_{n_{01}n_{21}} x^{n_{01}}z^{n_{21}}\\
\psi_3(x,y,z) &= \sum_{n_{02},n_{22}} p_{n_{02}n_{22}} x^{n_{02}}z^{n_{22}}
\end{align*}
Note that $\psi_1$ does not depend on $z$ while $\psi_2$ and $\psi_3$ do not depend on $y$.

We define $\vec{\alpha}_0=(0,0,0)$ and set $\vec{\alpha}_g = (\psi_1(\vec{\alpha}_{g-1}),\psi_2(\vec{\alpha}_{g-1}),\psi_3(\vec{\alpha}_{g-1}))$.  Then taking $\vec{\alpha}$ to be the limit as $g \to \infty$, the first entry of $\vec{\alpha}$ is the probability that the disease goes extinct starting from a single primary case.

\end{example}




\ifsolns
\else
\subsection{Continuous-time SIR dynamics}
\fi
Now we consider a continuous-time version of SIR dynamics in a heterogeneous population.  

Assume again that there are $M$ groups and let $\beta_{ij}$ be the
rate at which an individual in group $j$ causes transmissions that go
to group $i$.  Let $\xi_{i}$ be the expected number of transmissions
that an individual in group $i$ has received since time $0$.  Finally
assume that individuals in group $i$ recover at rate $\gamma_i$.  Then
the expected number of transmissions an individual in group $i$ has
received by time $t$ is Poisson distributed with mean $\xi_i$.  The
PGF for the number of transmissions received is thus
$e^{-\xi_i(1-x)}$.  Setting $x=0$, the probability of having received
zero transmissions is $e^{-\xi_i(t)}$.  Thus
$S_i = S_i(0) e^{-\xi_i(t)}$.  We have $I_i = N_i - S_i - R_i$ and
$\dot{R}_i = \gamma_i I_i$.  To find $\xi_i$, we simply note that the total rate that group $i$ is receiving infection is $\sum_j I_j \beta_{ij}$, and so 
\[
\dot{\xi}_i = \frac{\sum_j I_j \beta_{ij}}{N_i} \, .
\]
Thus:
\begin{thm}
If the rate of transmission from an infected individual in group $j$
to group $i$ is $\beta_{ij}$, then
\begin{subequations}
\label{sys:heterogeneous_dynamic}
\begin{align}
S_i &= S_i(0) e^{-\xi_i(t)}\\
I_i &= N_i - S_i - R_i\\
\dot{R}_i &= \gamma_i I_i\\
\dot{\xi}_i &= \frac{\sum_j I_j \beta_{ij}}{N_i} \,
\end{align}
\end{subequations}
with $\xi(0)=0$.
\end{thm}

\ifsolns
\else
\subsection{Exercises}
\fi
\begin{exercise}
Consider a vector-borne disease for which each infected individual
infects a Poisson-distributed number of vectors, with mean $\lambda$.
Each infected vector causes $i$ infections with probability $p_i =
\pi^i(1-\pi)$ for some $\pi \in [0,1]$.  This scenario corresponds to human infection lasting for a fixed time with some constant transmission rate to vectors, and each vector having probability $\pi$ of living to bite again after each bite and transmitting with probability $1$ if biting.
\begin{myenumerate}
\item Let $\alpha_{g|1}$ and $\alpha_{g|2}$ be the probability that an
  outbreak would go extinct in $g$ generations starting with an
  infected human or vector respectively.  Find the vector-valued
  function $\vec{\psi}(\vec{x}) = (\psi_1(\vec{x}), \psi_2(\vec{x}))$.
  That is, what are the PGFs $\psi_1(x_1,x_2)$ and $\psi_2(x_1,x_2)$?
\item Set $\lambda = 3$ and $\pi = 0.5$.  Find the probability of an
  epidemic if one infected human is introduced or if one infected
  vector is introduced.
\item For the same values, find the probability of an epidemic if one infected vector is introduced.
\item Find $\psi_2(\psi_1(0,x),0)$.  How should we interpret the terms of its Taylor Series expansion?
\end{myenumerate}
\end{exercise}
\begin{solution}
\mbox{}
\begin{myenumerate}
\item \begin{itemize}
\item $\psi_1(x_1,x_2) = e^{\lambda(x_2-1)}$  (independent of $x_1$)
\item $\psi_2(x_1,x_2) = \frac{1-\pi}{1-\pi x_1}$  (independent of
  $x_2$)
\end{itemize}
\item $\vec{\psi}(\vec{x}) = \left(e^{3(x_2-1)},
    \frac{1}{2- x_1}\right)$.  Starting with $\vec{x}=(0,0)$, and
  iterating until convergence we have $\vec{x} = (0.2868, 0.5837)$.
  So the epidemic probability is $0.7132$ starting from a human and
  $0.4163$ starting from a vector.
\end{myenumerate}
\end{solution}

\begin{exercise}
Starting from the equations
\begin{align*}
\dot{S}_i &= -\frac{S_i}{N_i} \sum_j \beta_{ij} I_j\\
\dot{I}_i &= - \gamma_i I_i + \frac{S_i}{N_i} \sum_j \beta_{ij}  I_j \\
\dot{R}_i &= \gamma_i I_i 
\end{align*}
use integrating factors to derive System~\eqref{sys:heterogeneous_dynamic}.
\end{exercise}
\begin{solution}
\mbox{}
\begin{myenumerate}
\item We have
\begin{align*}
\dot{S}_i &= - \frac{S_i}{N_i}\sum_j \beta_{ij} I_j\\
\dot{S}_i + \left (\frac{\sum_j \beta_{ij} I_j}{N_i}\right) S_i &= 0\\
\diff{}{t} S_i e^{\xi_i}
          &=0\\
S_i = S_i(0) e^{-\xi_i}
\end{align*}
where $\dot{\xi}_i = \sum_j \beta_{ij} I_j/ N_i$.

It is straightforward to add $\dot{R}_i = \gamma_i I_i$ and $I_i = N_i
- S_i - R_i$.
\end{myenumerate}
\end{solution}

\begin{exercise}
  Assume the population is grouped into subgroups of size $N_i$ with
  $N = \sum_i N_i$ and the $i$-th subgroup has a parameter $\kappa_i$
  representing their rate of contact with others.  Take
  \[
\beta_{ji} = \kappa_j \frac{\kappa_i N_i}{\sum_\ell N_\ell
  \kappa_\ell} \beta
\]
  to be the transmission rate from type $i$ individuals to a single
  type $j$ individual, and assume all infected individuals recover with the same rate
  $\gamma$.

Define $\theta = e^{-\beta\left(\sum_j \kappa_j\int_0^t I_j(\tau) \,
  \mathrm{d}\tau\right)/\sum_j \kappa_jN_j}$ and define the PGF $\psi(x)=\sum_i \frac{N_i}{N} x^i$.  Let $S = \sum_i S_i$, \ $I = \sum_i I_i$, and $R = \sum
R_i$.
\begin{myenumerate}
\item Explain what assumptions this model makes about interactions
  between individuals in group $i$ and $j$.
\item \label{b} Show that
\begin{align*}
S &= N\psi(\theta)\\
I &= N - S- R\\
\dot{R} &= \gamma I\\
\dot{\theta} &= -\beta\theta \frac{\sum_j \kappa_j I_j}{\sum_j \kappa_j N_j} 
\end{align*}
with $\theta(0)=1$.
\item \label{c} Explain why $\frac{\sum_j \kappa_j I_j}{\sum_j \kappa_j N_j} = 1 - \frac{\sum_j
  \kappa_j S_j}{\sum_j \kappa_j N_j} - \frac{\sum_j \kappa_j
  R_j}{\sum_j \kappa_j N_j}$.
\item \label{d} Show that $\frac{\sum_j \kappa_j S_j}{\sum_j \kappa_j N_j}  =
  \frac{\theta\psi'(\theta)}{\psi'(1)}$.
\item \label{e} Show that $\diff{}{t} \frac{\sum_j \kappa_j R_j}{\sum_j \kappa_j
    N_j} = - (\gamma/\beta) \frac{\dot{\theta}}{\theta}$, and 
    solve for $\frac{\sum_j \kappa_j R_j}{\sum_j \kappa_j N_j}$ in
    terms of $\theta$
    assuming $R_j = 0$ for all $j$.
\item Thus conclude that
\[
\dot{\theta} = -\beta \theta + \beta
\frac{\theta^2\psi'(\theta)}{\psi'(1)} - \theta \gamma \ln \theta
\]
\end{myenumerate}

\end{exercise}
\begin{solution}
\mbox{}
\begin{myenumerate}
\item People in group $i$ interact with others at rate $\kappa_i$, and they
  choose their partners at random at each time from the
  entire population with probability proportional to the partner's
  interaction rate.
\item This can be derived directly from an integrating factor or by
  simply substituting in and checking that the ODEs are satisfied.
\item If we move the $S$ and $R$ terms to the right hand side, the
  numerator becomes $\sum_j \kappa_j (I_j +S_j + R_j) = \sum_j \kappa_j
  N_j$.
\item We have $S_j = N_j \theta^\kappa_j$  So 
\begin{align*}
\kappa_j S_j &= \kappa_j
  \theta^{\kappa_j} \\
&= \theta \diff{}{\theta} \theta^{\kappa_j}
\end{align*}
Thus summing the terms in the numerator gives $\theta
\psi'(\theta)$.  The denominator can be foundsimilarly.
\item If we substitute $\gamma I_j$ for $\dot{R}_j$ and compare with
  the $\dot{\theta}$ equation the ODE is shown.  Then integrating both
  sides gives $\sum_j \kappa_j R_j/ \sum_j \kappa_j N_j =
  (\gamma/\beta) \ln \theta$.
\item We simply substitute the results of~\textbf{\ref{d}}
  and~\textbf{\ref{e}} into~\textbf{\ref{c}}.  Then substitute this
  into the $\dot{\theta}$ equation from~\textbf{\ref{b}}.
\end{myenumerate}
\end{solution}

\ifsolns
\else
\section{Discussion}
\fi
There are many contexts where we are interested in how a newly
introduced infectious disease would spread.  We encounter situations
like this in the spread of zoonotic infections such as Monkey Pox or
Ebola as well as the importation of novel diseases such as the Zika in
the Americas or the reintroduction of locally eliminated diseases such
as Malaria.  

PGFs are an important tool for the analysis epidemics, particularly at
early stages.  They allow us to relate the individual-level
transmission process to the distribution of outcomes.  This allows us
to take data about the transmission process and make predictions about
the possible outcomes, but it also allows us to take observed
outbreaks and use them to infer the individual-level transmission
properties.

For SIR disease PGFs also provide a useful alternative formulation to
the usual mass-action equations.  This formulation leads to a simple
derivation of final-size relations and helps explain why
previous studies have shown that a wide range of disease assumptions
give the same final size relation.

Our goal with this primer has been to introduce researchers to the
many applications of PGFs to disease spread.  We have used the
appendices to derive some of the more technical properties of PGFs.
Additionally we have developed a Python package \texttt{Invasion\_PGF}
which allows for quick calculation of the results in the first three
sections of this primer.  A detailed description of the package is in
Appendix~\ref{app:software}.  The software can be downloaded at
\url{https://github.com/joelmiller/Invasion_PGF}.  Documentation is
available within the repository, starting with the file
\texttt{docs/\_build/html/index.html}.  The supplementary information
includes code that uses \texttt{Invasion\_PGF} to generate the figures
of Section~\ref{sec:discrete}.

\appendix

\section{Important properties of PGFs}
\label{app:properties}
In this appendix, we give some theoretical background behind the important properties of PGFs which we use in the main part of the primer.  We attempt to make each subsection self-contained so that the reader has a choice of reading through the appendix in its entirety, or waiting until a property is used before reading that section.  Because we expect the appendix is more likely to be read piecemeal, the exercises are interspersed through the text where the relevant material appears.

A PGF has been described as ``a clothesline on which we hang up a sequence of numbers for display''~\cite{gf}.  Similarly~\cite{polya1990mathematics} says ``A generating function is a device somewhat similar to a bag. Instead of carrying many little objects detachedly, which could be embarrassing, we put them all in a bag, and then we have only one object to carry, the bag.''  Indeed for many purposes mathematicians use PGFs primarily because once we have the distribution put into this ``bag'', many more mathematical tools are available, allowing us to derive interesting and sometimes surprising identities~\cite{gf}.

However, for our purposes there is a meaningful direct interpretation
of a PGF.  Assume that we are interested in the probability that an
event does not happen given some unknown number $i$ of independent
identical Bernoulli trials with probability $\alpha$ the event does
not happen in any one trial.  Let $r_i$ represent the probability that
there are $i$ trials.  Then the probability that the event does not
occur in any trial is
\[
\sum_i r_i \alpha^i = f(\alpha) \, ,
\]
and so PGFs emerge naturally in this context.

In infectious disease, this context occurs frequently and many results in this primer can be expressed in this framework.  For reference, we make this property more formal:
\begin{property}
\label{property:notclothesline}
Assume we have a process consisting of a random number $i$ independent identical Bernoulli trials.  Let $r_i$ be the distribution of $i$ and $f(x)= \sum_i r_i x^i$ be its PGF.  If $\alpha$ is the probability that each trial fails, then $f(\alpha)$ is the probability all trials fail.
\end{property}

\ifsolns
\else
\subsection{Properties related to individual coefficients}
\fi
We start by investigating how to find the coefficients of a PGF if we can calculate the numeric value of the PGF at any point.

This section makes use of the imaginary number $i = \sqrt{-1}$, and so in this section we avoid using $i$ as an index in the sum of $f(x)$.
\begin{property}
\label{prop:coefficient}
Given a PGF $f(x) = \sum_n r_n x^n$, the coefficient of $x^n$ in its
expansion for a particular $n$ can be calculated by taking $n$ derivatives, evaluating the result at $x=0$, and dividing by $(n!)$.  That is
\[
r_n =  \frac{1}{n!}\left. \left(\diff{}{x}\right)^n f(x) \right|_{x=0}
\]
\end{property}
This result holds for any function with a Taylor Series (it does not use any special properties of PGFs).

\begin{exercise}
Prove Property~\ref{prop:coefficient} [write out the sum and show that
the derivatives eliminate any $r_m$ for $m<n$, the leading coefficient
of the result is
$n!r_n$, and the later terms are all zero].  
\end{exercise}
\begin{solution}
We first find
\begin{align*}
\left(\diff{}{x}\right)^n f(x) &= \left(\diff{}{x}\right)^n \sum_m r_m
                                 x^m\\
&= \sum_m \left(\diff{}{x}\right)^n r_m x^m\\
&= \sum_m m(m-1)\cdots(m-n+1) r_m x^{m-n}
\end{align*}
\begin{itemize}
\item if $m<n$, one of the terms in the product $m(m-1) \cdots (m-n+1)$
is zero. 
\item if $m=n$ then $m(m-1) \cdots(m-n+1)$ becomes $n!$, and $x^{m-n}$
  is simply $1$.
\item if $m>n$, then there is a nonzero factor in front and a factor
  of $x^{m-n}$.
\end{itemize}
When we evaluate this derivative at $x=0$, the terms which are
already zero remain zero, the $m=n$ term remains unchanged, and the
other terms all have $0$ raised to a positive power so they become
zero.

Thus we are left $r_n$.

\end{solution}

There are many contexts in which we can only calculate a function
numerically.  In
this case the calculation of these derivatives is likely to be difficult and
inaccurate.  An improved way to calculate it is given by a Cauchy
integral~\cite{moore2000exact}.  This is a standard result of Complex
Analysis, and initially we simply take it as given.
\[
r_n = \frac{1}{2\pi i} \oint  \frac{f(z)}{z^{n+1}} \, \mathrm{d}z
\]
This integral can be done on a closed circle around the origin $z = Re^{i\theta}$, in which case $\mathrm{d}z = iz \mathrm{d}\theta$.  Then $r_n$ can be rewritten as
\[
r_n = \frac{1}{2\pi}\int_0^{2\pi} \frac{f(Re^{i\theta})}{(Re^{i\theta})^n} \, \mathrm{d}\theta
\]
Using another substitution, $\theta = 2\pi u$, we find
  $\mathrm{d}\theta = 2 \pi \mathrm{d}u$ with $u$ varying from $0$ to
  $1$.  This integral becomes
\[
 r_n =
 \int_0^1 \frac{f(Re^{2 \pi i u})}{R^n e^{2 n \pi i u}} \, \mathrm{d}u
\] 
The integral on the right hand side can be approximated by a simple
summation and we find
\[
r_n \approx \frac{1}{M}\sum_{m=1}^{M} \frac{f(Re^{ 2 \pi im/M})}{R^n e^{2n\pi i m/M}}
\]
for large $M$.

A few technical steps show that the PGF $f(z)$ converges for
any $z$ with $|z| \leq 1$ (any PGF is analytic within the unit circle
$R=1$ and that the PGF converges everywhere on the unit circle [the
coefficients are all positive or zero and the sum converges for $z=1$,
so it converges absolutely on the unit circle]).  Thus this integral
can be performed for any positive $R\leq 1$.  We have found that the
unit circle ($R=1$) yields remarkably good accuracy, so we recommend using it unless there is a good reason not to.  Some discussion of identifying the optimal radius appears in~\cite{bornemann2011accuracy}.

Thus we have
\begin{property}
\label{prop:cauchy_coefficient}
Given a PGF $f(x)$, the coefficient of $x^n$ in its expansion can be
calculated by the integral
\begin{equation}
\label{eqn:rn_integral}
r_n = \int_0^1 \frac{f(Re^{2 \pi i u})}{R^n e^{2 n \pi i u}} \, \mathrm{d}u
\end{equation}
This is well-approximated by the summation
\begin{equation}
r_n \approx \frac{1}{M}\sum_{m=1}^{M} \frac{f(Re^{ 2 \pi im/M})}{R^n e^{2n\pi i m/M}}
\label{eqn:rn_sum}
\end{equation}
with $R=1$ and $M \gg 1$.
\end{property}

It turns out that this approach is closely related to the approach to get a
particular coefficient of a Fourier Series.  Once the variable is
changed from $z$ to $\theta$, our function is effectively a Fourier
Series in $\theta$, and the integral is the standard
approach to finding the $n$th coefficient of a Fourier Series.  

\begin{exercise}
\textbf{Verification of Equation~\eqref{eqn:rn_integral}:}

In this exercise we show that the formula in
Equation~\eqref{eqn:rn_integral} yields $r_n$.  Assume that the
integral is performed on a circle of radius $R\leq 1$ about the origin.
\begin{myenumerate}
\item Write $f(z) = \sum_m r_m z^m$ and rewrite $\int_0^1
  \frac{f(Re^{2 \pi i u})}{R^n e^{2 n \pi i u}} \, \mathrm{d}u$ as 
a sum 
\[
\int_0^1
  \frac{f(Re^{2 \pi i u})}{R^n e^{2 n \pi i u}} \, \mathrm{d}u = \sum_m
  r_m  \int_0^1 R^{m-n} e^{2(m-n)\pi i u} \, \mathrm{d}u
\]
\item Show that for $m=n$ the integral in the summation on the right
  hand side is $1$.
\item Show that for $m \neq n$, the integral in the summation on the
  right hand side  is $0$.
\item Thus conclude that the integral on the left hand side must yield $r_n$.
\end{myenumerate}
\end{exercise}
\begin{solution}\mbox{}
\begin{myenumerate}
\item Substituting we have
\begin{align*}
\int_0^1
  \frac{f(Re^{2 \pi i u})}{R^n e^{2 n \pi i u}} \, \mathrm{d}u &= \int_0^1
  \frac{\sum_m r_m (Re^{2 \pi i u})^m}{R^n e^{2 n \pi i u}} \,
                                                                 \mathrm{d}u\\
&= \sum_m  r_m\int_0^1 R^{m-n}e^{2(m-n)\pi i u} \, \mathrm{d}u
\end{align*}
\item For $m=n$, both $R^{m-n}$ and $e^{2(m-n)\pi i u}$ become $1$.
  Thus we have $\int_0^1 1 \mathrm{d}u = 1$.
\item For $m\neq n$, the integral becomes
\begin{align*}
\int_0^1 R^{m-n}e^{2(m-n)\pi i u} \, \mathrm{d}u &=
                                                   R^{m-n}\int_0^1e^{2(m-n)\pi
                                                   i u} \,
                                                   \mathrm{d}u\\
&=  R^{m-n}\int_0^1e^{2(m-n)\pi i u} \, \mathrm{d}u\\
&=  R^{m-n}\frac{1}{2(m-n)\pi i} \left( e^{2(m-n)\pi i 1} -
  e^{2(m-n)\pi i 0}\right)\\
&= R^{m-n} \frac{1}{2(m-n)\pi i} (1-1)\\
&=0
\end{align*}
where in the last step we use the fact that $e^{2N\pi i} = \cos 2N\pi
+ i \sin 2N\pi = 1$ for any integer $N$.
\item So the integral $\int_0^1 \frac{f(Re^{2 \pi i u})}{R^n e^{2 n \pi i u}} \, \mathrm{d}u$
  becomes
\begin{align*}
\int_0^1 \frac{f(Re^{2 \pi i u})}{R^n e^{2 n \pi i u}} \, \mathrm{d}u
&= 0r_0 + 0r_1 + \cdots + 0r_{n-1} + 1r_n + 0r_{n+1} + \cdots\\
&= r_n
\end{align*}

\end{myenumerate}
\end{solution}

\begin{exercise}
Let $f(z) = e^z=1 + z + z^2/2 + z^3/6 + z^4/24 + z^5/120 + \cdots$.  Write a program that estimates $r_0$, $r_1$, \ldots, $r_5$ using Equation~\eqref{eqn:rn_sum} with $R=1$.  Report the values to four significant figures for
\begin{myenumerate}
\item $M=2$
\item $M=4$
\item $M=5$
\item $M=10$
\item $M=20$.
\item How fast is convergence for different $r_n$?
\end{myenumerate}
\end{exercise}
\begin{solution}
The script \texttt{integral\_exercise.py} which is given as
a supplementary file perfoms these calculations.
\begin{myenumerate}
\item $M=2$:
\begin{align*}
r_0 &\approx 1.543\\
r_1 &\approx 1.175\\
r_2 &\approx 1.543\\
r_3 &\approx 1.175\\
r_4 &\approx 1.543\\
r_5 &\approx 1.175
\end{align*}
\item $M=4$:
\begin{align*}
r_0 &\approx 1.042\\
r_1 &\approx 1.008\\
r_2 &\approx 0.5013\\
r_3 &\approx 0.1669\\
r_4 &\approx 1.042\\
r_5 &\approx 1.008
\end{align*}
\item $M=5$:
\begin{align*}
r_0 &\approx 1.008\\
r_1 &\approx 1.001\\
r_2 &\approx 0.5002\\
r_3 &\approx 0.1667\\
r_4 &\approx 0.04167\\
r_5 &\approx 1.008
\end{align*}
\item $M=10$:
\begin{align*}
r_0 &\approx 1.000\\
r_1 &\approx 1.000\\
r_2 &\approx 0.500\\
r_3 &\approx 0.1667\\
r_4 &\approx 0.04167\\
r_5 &\approx 0.008333
\end{align*}
\item $M=20$:
\begin{align*}
r_0 &\approx 1.000\\
r_1 &\approx 1.000\\
r_2 &\approx 0.5000\\
r_3 &\approx 0.1667\\
r_4 &\approx 0.04167\\
r_5 &\approx 0.008333
\end{align*}
\item Convergence is quite fast, and predicts $r_n$ quite well once $M > n$.
\end{myenumerate}
\end{solution}

\ifsolns
\else
\subsection{Properties related to distribution moments}
\fi

We next look at two straightforward properties about the moments of
the distribution $r_i$ having PGF $f(x)$.  We return to using
$i=0,1,\ldots$ as an indexing variable, so $i$ is no longer $\sqrt{-1}$.  We have 
\begin{align*}
f(1) &= \sum_i r_i 1^i\\
     &= \sum_i r_i\\
     &= 1
\end{align*}
where the final equality is because the $r_i$ determine a probability distribution.

With mildly more effort, we have
\begin{align*}
f'(1) &= \sum_i r_i i 1^{i-1}\\
      &= \sum_i i r_i\\
      &= \mathbb{E}(i)
\end{align*}
where $\mathbb{E}(i)$ denotes the expected value of $i$.  
These arguments show
\begin{property}
\label{property:1}
Any PGF $f(x)$ must satisfy $f(1)=1$.
\end{property}
\begin{property}
\label{property:mean}
The expected value of a random variable $i$ whose distribution has PGF $f(x)$ is given by $\mathbb{E}(i) = f'(1)$.
\end{property}
It is straightforward to derive relationships for $\mathbb{E}(i^2)$ and higher order moments by repeated differentiation of $f$ and evaluating the result at $1$.

\ifsolns
\else
\subsection{Properties related to function composition}
\fi
To motivate function composition, we start with an example.
\begin{example}
\label{example:compose}
Consider a weighted coin which comes up `Success' with probability $p$ and with `Failure' with probability $1-p$.  We play a game in which we stop at the first failure, and otherwise flip it again.  Define $f(x) = px + 1-p$

Let $\alpha_g$ be the probability of failure within the first $g$ flips.  Then $\alpha_0=0$ and $\alpha_1 = 1-p = f(0)$ are easily calculated.

More generally the probability of starting the game and failing immediately is $\alpha_0=1-p = f(0)$ while the probability of having a success and flipping again is $p$, at which point the probability of failure within $g-1$ flips is $\alpha_{g-1}$.  So we have $\alpha_g = (1-p) + p \alpha_{g-1} = f(\alpha_{g-1})$.  So using induction we can show that the probability of failure within $g$ generations is $f^{[g]}(0)$.
\end{example}

\begin{exercise}
The derivation in example~\ref{example:compose} was based on looking
at what happened after a single flip and then looking $g-1$ flips into
the future in the inductive step. Derive $\alpha_g = f(\alpha_{g-1})$
by instead looking $g-1$ flips into the future and then considering
one additional step.  [the distinction between this argument and the
previous one becomes useful in the continuous-time case where we use
the `backward' or `forward' Kolmogorov equations.]
\end{exercise}
\begin{solution}
As in the example, we take $\alpha_g$ to be the probability of failure
within the first $g$ flips, with $\alpha_0=0$ and $\alpha_1=1-p =
f(0)$.

If the first flip does not come up as ``failure'', then the
probability that failure occurs within the following $g-1$ flips is
(by definition) $\alpha_{g-1}$.  So the probability of failure within
the first $g$ flips is the probability of failure in the first flip
plus the probability of success time $\alpha_{g-1}$. That is:
\[
\alpha_{g} = (1-p) + p \alpha_{g-1}  = f(\alpha_{g-1})
\]
\end{solution}


\begin{exercise}
\label{exercise:die2disease}
Consider a fair six-sided die with numbers $0$, $1$, \ldots, $5$, rather
than the usual $1$, \ldots, $6$.  We roll the die once.  Then we look at the result, and roll that many copies (if  zero, we stop), then we look at the sum of the result and repeat.  Define
\[
f(x) = \frac{1+x+\cdots + x^5}{6} = \begin{cases} \frac{x^6-1}{6(x-1)} & x \neq 1\\
1 & x=1
\end{cases}
\]
Define $\alpha_g$ to be the probability the process stops after $g$
iterations  (with $\alpha_0=0$ and $\alpha_1 = 1/6$).
\begin{myenumerate}
\item Find an expression for $\alpha_g$, the probability that by the
  $g$'th iteration the process has stopped, in terms of  $f(x)$.
\item Rephrase this question in terms of the extinction probability for an infectious disease.
\end{myenumerate}
\end{exercise}
\begin{solution}
\mbox{}
\begin{myenumerate}
\item The probability of dying out after $g$ iterations is the sum
  over all $i$ of the probability that the first roll is an $i$ and
  the process dies out after $g-1$ iterations.  By thinking of each
  die that rolls an $i$ as having $i$ ``offspring'' we can assign each
  individual a collection of descendants.  The process goes extinct
  after $g$ iterations if
  all the dice in the second roll have no offspring after $g-1$ iterations.
\[
\sum_{i=0}^6 \frac{1}{6} \alpha_{g-1}^i = f(\alpha_{g-1}) = f^{[g]}(0)
\]
\item This is equivalent to a disease for which each individual causes
  $0$, $1$, $2$, $3$, $4$, or $5$ new infections with equal
  probability.  So the extinction probability for such a disease is $f^{[g]}(0)$.
\end{myenumerate}
\end{solution}

Processes like that in Exercise~\ref{exercise:die2disease} can be thought of as ``birth-death'' processes where each event generates a discrete number of new events.  Our examples above show that function composition arises naturally in calculating the probability of extinction in a birth-death process.  We show below that it also arises naturally when we want to know the distribution of population sizes after some number of generations rather than just the probability of $0$.

Specifically, we often assume an initially infected individual causes some random number of new infections $i$ from some distribution.  Then we assume that each of those new infections independently causes an additional random number of infections from the same distribution.  We will be interested in how to get from the one-generation PGF to the PGF for the distribution after $g$ generations.

We derive this in a few stages.  
\begin{itemize}
\item We first show that if we take two numbers from different distributions with PGFs $f(x)$ and $h(x)$, then their sum has distribution $f(x)h(x)$ [Property~\ref{property:sum2product}].  Then inductively applying this we conclude that the distribution of the sum of $n$ numbers from a distribution with PGF $f(x)$ has PGF $[f(x)]^n$.
\item We also show that if the probability we take a number from the distribution with PGF $f(x)$ is $\pi_1$ and the probability we take it from the distribution with PGF $h(x)$ is $\pi_2$, then the PGF of the resulting distribution is $\pi_1 f(x) + \pi_2 h(x)$ [Property~\ref{property:choice2sum}].
\item Putting these two properties together, we can show that if we choose $i$ from a distribution with PGF $f(x)$ and then choose $i$ different values from a distribution with PGF $h(x)$, then the sum of the $i$ values has PGF $f(h(x))$ [Property~\ref{prop:composition}].  
\end{itemize}
Our main use of Properties~\ref{property:sum2product} and~\ref{property:choice2sum} is as stepping stones towards Property~\ref{prop:composition}.

Consider two probability distributions, let $r_i$ be the probability of $i$ for the first distribution and $q_j$ be the probability of $j$ for the second distribution.  Assume they have PGFs $f(x) = \sum_i r_i x^i$ and $h(x)= \sum_j q_j x^j$ respectively.  

We are first interested in the process of choosing $i$ from the first distribution, $j$ from the second, and adding them.  In the disease context this arises where the two distributions give the probability that one individual infects $i$ and the other infects $j$ and we want to know the probability of a particular sum.

The probability of obtaining a particular sum $k$ is
\[
\sum_i r_i q_{k-i}
\]
So the PGF of the sum is $\sum_k \sum_{i=0}^k r_i q_{k-i} x^k$.  By inspection, this is equal to the product $f(x) h(x)$.  This means that the PGF of the process where we choose $i$ from the first and $j$ from the second and look at the sum is the product $f(x) h(x)$.

We have shown
\begin{property}
\label{property:sum2product}
Consider two probability distributions, $r_0$, $r_1$, $\ldots$ and $q_0$, $q_1$, $\ldots$ with PGFs $f(x) = \sum_i r_i x^i$ and $h(x) = \sum_j q_j x^j$.  Then if we choose $i$ from the distribution $r_i$ and $j$ from the distribution $q_j$, the PGF of their sum is $f(x)h(x)$.
\end{property}

Usually we want the special case where we choose two numbers from the same distribution having PGF $f(x)$.  The PGF for the sum is $[f(x)]^2$.  The PGF for the sum of three numbers from the same distribution can be thought of as the result of $[f(x)]^2$ and $f(x)$, yielding $[f(x)]^3$.  By induction, it follows that the PGF for the sum of $i$ numbers sum is $[f(x)]^i$.

Now we want to know what happens if we are not sure what the current system state is.  For example, we might not know if we have $1$ or $2$ infected individuals, and the outcome at the next generation is different based on which it is. 

We use the distributions $r_i$ and $q_j$.  We assume that with probability $\pi_1$ we choose a random number $k$ from the $r_i$ distribution, while with probability $\pi_2=1-\pi$ it is chosen from the $q_j$ distribution.  Then the probability of a particular value $k$ occurring is $\pi_1 r_k + \pi_2 q_k$, and the resulting PGF is $\sum_k (\pi_1 r_k + \pi_2 q_k)x^k = \pi_1 f(x) + \pi_2 h(x)$.  This becomes:
\begin{property}
\label{property:choice2sum}
Consider two probability distributions, $r_0$, $r_1$, $\ldots$ and $q_0$, $q_1$, $\ldots$ with PGFs $f(x) = \sum_i r_i x^i$ and $h(x) = \sum_j q_j x^j$.  We consider a new process where with probability $\pi_1$ we choose $k$ from the $r_i$ distribution and with probability $\pi_2=1-\pi_1$ we choose $k$ from the $q_j$ distribution.  Then the PGF of the resulting distribution is $\pi_1 f(x) + \pi_2 g(x)$.
\end{property}

We finally consider a process in which we have two distributions with PGFs $f(x)=\sum_i r_i x^i$ and $h(x)=\sum_j q_j x^j$.  We choose the number $i$ from the distribution $r_i$ and then take the sum of $i$ values chosen from the $q_j$ distribution, $\sum_{\ell=1}^i j_\ell$.  Both the number of terms in the sum and their values are random variables.  Using the results above, the PGF of the resulting sum is $\sum_i r_i h(x)^i = f(h(x))$.  Thus we have
\begin{property}
Consider two probability distributions, $r_0$, $r_1$, $\ldots$ and $q_0$, $q_1$, $\ldots$ with PGFs $f(x) = \sum_i r_i x^i$ and $h(x) = \sum_j q_j x^j$.  Then if we choose $i$ from the distribution $r_i$ and then take the sum of $i$ values chosen from the distribution $q_j$, the PGF of the sum of those $i$ values is $f(h(x))$.
\label{prop:composition}
\end{property}
This property is closely related to the spread of infectious disease.  An individual may infect $i$ others, and then each of them causes additional infections.  The number of these second generation cases is the sum of $i$ random numbers $\sum_{\ell = 1}^i j_\ell$ where $j_{\ell}$ is the number of additional infections caused by the $\ell$-th infection caused by the initial individual.  So if $f(x)$ is the PGF for the distribution of the number of infections caused by the first infection and $h(x)$ is the PGF for the distribution of the number of infections caused by the offspring, then $f(h(x))$ is the PGF for the number infected in the second generation [and if the two distributions are the same this is $f^{[2]}(x)$].  Repeated iteration gives us the distribution after $g$ generations.

\begin{exercise}
Note that if we interchange $p$ and $q$ in the PGF of the negative
binomial distribution in Table~\ref{tab:example_PGFs}, it is simply the PGF of the
geometric distibution raised to the power $\hat{r}$.  A number chosen
from the negative binomial can be defined as the number of successful
trials (each with success probability $p$) before the $\hat{r}$th
failure.

Using this and Property~\ref{prop:composition}, derive the PGF of the
negative binomial.
\end{exercise}

\begin{solution}
We re-express the negative binomial by interchanging what we count as a
failure and success.  We seek the number of failures that occur before
the $\hat{r}$th success.  We say a success occurs with probability $\hat{p}$ and
failure with probability $\hat{q}=1-\hat{p}$ (which satisfy $\hat{p} =
q$, \ $\hat{q}=p$).  

The result for the geometric distribution says that the number of
failures between each success has PGF $\hat{p}/(1-\hat{q}x)$.  We need
the sum of $\hat{r}$ of these.  So by Property~\ref{prop:composition},
we have that the negative binomial has PGF
$[\hat{p}/(1-\hat{q}x)]^{\hat{r}}$.  Replacing $\hat{p}$ by $q$ and
$\hat{q}$ by $p$ completes the result.
\end{solution}

\begin{exercise}\textbf{Sicherman dice}~\cite{gardner1978sicherman,gallian1979cyclotomic}.

To motivate this exercise consider two tetrahedral dice, numbered $1,
2, 3, 4$.  When we roll them we get sums from $2$ to $8$, each with
its own probability, which we can infer from this table:
\begin{center}
\begin{tabular}{Sc||Sc|Sc|Sc|Sc|}
& $\die{1}$ & $\die{2}$ & $\die{3}$ & $\die{4}$ \\
\hline\hline
$\die{1}$ &$2$ &$3$&$4$&$5$\\
\hline 
$\die{2}$ &$3$&$4$&$5$&$6$\\
\hline 
$\die{3}$ &$4$&$5$&$6$&$7$\\
\hline 
$\die{4}$ &$5$&$6$&$7$&$8$\\
\hline 
\end{tabular}
\end{center}
However another pair of tetrahedral dice, labelled $1,2,2,3$ and
$1,3,3,5$ yields the same sums with the same probabilities:
\begin{center}
\begin{tabular}{Sc||Sc|Sc|Sc|Sc|}
& $\die{1}$ & $\die{2}$ & $\die{2}$ & $\die{3}$ \\
\hline\hline
$\die{1}$ &$2$ &$3$&$3$&$4$\\
\hline 
$\die{3}$ &$4$&$5$&$5$&$6$\\
\hline 
$\die{3}$ &$4$&$5$&$5$&$6$\\
\hline 
$\die{5}$ &$6$&$7$&$7$&$8$\\
\hline 
\end{tabular}
\end{center}

We now try to find a similar pair for $6$-sided dice.  First consider
a pair of standard $6$-sided dice.
\begin{myenumerate}
\item Show that the PGF of each die is $f(x) = (x +x^2+x^3 + x^4 + x^5 + x^6)/6$.
\item Fill in the tables showing the possible sums from rolling two dice (fill in each square with the sum of the two entries) and multiplication for two polynomials (fill in each square with the product of the two entries):\\
\begin{center}
\begin{tabular}{Sc||Sc|Sc|Sc|Sc|Sc|Sc|}
& $\die{1}$ & $\die{2}$ & $\die{3}$ & $\die{4}$ & $\die{5}$ & $\die{6}$\\
\hline\hline
$\die{1}$ & &&&&&\\
\hline 
$\die{2}$ &&&&&&\\
\hline 
$\die{3}$ &&&&&&\\
\hline 
$\die{4}$ &&&&&&\\
\hline 
$\die{5}$ &&&&&&\\
\hline 
$\die{6}$ &&&&&&\\
\hline
\end{tabular}
\hspace{1in}
\begin{tabular}{Sc||Sc|Sc|Sc|Sc|Sc|Sc|}
& $x^1$ & $x^2$ & $x^3$ & $x^4$ & $x^5$ & $x^6$\\
\hline\hline
$x^1$ & &&&&&\\
\hline 
$x^2$ &&&&&&\\
\hline 
$x^3$ &&&&&&\\
\hline 
$x^4$ &&&&&&\\
\hline 
$x^5$ &&&&&&\\
\hline 
$x^6$ &&&&&&\\
\hline
\end{tabular}.
\end{center}
\item Explain the similarity.
\item Show that each step of the following factorization is correct: 
\begin{align*}
f(x) &= \frac{x(1+x+x^2+x^3 + x^4 + x^5)}{6}\\
& = \frac{x(1+x+x^2)(1+x^3)}{6}\\
&= \frac{x(1+x+x^2)(1+x)(1-x+x^2)}{6}.
\end{align*}
\end{myenumerate}
This cannot be factored further, and indeed it can be shown that a property similar to prime numbers holds.  Namely, any factorization of $f(x)f(x)$ as $h_1(x)h_2(x)$ has the property that each of $h_1$ and $h_2$ can be factored into some powers of these ``prime'' polynomials times a constant.

We seek two new six-sided dice (each different) such that the sum of a roll of the two dice has the same probabilities as the normal dice.  The two dice have positive integer values on them (so no fair adding a constant $c$ to everything on one die and subtracting $c$ on the other).  Let $h_1(x)$ and $h_2(x)$ be their PGFs.
\begin{continuemyenumerate}
\item Explain why we must have $h_1(x)h_2(x)=[f(x)]^2$.
\item If the dice have numbers $a_1, \ldots, a_6$ and $b_1, \ldots,
  b_6$, show that their PGFs are of the form $h_1(x) = \sum_i x^{a_i}/6$
  and $h_2(x) = \sum_i x^{b_i}/6$ where all $a_i$ and $b_i$ are positive integers.
\item Given the properties we want for the dice, find $h_1(0)$ and $h_2(0)$.
\item Given the properties we want for the dice, find $h_1(1)$ and $h_2(1)$.
\item Using the values at $x=0$ and $x=1$, explain why $h_1(x) = x(1+x+x^2)(1+x)(1-x+x^2)^b/6$ and $h_2(x) = x(1+x+x^2)(1+x)(1-x+x^2)^{2-b}/6$ where $b$ is $0$, $1$, or $2$.
\item The case $b=1$ gives the normal dice.  Conside $b=0$ ($b=2$
  gives the same final result).  Find $h_1(x)$.
  $h_2(x) = \frac{1}{6}(x + x^3 +x^4+x^5+x^6+x^8) $
\item Create the table for the two dice corresponding to $h_1(x)$ and
  $h_2(x)$  and verify that the sums
  occur with the same frequency as a normal pair: \hspace{1in} 
\begin{tabular}{Sc||Sc|Sc|Sc|Sc|Sc|Sc|}
\phantom{$1$}&\phantom{$1$} & \phantom{$1$} &\phantom{$1$} & \phantom{$1$}& \phantom{$1$}&\phantom{$1$} \\
\hline\hline
 & &&&&&\\
\hline 
 &&&&&&\\
\hline 
 &&&&&&\\
\hline 
 &&&&&&\\
\hline 
 &&&&&&\\
\hline 
 &&&&&&\\
\hline
\end{tabular}
\end{continuemyenumerate}
\end{exercise}
\begin{solution}\mbox{}
\begin{myenumerate}
\item Because each $i$ in $1, 2, \ldots, 6$ has probability $1/6$, we
  have $f(x) = \sum_{i=1}^6 (1/6) x^i$.
\item 
\begin{center}
\begin{tabular}{Sc||Sc|Sc|Sc|Sc|Sc|Sc|}
& $1$ & $2$ & $3$ & $4$ & $5$ & $6$\\
\hline
\hline
$1$ &$2$&$3$&$4$&$5$&$6$&$7$\\
\hline 
$2$ &$3$&$4$&$5$&$6$&$7$&$8$\\
\hline 
$3$ &$4$&$5$&$6$&$7$&$8$&$9$\\
\hline 
$4$ &$5$&$6$&$7$&$8$&$9$&$10$\\
\hline 
$5$ &$6$&$7$&$8$&$9$&$10$&$11$\\
\hline 
$6$ &$7$&$8$&$9$&$10$&$11$&$12$\\
\hline
\end{tabular}
\hspace{1in}
\begin{tabular}{Sc||Sc|Sc|Sc|Sc|Sc|Sc|}
& $x^1$ & $x^2$ & $x^3$ & $x^4$ & $x^5$ & $x^6$\\
\hline\hline
$x^1$ &$x^2$&$x^3$&$x^4$&$x^5$&$x^6$&$x^7$\\
\hline 
$x^2$ &$x^3$&$x^4$&$x^5$&$x^6$&$x^7$&$x^8$\\
\hline 
$x^3$ &$x^4$&$x^5$&$x^6$&$x^7$&$x^8$&$x^9$\\
\hline 
$x^4$ &$x^5$&$x^6$&$x^7$&$x^8$&$x^9$&$x^{10}$\\
\hline 
$x^5$ &$x^6$&$x^7$&$x^8$&$x^9$&$x^{10}$&$x^{11}$\\
\hline 
$x^6$ &$x^7$&$x^8$&$x^9$&$x^{10}$&$x^{11}$&$x^{12}$\\
\hline
\end{tabular}.
\end{center}
\item Multiplication of two powers of $x$ results in addition of the exponents.
\item We have
\begin{align*}
x(1+x+x^2)(1+x^3) &=  x(1+x+x^2+x^3+x^4+x^5)\\
&= x+x^2+x^3+x^4+x^5+x^6
\end{align*}
and
\begin{align*}
(1+x)(1-x+x^2) &= 1 -x + x^2 +x-x^2+x^3\\
&= 1+x^3
\end{align*}
\item Since the PGF of the sum of the two dice is equal to the product
  of the PGFs of the two dice, we must have $h_1(x)h_2(x)=[f(x)]^2$.
\item For each die, the probabilities of the sides are $1/6$.  So they
  take the form $\sum_i x^{a_i}/6$ and $\sum_i x^{b_i}/6$.
\item Because each $a_i$ and $b_i$ is positive $\sum_i 0^{a_i}/6$ and
  $\sum_i 0^{b_i}/6$ are both $0$.
\item $\sum_i 1^{a_i}/6 = 1$ and $\sum_i 1^{b_i}/6=1$.
\item Because $h_1(0)=h_2(0)=0$, both of them must have a factor of
  $x$.  Because $h_1(1)=h_2(1)=1$ and the denominators are $6$, the
  numerators when $x=1$ must be $6$.  So there must be a factor of $2$
  and $3$ in the numerator.  This requires that each has a single factor of
  $(1+x+x^2)$ and $(1+x)$.  Because $1-x+x^2$ is $1$ when $x=1$,
  they could have a factor of $(1-x+x^2)^{2-b}$ for $b=0$, $1$, or
  $2$.
\item Taking $b=0$ we have 
\begin{align*}
h_1(x) &= x(1+x+x^2)(1+x)/6\\
 &=\frac{x(1+2x+2x^2+x^3)}{6}\\
&= \frac{x+x^2+x^2+x^3+x^3+x^4}{6}\\
\end{align*}
The calculation for $h_2(x)$ is not asked for in the question, but it is
\begin{align*}
h_2(x) &= h_1(x) (1-x+x^2)(1-x+x^2)\\
&= h_1(x)(1 -x+x^2-x+x^2-x^3+x^2-x^3+x^4)\\
&= \frac{x+2x^2+2x^3+x^4}{6}(1-2x+3x^2-2x^3+x^4)\\
&= \frac{1}{6} \big( x -2x^2+3x^3-2x^4+x^5\\
&\qquad\quad +2x^2- 4x^3+6x^4-4x^5+2x^6 \\
&\qquad\qquad\qquad+ 2x^3-4x^4+6x^5-4x^6+2x^7\\
&\qquad\qquad\qquad\qquad\quad + x^4 -2x^5 + 3x^6 - 2x^7 + x^8\big)\\
&=\frac{1}{6}(x + x^3 +x^4+x^5+x^6+x^8)
\end{align*}
\item \begin{tabular}{Sc||Sc|Sc|Sc|Sc|Sc|Sc|}
 &$1$&$2$&$2$&$3$&$3$&$4$\\
\hline\hline
 $1$&$2$&$3$&$3$&$4$&$4$&$5$\\
\hline 
 $3$&$4$&$5$&$5$&$6$&$6$&$7$\\
\hline 
 $4$&$5$&$6$&$6$&$7$&$7$&$8$\\
\hline 
 $5$&$6$&$7$&$7$&$8$&$8$&$9$\\
\hline 
 $6$&$7$&$8$&$8$&$9$&$9$&$10$\\
\hline 
 $8$&$9$&$10$&$10$&$11$&$11$&$12$\\
\hline 
\end{tabular}
The sums all occur with the same frequency as a normal pair.

If we didn't calculate $h_2(x)$ in the previous part, we could have
inferred the values from knowing what the first die was and what
values would be needed to match the normal pair.
\end{myenumerate}
\end{solution}

\begin{exercise}\textbf{Early-time outbreak dynamics}
\begin{myenumerate}
\item Consider normal dice.  The PGF is $f(x) = (x +x^2+x^3 + x^4 + x^5 + x^6)/6$. Consider the process where we roll a die, take the result $i$, and then roll $i$ other dice and look at their sum.  What is the PGF of the resulting sum in terms of $f$?
\item If an infected individual causes anywhere from $1$ to $6$ infections, all with equal probability, find the PGF for the number of infections in generation $2$ if there is one infection in generation $0$. [you can express the result in terms of $f$]  
\item And in generation $g$ (assuming depletion of susceptibles is unimportant)?
\end{myenumerate}
\end{exercise}
\begin{solution}\mbox{}
\begin{myenumerate}
\item By Property~\ref{prop:composition}, the PGF is $f(f(x))$.  
\item This is equivalent to the dice-rolling example.  It is
  $f(f(x))$.
\item More generally for generation $g$, we have $f^{[g]}(x)$.
\end{myenumerate}
\end{solution}

\ifsolns
\else
\subsection{Properties related to iteration of PGFs}
\fi
There are various contexts in which we might iterate to calculate $f^{[n]}(x)$ (the result of applying $f$ $n$ times to $x$).  

In the disease context, this occurs most frequently in calculating the
probability of outbreak extinction.  If we think of $\alpha$ as the
probability that the outbreak goes extinct from a single individual, then from Property~\ref{property:notclothesline} we would expect that $\alpha = f(\hat{\alpha})$ where $\hat{\alpha}$ is the probability that an offspring of the individual fails to produce an epidemic.  However, under common assumptions, the number of infections from the offspring should be from the same distribution as from the parent.  In this case we would conclude $\alpha = \hat{\alpha}$ and so $\alpha=f(\alpha)$. 

It turns out that a good way to solve for $\alpha$ is iteration, starting with the guess $\alpha_0=0$.  We will show that this converges to the correct value [$x=f(x)$ can have multiple solutions, only one of which is the correct $\alpha$].

\begin{figure}
\begin{center}
\includegraphics[width = 0.4\textwidth]{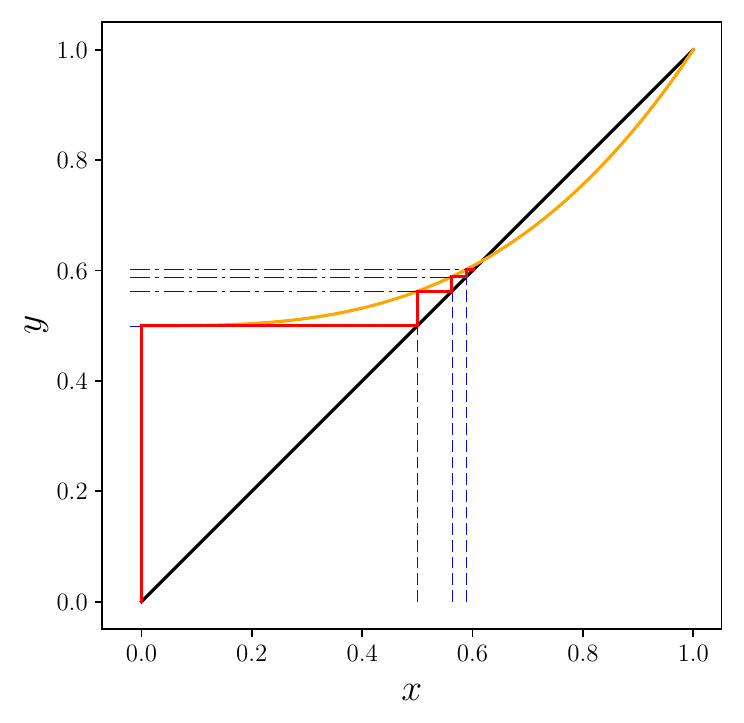}
\end{center}
\caption{\textbf{Cobweb diagrams}: We take the function $f(x) = (1+x^3)/2$.  A cobweb diagram is built by alternately drawing vertical lines from the diagonal to $f(x)$ and then horizontal lines from $f(x)$ to the diagonal.  The dashed lines show $\alpha_g = f(\alpha_{g-1})$ starting with $\alpha_0=0$ and highlight the relation to the iterative process.}
\label{fig:cobweb}
\end{figure}

Figure~\ref{fig:cobweb} demonstrates how the iterative process can be
represented by a ``cobweb diagram''~\cite{peitgen2006chaos,may1976simple}  To use a cobweb diagram
to study the behavior of $f^{[g]}(x_0)$, we draw the line $y=x$ and
the curve $y=f(x)$.  Then at $x_0$ we draw a vertical line to the
curve $y=f(x)$.  We draw a horizontal line to the line $y=x$ [which
will be at the point $(x_1,x_1)$].  We then repeat these steps,
drawing a vertical line to $y=f(x)$ and a horizontal line to $y=x$.
Cobweb diagrams are particularly useful in studying behavior near
fixed or periodic points.

\begin{exercise}
\textbf{Understanding cobweb diagrams}

From figure~\ref{fig:cobweb} the origin of the term ``cobweb'' may be
unclear.  Because of properties of PGFs, the more interesting behavior
does not occur for our applications.  Here we investigate cobweb
diagrams in more detail for non-PGF functions.  Since we use $f(x)$ to
denote a PGF, in this exercise we use $z(x)$ for an arbitrary function.
\begin{myenumerate}
\item Consider the line $z(x) = 2(1-x)/3$.  Starting with $x_0 = 0$, show how the first few iterations of $x_i = z(x_{i-1})$ can be found using a cobweb diagram (do not explicitly calculate the values).
\item Now consider the line $z(x) = 2(1-x)$.  The solution to $z(x)=x$
  is $x=2/3$.  Starting from an initial $x_0$ close to (but not quite
  equal to) $2/3$, do several iterations of the cobweb diagram graphically.
\item Repeat this with the lines $z(x) = 1/4 + x/2$ starting at
  $x_0=0$ and $z(x) = -1 +
  3x$ starting close to where $x=z(x)$.
\item What is different when the slope is positive or negative?
\item Can you predict what condition on the slope's magnitude leads to convergence to or divergence from the solution to $x=z(x)$ when $z$ is a line?
\end{myenumerate}
So far we have considered lines $z(x)$.  Now assume $z(x)$ is nonlinear and consider the behavior of cobweb diagrams close to a point where $x=z(x)$.
\begin{continuemyenumerate}
\item Use Taylor Series to argue that (except for degenerate cases
  where $z'$ is $1$ at the intercept) it is only the slope at the
  intercept that determines the behavior sufficiently close to the intercept.
\end{continuemyenumerate}
\end{exercise}
\begin{solution}\mbox{}
\begin{myenumerate}
\item For $z(x)=2(1-x)/3$ with $x_0=0$, the cobweb diagram spirals in
  towards $0.4$,  the solution to $x=z(x)$.  \raisebox{-0.5\height}{\includegraphics[scale=0.3]{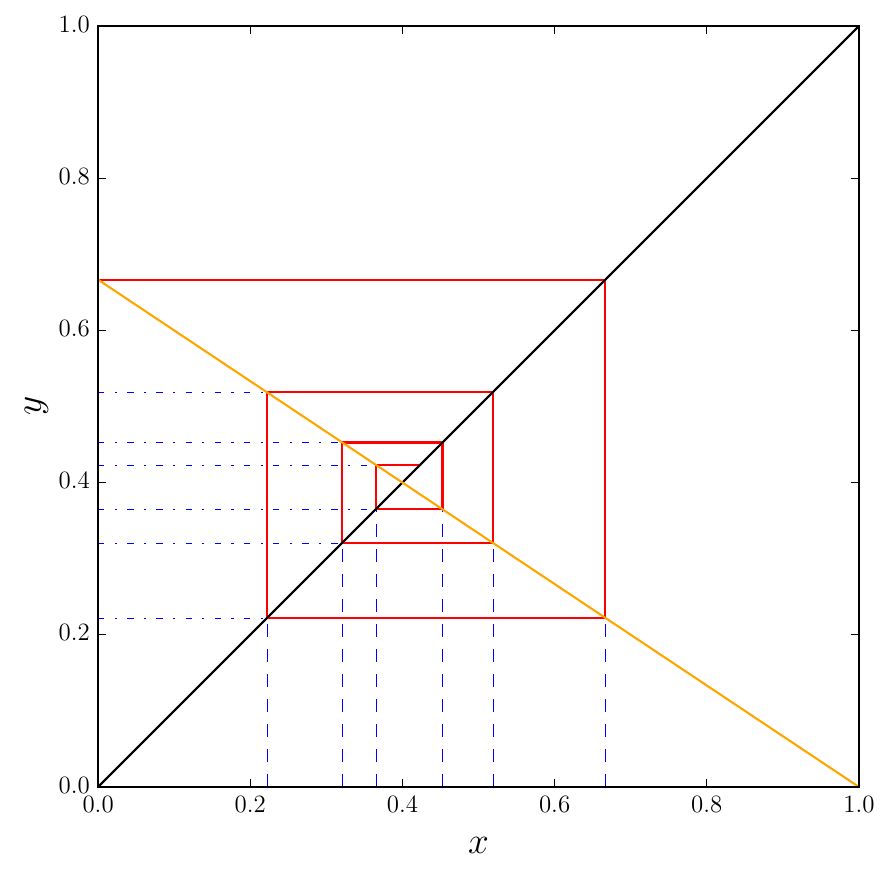}}
\item For $z(x)=2(1-x)$ with $x_0=0.664$, the cobweb diagram spirals
  away from $2/3$, the solution to  $x=z(x)$.  \raisebox{-0.5\height}{\includegraphics[scale=0.3]{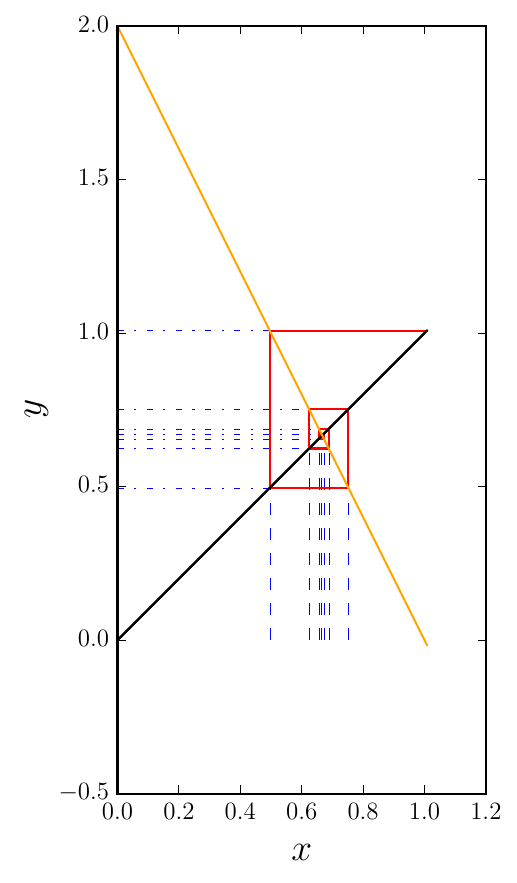}}
\item For $z(x)=1/4 + x/2$, the cobweb diagram steps in to $x=1/2$, the solution $x=z(x)$. \\ \raisebox{-0.5\height}{\includegraphics[scale=0.3]{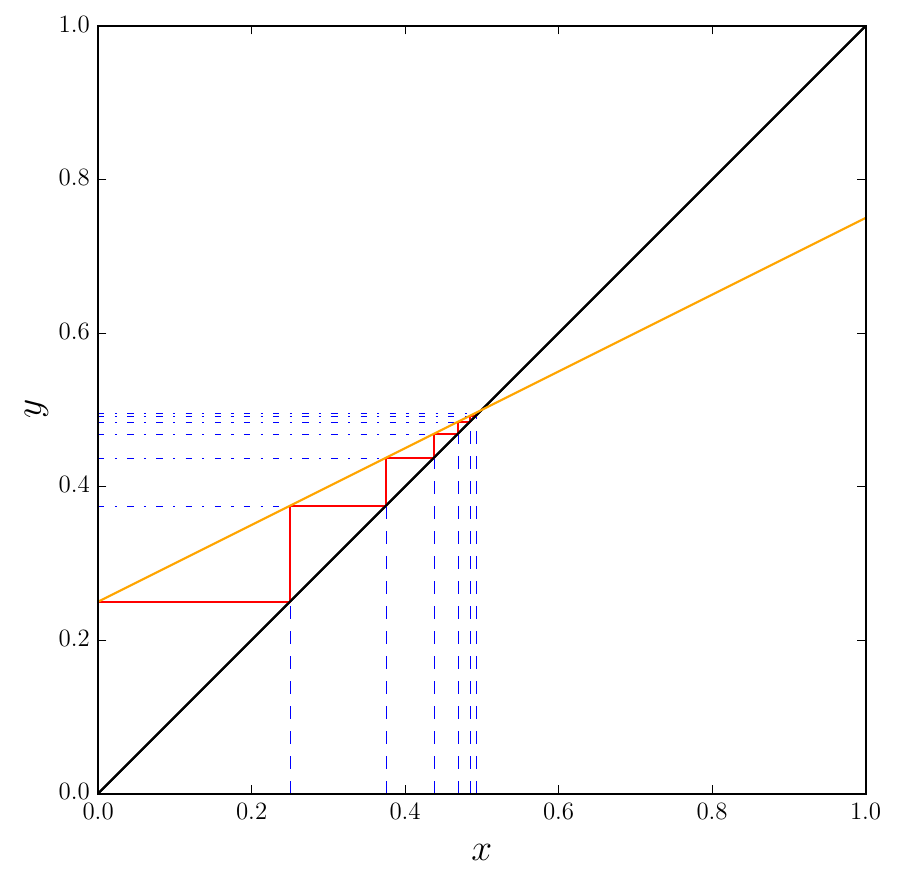}}\\
For $z(x)=-1+3x$, the cobweb diagram steps away from $x=1/2$, the solution $x=z(x)$.  \raisebox{-0.5\height}{\includegraphics[scale=0.3]{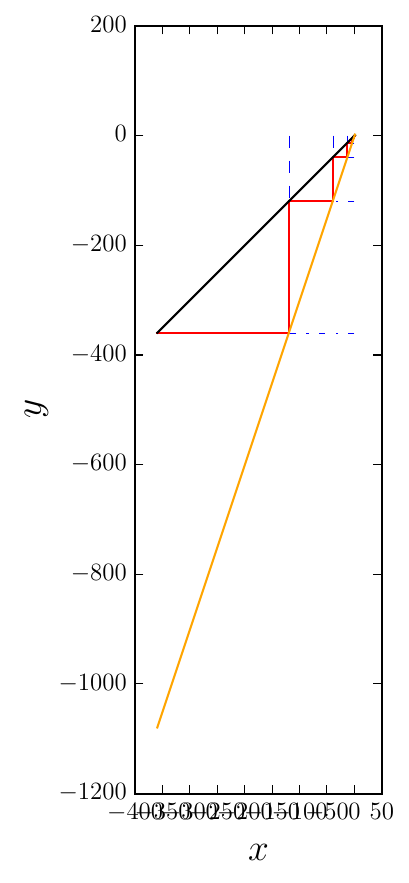}}
\item When the slope is positive, each successive value is on the same
  side of the intercept.  When it is negative the values alternate.
\item If we take $x^*$ to be the solution $x^*=z(x^*)$, then when $|z'(x^*)|>1$ the values diverge from $x^*$, while
  when $|z'(x^*)|<1$ they converge.
\item Sufficiently close to the intersection where $x^*=z(x^*)$,
  we can locally treat $z(x)$ as a
  line.  If $z'(x^*)$ is $1$ at this intersection, this line lies on top
  of $z(x)=x$ and the correction to the line makes a difference.  For all other values
  the nonlinear terms can be neglected.
\end{myenumerate}
\end{solution}

\begin{exercise} \textbf{Structure of fixed points of $f(x)$.}

Consider a PGF $f(x) = \sum_i r_i x^i$, and assume $r_0>0$.  
\begin{myenumerate}
\item Show that $f(1) = 1$ and $f(0)>0$.
\item Show that $f(x)$ is convex (that is $f''(x)\geq 0$) for $x>0$. [hint $r_i \geq 0$ for all $i$]
\item Thus argue that if $f'(1) \leq 1$, then $x=f(x)$ has only one
  solution to $x=f(x)$ in $[0, 1]$, namely $f(1)=1$.  It may help to draw pictures of $f(x)$ and the function $y=x$ for $x$ in $[0,1]$.
\item Explain why if there is a point $x_0\neq 1$ where $f(x_0) =x_0$
  and $f(x)>x$ for $x$ in some region $(x_0,x_1)$
  then  $0<f'(x_0)<1$. 
\item Thus  show that if $f'(1)>1$ then there are exactly two solutions to $x=f(x)$ in $[0,1]$, one of which is $x=1$.
\end{myenumerate}
\end{exercise}
\begin{solution}
\mbox{}
\begin{myenumerate}
\item $f(1) = \sum_i r_i 1^i = \sum_i r_i = 1$ because the $r_i$ form
  a probability distribution and $f(0) = \sum_i r_i 0^i = r_0 >0$.
\item $f''(x) = \sum_i i(i-1) r_i x^{i-2}$.  Because $r_i \geq 0$ and
  $x \geq 0$  this sum is $\geq 0$.  [the inequality is strict if any
  $r_i >0$ for $i \geq 2$].
\item Note first that $f(1)=1$.  If $f'(1)\leq 1$, then for all $x$
  in $0 \leq x < 1$, we have $f'(1) < 1$.  So because the line $y=x$
  has a larger slope, it lies strictly below $f(x)$ for all $x$ in
  $[0,1)$.  [a more rigorous argument can be made using the mean value
  theorem: assume some $x_0$ in $[0,1)$ has $f(x_0)=x_0$.  Then by the
  mean value theorem there is an $\hat{x} \in (x_0, 1)$ where
  $f'(\hat{x})$ is equal to the slope of the line connecting the
  points $(x_0, f(x_0))$ and $(1, f(1))$.  This line is $y=x$, and has
  slope $1$, so $f'(\hat{x})=1$.  No such $\hat{x}$ exists, so the
  assumption that such an $x_0$ exists must be false.]
\item $f'(x_0)>0$ because $f'(x_0) = \sum_i i r_i x^{i-1}$ and unless
  $r_0=1$, this is strictly positive.  If $r_0=1$, then $f(x)=1$ and
  no such $x_0$ exists.  To show $f'(x_0) <1$, note that for $f(x)$ to
  be less than $x$ for $x>x_0$, but equal to $x$ for $x=x_0$, it
  must have smaller slope than that of $y=x$ [a more rigorous proof uses the
  mean value theorem to show that there is an $\hat{x} \in (x_0, x_1)$
  with slope $<1$ and so since $f''(x)>0$, we have $f'(x_0) < f'(\hat{x})<1$.].
\item Again we start with the observation that $f(1)=1$.  Because
  $f'(x) >1$ at $x=1$, there must be a region $(\hat{x}, 1)$ such that if
  $x \in (\hat{x}, 1)$ then $f(x) < x$ as it has greater slope than the
  line $y=x$ [again, a more rigorous proof would use the mean-value
  theorem, and the fact that $f'(x)$ is continuous at $1$].  However,
  for the curve $f(x)$ to reach $f(0)>0$ at $x=0$, it must somewhere
  cross the line [rigorously, we can use the intermediate value
  theorem applied to the function $f(x)-x$ with the point $x=0$ and
  some $x  = x_1 \in (x_0, 1)$].  At the point $x_0$ where $f(x)$ crosses
  $y=x$, our previous result shows that $f'(x_0)< 1$.  So for $x \in
  (0, x_0)$ we have $f'(x)<1$ as well.  Thus $f(x)$ cannot cross $y=x$
  at any other point in $(0,x_0)$ [again, the mean value theorem does
  this more rigorously: if it did cross again, then we could prove
  some $\hat{x} \in (0, x_0)$ where $f'(\hat{x})=1$].
\end{myenumerate}
\end{solution}

These results suggest:
\begin{property}
\label{property:PGFcobweb}
Assume $f(x) = \sum_i r_i x^i$ is a PGF, and $f(0)>0$.
\begin{itemize}
\item If
$f'(1)\leq 1$ then the only intercept of $x=f(x)$ in $[0,1]$ is at $x=1$.
\item Otherwise, there is another intercept $x^*$, $0 < x^* < 1$,
and if $x< x^*$ then $x<f(x)<x^*$ while if $x>x^*$ then $x>f(x)>x^*$
and for $0\leq x_0<1$, $f^{[g]}(x_0)$ converges monotonically to $x^*$.
\end{itemize}
\end{property}
The assumption $r_0>0$ was used to rule out $f(x)=x$.  Excluding this degenerate case, these results hold even if $r_0=0$, in which case we can show $f'(1)>1$ and $x^*=0$.

To sketch the proof of this property, we note that clearly $f(1)=1$, so if $f(0)>0$ then either $f(x)$ crosses $y=x$ at some intermediate $0<x^*<1$ or it does not cross until $x=1$.  Then using the fact that for $x>0$ the slope of $f$ is positive and increasing, we can inspect the cobweb diagram to see these results.

\ifsolns
\else
\subsection{Finding the Kolmogorov Equations}
\fi
To study continuous-time dynamics, we will want to have partial differential equations (PDEs) where we write the time derivative of a PGF $f(x,t)$ or $f(x,y,t)$ in terms of $f$ and its spatial derivatives.

We will use two approaches to find the derivative.  Both start with the assumption that we know $f(x,t)$, and calculate the derivative by finding $f(x,t+\Delta t)$ and use  the definition of the derivative: 
\[
\pd{}{t} f(x,t) = \lim_{\Delta t\to 0} \frac{f(x,t+\Delta t)- f(x,t)}{\Delta t}
\]
The methods differ in how they find $f(x,t+\Delta t)$.  The distinction is closely related to the observation in Exercise~\ref{exercise:discrete_backward_and_forward} that $\mu^{[g]}(x)$ can be written as either $\mu^{[g-1]}(\mu(x))$ or $\mu(\mu^{[g-1]}(x))$.

\begin{itemize}
\item The first involves assuming we know $f(x,t)$ and then looking through all of the possible transitions to find how the system changes going from $t$ to $t+\Delta t$.  This will yield the forward Kolmogorov Equations.
\item The second involves starting from the initial condition $f(x,0)$ and finding $f(x,\Delta t)$ by investigating all of the possible transitions.  Then taking $f(x,\Delta t)$ and $f(x,t)$ we are able to find $f(x,t+\Delta t)$.  This will yield the backward Kolmogorov Equations.
\end{itemize}

\ifsolns
\else
\subsubsection{Forward Kolmogorov Equations}
\fi

We start with the Forward Kolmogorov Equations.  We let $r_i(t)$ denote the probability that at time $t$ there are $i$ individuals, and define the PGF
\[
f(x,t) = \sum_i r_i(t) x^i
\]
We begin by looking at events that can be treated as if they remove one individual and replace it with $m$ individuals.  Thus $i$ is replaced by $i+m-1$:
\[
i \mapsto i+m-1 \, .
\]
For example early in an epidemic, we may assume that an infected individual causes new infections at rate $\beta$.  The outcome of an infection event is equivalent to the removal of the infected individual and replacement by two infected individuals.  Similarly, a recovery event occurs with rate $\gamma$ and is equivalent to removal with no replacement.  So $\lambda_2 = \beta$, \ $\lambda_0=\gamma$, and all other $\lambda_m$ are $0$.

Our events happen at a per-individual rate $\lambda_m$, so the total
rate an event occurs across the population of $i$ individuals is $\lambda_m i$.  Events that can be modeled like this include decay of a radioactive particle, recovery of an infected individual, or division of a cell.  We assume that different events may be possible, each having a different $m$.  If multiple events have the same effect on $m$ (for example emigration or death), we can combine their rates into a single $\lambda_m$.

It will be useful to define 
\[
\Lambda = \sum_m \lambda_m
\]
to be the combined per-capita rate of all possible events and 
\[
h(x) = \sum_m \lambda_m x^m/\Lambda
\]
We can think of $h(x)$ as the PGF for the number of new individuals given that a random event happens (since $\lambda_m/\Lambda$ is the probability that the random event introduces $m$ individuals).

We start with one derivation of the equation for $\dot{f}(x,t)$  based on directly
calculating $f(x,t+\Delta t)$ and using the definition of the derivative.  An alternate way is shown in exercise~\ref{exercise:derivative_sum}.
For small $\Delta t$ the probability that multiple events occur in the
same time interal is $\littleo(\Delta t)$, and we will see that this
is negligible.  Let us assume the system has $i$ individuals at time
$t$, whch occurs with probability $r_i(t)$.  For a given $m$, the
probability that the event occurs in the time interval given $i$ is
$\lambda_m i \Delta t + \littleo(\Delta t)$ and $(1-\sum_m \lambda_m i
\Delta t)+\littleo(\Delta t)$ measures the probability that none of the the events occur
in the time interval and the system remains in state $i$.  If the event occurs, the system leaves the
state corresponding to $x^i$ and enters the state corresponding to
$x^{i+m-1}$.  Summing over $m$ and $i$, we have
\[
f(x,t+\Delta t) = \sum_i  \left(r_i(t) \left[\sum_m (\lambda_m i \Delta t)x^{i+m-1} + \left(1-\sum_m  \lambda_m i \Delta t\right) x^i \right] \right)+ \littleo(\Delta t)
\]
The $\littleo(\Delta t)$ corrects for the possibility of multiple events happening in the time interval.

A bit of algebra and separating the $i$ and $m$ summations shows that
\begin{align*}
f(x,t+\Delta t) &= \sum_i r_i(t) x^i +  \sum_m  \lambda_m (\Delta t) (x^m-x)\sum_i r_i(t) ix^{i-1} + \littleo(\Delta t)\\
&= f(x,t) + \sum_m \lambda_m (x^m-x) \Delta t \sum_i r_i(t) \pd{}{x} x^i+ \littleo(\Delta t)\\
&= f(x,t) + (\Delta t)\left(\sum_m \lambda_m x^m - x\sum_m \lambda_m \right) \pd{}{x} \sum_i r_i(t) x^i + \littleo(\Delta t)\\
&= f(x,t) + \Lambda(\Delta t) [h(x) -x] \pd{}{x} f(x,t)+ \littleo(\Delta t)
\end{align*}

So we now have
\begin{align*}
\pd{}{t} f(x,t) &=  \lim_{\Delta t \to 0} \frac{f(x,t+\Delta t)-f(x,t)}{\Delta t}\\
&= \lim_{\Delta t \to 0} \frac{f(x,t) + \Lambda\Delta t [h(x) -x] \pd{}{x} f(x,t)+ \littleo(\Delta t)-f(x,t)}{\Delta t}\\
&= \Lambda [h(x)-x] \pd{}{x} f(x,t)
\end{align*}

We finally have
\begin{property}
\label{prop:derivative_sum}
Let $f(x,t) = \sum_i r_i(t) x^i$ be the PGF for the probability of having $i$ individuals at time $t$.  Assume several events indexed by $m$ can occur, each with rate $\lambda_m i$, that remove one individual and replace it with $m$.  Let $\Lambda = \sum_m \lambda_m$ be the total per-capita rate and $h(x) = \sum_m \lambda_m x^m/\Lambda$ be the PGF of the outcome of a random event.  Then
\begin{equation}
\label{eqn:derivative_sum}
\pd{}{t}f(x,t) = \Lambda [h(x)-x] \pd{}{x} f(x,t)
\end{equation}
\end{property}

We look at a heuristic way to interpret this.  We can rewrite Equation~\eqref{eqn:derivative_sum} as
\[
\dot{f}(x,t) = \left[\sum_m (\lambda_m x^m - \lambda_m x)\right] \pd{}{x} f(x,t)
\]
Then if we expand $f$ on the right hand side, we have 
\[
\sum_m \sum_i \lambda_m (x^m - x)i r_i x^{i-1}
\]
The derivative serves the purpose of getting the factor $i$ into the coefficient of each term which addresses the fact that the rate events happen is proportional to the total count.  The derivative has the additional effect of reducing the exponent by $1$, corresponding to the removal of one individual.  The  $\lambda_m$ in the remaining factor gives the per-capita rate of changing state.  The $x^m - x$ captures the fact that when moving to that new state $x^m$ individuals are added but the system is leaving the current state (which has an exponent of $x^i$) with the same rate.

\begin{exercise}
\label{exercise:derivative_sum}
\textbf{Alternate derivation of Equation~\eqref{eqn:derivative_sum}}

An alternate way to derive Equation~\eqref{eqn:derivative_sum} is
through directly calculating $\dot{r}_i$.

\begin{myenumerate}
\item Explain why $\dot{r}_i= - \sum_m \lambda_m i r_i + \sum_m \lambda_m
  (i-m+1) r_{i-m+1}$.
\item Taking $\dot{f}(x,t) = \sum_i \dot{r}_i x^i$, derive Equation~\eqref{eqn:derivative_sum}.
\end{myenumerate}
\end{exercise}
\begin{solution}
\mbox{}
\begin{myenumerate}
\item We simply look at the rate $r_i$ is decreasing due to events
  that move the system out of state $i$ plus the rate it increases due
  to events that move the system into state $i$.  The total rate
  moving out is $\sum_m \lambda_m i r_i$.  The total rate in is
  $\sum_m \lambda_m (i-m+1) r_{i-m+1}$.
\item 
 \begin{align*}
 \dot{f}(x,t) &= \sum_i \dot{r}_i x^i\\
&= \sum_i \left(- \sum_m \lambda_m i r_i + \sum_m \lambda_m
  (i-m+1) r_{i-m+1}\right) x^i\\
&= - \sum_{i}\sum_m \lambda_m i r_i x^i + \sum_i\sum_m \lambda_m
(i-m+1)r_{i-m+1}x^i\\
&= -\Lambda \sum_i i r_i x^i + \sum_m \lambda_m \sum_i (i-m+1)
r_{i-m+1}x^i\\
&= -\Lambda x \sum_i i r_i x^{i-1} + \sum_m \lambda_m x^{m}\sum_i (i-m+1)
r_{i-m+1} x^{i-m}\\
&= -\Lambda x \pd{}{x} \sum_i r_i x^i + \sum_m \lambda_m x^m \pd{}{x} \sum_i
r_{i-m+1}x^{i-m+1}\\
&= -\Lambda x \pd{}{x} f(x,t) + \sum_m \lambda_m x^m \pd{}{x} f(x,t)\\
&= -\Lambda x \pd{}{x} f(x,t)+ \Lambda h(x) \pd{}{x} f(x,t)\\
&= \Lambda(h(x)-x) \pd{}{x} f(x,t)
\end{align*}

\end{myenumerate}
\end{solution}
We can generalize this to the case where there are multiple types of individuals.  For the Forward Kolmogorov equations, it is relatively straightforward to allow for interactions between individuals.  We may be interested in this generalization when considering predator-prey interactions or interactions between infected and susceptible individuals if we are interested in depletion of susceptibles.  We assume that there are two types of individuals $A$ and $B$ with counts $i$ and $j$ respectively, and we let $r_{ij}(t)$ denote the probability of a given pair $i$ and $j$.  We define the PGF
\[
f(x,y,t) = \sum_{i,j} r_{i,j}(t) x^i y^j
\]

We assume that interactions between an $A$ and a $B$ individual occur with some rate proportional to the product $ij$  We assume that the interaction removes both individuals and replaces them by $m$ of type $A$ and $n$ of type $B$.  We denote the rate as $\mu_{m,n}ij$, and the sum 
\[
\mathfrak{M} = \sum_{m,n} \mu_{m,n} \, .
\]

We also assume that individuals of type $A$ spontaneously undergo changes as they did above, but they can be replaced by type $A$ and/or type $B$ individuals.  So one individual of type $A$ is removed and replaced by $m$ individuals of type $A$ and $n$ of type $B$ with rate $\lambda_{m,n}$, and the combined rate for one specific transition over the entire set of individuals is $\lambda_{m,n} i$.  We define 
\[
\Lambda = \sum_{m,n} \lambda_{m,n}\,.
\]
We will ignore spontaneous changes by nodes of type $B$, but the generalization to include these can be found by following the same method.

Finally, let 
\[
h(x,y) = \sum_{m,n} \lambda_{m,n} x^my^n/\Lambda
\]
and 
\[
g(x,y) = \sum_{m,n} \mu_{m,n} x^my^n/\mathfrak{M}
\]
be the PGFs for the outcomes of the two types of events.

Then
\begin{align*}
f(x,y,t+\Delta t) &= \sum_{i,j} r_{i,j}(t) \left[\sum_{m,n} [(\lambda_{m,n} i \Delta t)x^{i+m-1}y^j + (\mu_{m,n}ij \Delta t)x^{i+m-1}y^{j+n-1}  \right.
\\
&\qquad\qquad\qquad \left. + \left(1-\sum_{m,n}[\lambda_{m,n}i \Delta t +  \mu_{m,n} ij \Delta t]\right) x^iy^j \right] + \littleo(\Delta t)\\
&= \sum_{i,j} r_{i,j}(t) x^iy^j + \sum_{m,n}\lambda_{m,n}(x^my^n-x)\Delta t \sum_{i,j} r_{i,j}(t) ix^{i-1}y^j\\
&\quad +\sum_{m,n} \mu_{m,n}(x^my^n-xy)\Delta t \sum_{i,j} r_{i,j}(t) ij x^{i-1}y^{j-1}+ \littleo(\Delta t)\\
&= f(x,y,t) + \sum_{m,n}\lambda_{m,n}(x^my^n-x)\Delta t \pd{}{x} f(x,y,t)\\
&\quad +\sum_{m,n} \mu_{m,n}(x^my^n-xy)\Delta t \pd{}{x}\pd{}{y} f(x,y,t)+ \littleo(\Delta t)\\
&= f(x,y,t) + (\Delta t)\left( \Lambda[h(x,y)-x]\pd{}{x}f(x,y,t) + \mathfrak{M}[g(x,y)-xy]\pd{}{x}\pd{}{y} f(x,y,t)\right)+ \littleo(\Delta t)
\end{align*}
So 
\begin{align*}
\pd{}{t} f(x,y,t) &= \lim_{\Delta t \to 0} \frac{f(x,y,t+\Delta t)-f(x,y,t)}{\Delta t}\\
&= \lim_{\Delta t \to 0} \frac{(\Delta t) \Lambda [h(x,y)-x] \pd{}{x} f(x,y,t) + (\Delta t) \mathfrak{M} [g(x,y)-xy] \pd{}{x}\pd{}{y}f(x,y,t) + \littleo(\Delta t)}{\Delta t}\\
&= \Lambda[h(x,y)-1]x \pd{}{x} f(x,y,t) + \mathfrak{M} [g(x,y)-1]xy \pd{}{x} \pd{}{y} f(x,y,t)
\end{align*}
We have shown:
\begin{property}
\label{prop:2Dderivative_sum}
Let $f(x,y,t) = \sum_{i,j} r_{ij}(t) x^iy^j$ be the PGF for the probability of having $i$ type $A$ and $j$ type $B$ individuals.  Assume that events occur with rate $\lambda_{m,n} i$ or $\mu_{m,n}ij$ to replace a single type $A$ individual or one of each type with $m$ type $A$ and $n$ type $B$ individuals.  Let $\Lambda = \sum_{m,n} \lambda_{m,n}$ and $\mathfrak{M} = \sum_{m,n} \mu_{m,j}$.  Then
\begin{equation}
\label{eqn:2Dderivative_sum}
\pd{}{t} f(x,y,t) = \Lambda[h(x,y)-x] \pd{}{x} f(x,y,t) + \mathfrak{M}[g(x,y)-xy] \pd{}{x} \pd{}{y} f(x,y,t)
\end{equation}
where $h(x,y) = \sum_{m,n} \lambda_{m,n}x^my^n/\Lambda$ is the PGF for the outcome of a random event whose rate is proportional to $i$ and $g(x,y) = \sum_{m,n} \mu_{m,n}x^my^n/\mathfrak{M}$ is the PGF for the outcome of a random event whose rate is proportional to $ij$.
\end{property}
This can be generalized if there are events whose rates are proportional only to $j$ or if there are more than two types.  The exercise below shows how to generalize this if the rate of events depend on $i$ in a more complicated manner.

\begin{exercise}
In many cases interactions between two individuals of the same type are important.  These may occur with rate $i(i-1)$ or $i^2$ depending on the specific details.  Assume we have only a single type of individual with PGF $f(x,t) = \sum_i r_i(t)x^i$.
\begin{myenumerate}
\item If a collection of  events to replace two individuals with $m$ individuals
  occur with rate $\beta_mi(i-1)$, find how write a PDE for $f$.  Your final result should
  contain $\pds{}{x} f(x,t)$.  Use $\mathfrak{B} = \sum_m \beta_m$ and
  $g(x) = \sum_m \beta_m x^m/\mathfrak{B}$.  Follow the derivation of Equation~\eqref{eqn:derivative_sum}.
\item If instead the events replace two individuals with $m$
  individuals and occur with rate $\beta_m i^2$, find how to incorporate them into a PDE for $f$.  Your final result should contain $\pd{}{x} \left( x \pd{}{x} f(x,t)\right)$ or equivalently $\pd{}{x} f(x,t) + x \pds{}{x} f(x,t)$.
\end{myenumerate}
\end{exercise}
\begin{solution}
\mbox{}
\begin{myenumerate}
\item Let $\beta_{m} i(i-1)$ denote the rate at which the system
  goes from a state with $i$ individuals to $i-2+m$ individuals (that
  is, two individuals are replaced by $m$).  Then
\begin{align*}
f(x,t+\Delta t) &= \sum_i r_i(t) \left[ \sum_m (\beta_m i(i-1) \Delta t
  )x^{i+m-2} + \left(1- \sum_m \beta_m i(i-1) \Delta t\right)
  x^i\right]  + \littleo(\Delta t)\\
&= \sum_i r_i x^i + \sum_m \beta_m (\Delta t) (x^m-x^2) \sum_i r_ii(i-1)
x^{i-2} + \littleo(\Delta t)\\
&= f(x,t) + \mathfrak{B}\Delta t [g(x) - x^2] \sum_i r_i i(i-1)
x^{i-2} + \littleo(\Delta t)\\
&= f(x,t) + \mathfrak{B} \Delta t[g(x)-x^2] \pds{}{x} f(x,t) +
\littleo(\Delta t)
\end{align*}
Plugging this into 
\[
\pd{}{t} f(x,t) = \lim_{\Delta t \to 0} \frac{f(x,t+\Delta t)-f(x)}{\Delta t}
\]
yields
\[
\pd{}{t} f(x,t) = \mathfrak{B} [g(x)-x^2] \pds{}{x} f(x,t)
\]
\item The proof is almost the same as the previous case:
Let $\beta_{m} i^2$ denote the rate at which the system
  goes from a state with $i$ individuals to $i-2+m$ individuals (that
  is, two individuals are replaced by $m$).  Then
\begin{align*}
f(x,t+\Delta t) &= \sum_i r_i(t) \left[ \sum_m (\beta_m i^2 \Delta t
  )x^{i+m-2} + \left(1- \sum_m \beta_m i^2 \Delta t\right)
  x^i\right]  + \littleo(\Delta t)\\
&= \sum_i r_i x^i + \sum_m \beta_m (\Delta t) (x^m-x^2) \sum_i r_ii^2
x^{i-2} + \littleo(\Delta t)\\
&= f(x,t) + \mathfrak{B}\Delta t [g(x) - x^2] \sum_i r_i i^2
x^{i-2} + \littleo(\Delta t)\\
&= f(x,t) + \mathfrak{B} \Delta t[g(x)-x^2] \frac{1}{x}\pd{}{x} x\pd{}{x} f(x,t) +
\littleo(\Delta t)
\end{align*}
Plugging this into 
\[
\pd{}{t} f(x,t) = \lim_{\Delta t \to 0} \frac{f(x,t+\Delta t)-f(x)}{\Delta t}
\]
yields
\[
\pd{}{t} f(x,t) = \mathfrak{B} [g(x)-x^2] \frac{1}{x}\pd{}{x} x \pd{}{x}f(x,t)
\]
\end{myenumerate}
\end{solution}

\begin{exercise}
Consider a chemical system that begins with some initial amount of chemical $A$.  Let $i$ denote the number of molecules of species $A$.  A molecule of $A$ spontaneously degrades into a molecule of $B$, with rate $\xi$ per molecule.  Let $j$ denote the number of molecules of species $B$.  Species $B$ reacts with $A$ at rate $\eta i j$ to produce new molecules of species $B$.  The reactions are denoted
\begin{align*}
A &\mapsto B\\
A + B &\mapsto 2B
\end{align*}
Let $r_{i,j}(t)$ denote the probability of $i$ molecules of $A$ and $j$ molecules of $B$ at time $t$.  Let $f(x,y,t) = r_{i,j}(t)x^iy^j$ be the PGF.  Find the Forward Kolmogorov Equation for $f(x,y,t)$.
\end{exercise}
\begin{solution}
\begin{align*}
\pd{}{t} f(x,y,t) &= \xi (x^{-1}y - 1)x \pd{}{x} f(x,y,t) + \eta (x^{-1}y-1) xy\pd{}{x}\pd{}{y}f(x,y,t)\\
&= \left(\xi (y-x) \pd{}{x} + \eta (y^2-xy) \pd{}{x}\pd{}{y}\right) f(x,y,t)
\end{align*}
\end{solution}
\ifsolns
\else
\subsubsection{Backward Kolmogorov equations}
\fi
We now look for another derivation of $\pd{}{t} f(x,t)$, and as before we find it by first finding $f(x,t+\Delta t)$ for small $\Delta t$ and then using the definition of the derivative.  We will assume that each individual acts independently, and at rate $\lambda_m$ an individual may be removed and replaced by $m$ new individuals.  So if there are $i$ total individuals, at rate $\lambda_m i$ the count $i$ is replaced by $i -1+m$.

Property~\ref{prop:composition} plays an important role in our
derivation.  We define $f_1(x,t) = \sum_i r_i(t) x^i$ where we assume
that $r_1(0)=1$, that is we start with exactly one individual at time
$0$.  Then Property~\ref{prop:composition} shows that $f_1(x,t_1+t_2)
= f_1(f_1(x,t_2),t_1)$.  Then from our initial condition $f_1(x,0)=x$, and
\begin{equation}
\label{eqn:f1_comp}
f_1(x,\Delta t + t) = f_1(f_1(x, t), \Delta t)
\end{equation}
We need to find $f_1(x,\Delta t)$.  We have
\begin{align*}
f_1(x, \Delta t) &= \sum_i r_i(0) x^i \left(1- \sum_m i \lambda_m (\Delta t) + \sum_m i \lambda_m (\Delta t) x^{m-1}\right) +\littleo(\Delta t)\\
&= x\left(1- \sum_m \lambda_m (\Delta t) + \sum_m \lambda_m (\Delta t) x^{m-1}\right)+\littleo(\Delta t)\\
&= x- x(\Delta t)\sum_m \lambda_m +\sum_m \lambda_m x^m+\littleo(\Delta t)\\
&= x + (\Delta t) \Lambda [h(x)-x]+\littleo(\Delta t)
\end{align*}
where, as in the forward Kolmogorov case, $\Lambda = \sum_m \lambda_m$
and $h(x)= \sum_m \lambda_m x^m/\Lambda$ is the PGF of the number of
new individuals created given that an event occurs.  In the first step
we used the fact that for $f_1(x,t)$, \ $r_i(0) = 1$ if $i=1$ and
otherwise it is $0$.  Thus Equation~\eqref{eqn:f1_comp} implies 
\[
f_1(x,t+\Delta t) = f_1(x,t) + (\Delta t) \Lambda[h(f_1(x,t))- f_1(x,t)] + \littleo(\Delta t) \, .
\]

Now taking the definition of the derivative, we have
\begin{align*}
\pd{}{t} f_1(x,t) &= \lim_{\Delta t \to 0} \frac{f_1(x,t+\Delta t)- f_1(x,t)}{\Delta t}\\
&= \lim_{\Delta t \to 0} \frac{f_1(x,t) +(\Delta t) \Lambda[h(f_1(x,t))- f_1(x,t)] + \littleo(\Delta t)-f_1(x,t)}{\Delta t}\\
&= \Lambda [ h(f_1(x,t))-f_1(x,t)]
\end{align*}
Thus we have an ODE for $f_1(x,t)$.  

In general, our initial condition may not be a single individual, but some other number (or perhaps a value chosen from a distribution).  Let the initial condition have PGF $f(x,0)$.  Then it follows from Property~\ref{prop:composition} that 
\[
f(x,t) = f(f_1(x,t),0)
\]

So we have
\begin{property}
\label{prop:bKe2PGF}
Consider a process in which the number of individuals change in time such that when an event occurs one individual is destroyed and replaced with $m$ new individuals.  The associated rate associated with an event that changes the population size by $m$ is $\lambda_m i$ where $i$ is the number of individuals.  Let $f_1(x,t)$ be the PGF for this process beginning from a single individual and $\Lambda = \sum_m \lambda_m$.
Then
\begin{equation}
\label{eqn:bKe2PGF_single_initial}
\dot{f}_1(x,t) = \Lambda \left[h(f_1(x,t))-f_1(x,t)\right]
\end{equation}
where $h(x)$ is the PGF for the number of new individuals created in a random event.
If the initial number of individuals is not $1$, let $f(x,0)$ denote the PGF for the initial condition.  Then
\begin{equation}
\label{eqn:bKE2PGF}
f(x,t) = f(f_1(x,t),0)
\end{equation}
is the PGF at arbitrary positive time.
\end{property}
This is fairly straightforward to generalize to multiple types as long as none of the events involve interactions.  

\begin{exercise}
\label{exercise:bkwds_2type}
In this exercise we generalize Property~\ref{prop:bKe2PGF} for the
case where there are two types of individuals $A$ and $B$ with counts
$i$ and $j$.  

Assume events occur spontaneously with rate $\lambda_{m,n}i$ to remove
an individual of type $A$ and replace it with $m$ of type $A$ and $n$
of type $B$, or they occur spontaneously with rate $\zeta_{m,n}j$ to
remove an individual of type $B$ and replace it with $m$ of type $A$
and $n$ of type $B$.

Set $\Lambda = \sum_{m,n} \lambda_{m,n}$ and $\mathfrak{Z} =
\sum_{m,n} \zeta_{m,n}$.  Let $f_{1,0}(x,y,t)$ denote the outcome
beginning with one individual of type $A$ and $f_{0,1}(x,y,t)$ denote
the outcome beginning with one individual of type $B$.

\begin{myenumerate}
\item Write $f_{1,0}(x,y, \Delta t)$ and $f_{0,1}(x,y,\Delta t)$ in
  terms of $h(x,y) = \sum_{m,n} \lambda_{m,n} x^my^n/\Lambda$ and $g(x,y) =
  \sum_{m,n} \zeta_{m,n} x^m y^n/\mathfrak{Z}$.

 \item Use Property~\ref{prop:composition}, write $f_{1,0}(x,\Delta t + t)$ and $f_{0,1}(x, \Delta t + t)$
  in terms of $f_{1,0}$ and $f_{0,1}$ evaluated at $t$ and $\Delta
  t$.  The answer should resemble Equation~\eqref{eqn:f1_comp}.

\item Derive expressions for $\pd{}{t} f_{1,0}(x,y,t)$ and $\pd{}{t}
  f_{0,1}(x,y,t)$.
\item Use this to derive Equation~\eqref{eqn:backward_result}.
\end{myenumerate}

\end{exercise}
\begin{solution}
\mbox{}

\begin{myenumerate}
\item \begin{align*}
f_{1,0}(x,y,\Delta t) &= \sum_i r_{i,j}(0) x^i \left(1 - \sum_{m,n} i
  \lambda_{m,n} \Delta t + \sum_{m,n}  i \lambda_{m,n} x^{m-1}y^n\right) +
  \littleo(\Delta t)\\
&= x\left( 1 - \sum_{m,n} 
  \lambda_{m,n} \Delta t + \sum_{m,n}   \lambda_{m,n} x^{m-1}y^n\right) +
  \littleo(\Delta t)\\
&= x - x(\Delta t) \sum_{m,n} \lambda_{m,n} + \sum_{m,n}\lambda_{m,n} x^m y^n +
\littleo(\Delta t)\\
&= x + (\Delta t) \Lambda [h(x,y) - x] + \littleo(\Delta t)
\end{align*}
Similarly
\[
f_{0,1}(x,y,\Delta t) = y + (\Delta t) \mathfrak{Z} [g(x,y)-y] +
\littleo(\Delta t)
\]
\item We find 
\[
f_{1,0}(x,y,\Delta t + t) =
f_{1,0}(f_{1,0}(x,y,t),f_{0,1}(x,y,t), \Delta t)
\]
and
\[
f_{0,1}(x,y,\Delta t + t) = f_{0,1}(f_{1,0}(x,y,t), f_{0,1}(x,y,t),
\Delta t)
\]

\item We now have
\begin{align*}
\pd{}{t} f_{1,0}(x,y,t) &= \lim_{\Delta t \to 0}
\frac{f_{1,0}(x,y,\Delta t + t)-f_{1,0}(x,y,t)}{\Delta t} \\
&= \lim_{\Delta t \to 0}
\frac{f_{1,0}(f_{1,0}(x,y,t),f_{0,1}(x,y,t),
\Delta t) -f_{1,0}(x,y,t)}{\Delta t} \\
&=\lim_{\Delta t \to 0}
\frac{ f_{1,0}(x,y,t) + (\Delta t) \Lambda[h(f_{1,0}(x,y,t),f_{0,1}(x,y,t))
- f_{1,0}(x,y,t)]  + \littleo(\Delta t) -f_{1,0}(x,y,t)}{\Delta t} \\
&= \Lambda [h(f_{1,0}(x,y,t), f_{0,1}(x,y,t)) - f_{1,0}(x,y,t)] + \lim_{\Delta t \to 0}
\frac{\littleo(\Delta t)}{\Delta t}\\
&=\Lambda [h(f_{1,0}(x,y,t), f_{0,1}(x,y,t)) - f_{1,0}(x,y,t)]
\end{align*}
Similarly
\[
\pd{}{t} f_{0,1}(x,y,t) = \mathfrak{Z} [g(f_{1,0}(x,y,t),
f_{0,1}(x,y,t))]
\]
\item To derive Equation~\eqref{eqn:backward_result}, we take infected
  individuals to be type $A$ and recovered individuals to be type
  $B$.  We replace $x$ with $\tilde{y}$, and $y$ with $\tilde{z}$.  We will drop the
  tildes later.  

  The events that can happen are that a single infected individual can
  be replaced by $2$ infected individuals (with rate $\beta$ per
  infected individual) or by $1$ recovered individual (with rate
  $\gamma$ per infected individual).  So $\Lambda = (\beta + \gamma)$,
  and $h(\tilde{y}, \tilde{z}) = (\beta \tilde{y}^2 +
  \gamma\tilde{z})/(\beta+\gamma)$.  There are no events that can
  happen to recovered individuals, so $\mathfrak{Z} = 0$ and $g$ could
  be anything.  

  So
\begin{align*}
\pd{}{t} f_{1,0}(\tilde{y},\tilde{z}, t) &= (\beta+\gamma)
\left[\frac{\beta f_{1,0}(\tilde{y},\tilde{z},t)^2 + \gamma
    f_{0,1}(\tilde{y},\tilde{z},t)}{\beta+\gamma} -
  f_{1,0}(\tilde{y},\tilde{z},t)\right]\\
\pd{}{t} f_{0,1}(\tilde{y},\tilde{z},t) &= 0
\end{align*}
Since $f_{0,1}(\tilde{y},\tilde{z},t)$ is constant and initially it is
simply $\tilde{z}$, we conclude that it is always $\tilde{z}$.  Thus
we get
\[
\pd{}{t} f_{1,0}(\tilde{y},\tilde{z},t) = (\beta+\gamma)
\left[\frac{\beta f_{1,0}(\tilde{y},\tilde{z},t)^2 + \gamma
\tilde{z}}{\beta+\gamma} -
  f_{1,0}(\tilde{y},\tilde{z},t)\right]
\]
Then replacing $f_{1,0}$ by $\Pi$, \ $\tilde{y}$ by $y$ and
$\tilde{z}$ by $z$ and replacing $\frac{\beta f_{1,0}(\tilde{y},\tilde{z},t)^2 + \gamma
\tilde{z}}{\beta+\gamma}$ by $\hat{\mu}(\Pi(y,z,t),z)$ completes the result.
\end{myenumerate}

\end{solution}

\section{Proof of Theorems~\ref{thm:power_magic} and~\ref{thm:cts_power_magic}}
\label{app:power_magic}
We now prove Theorems~\ref{thm:power_magic}
and~\ref{thm:cts_power_magic}.  

We first sketch out the idea behind the method of proof of Theorem\ref{thm:power_magic}.  The idea is that if an outbreak dies out with exactly $j$ infections, then there must be a transmission tree that corresponds to exactly $j$ infections.  In the construction of the tree, each successive number of downward links was chosen from the offspring distribution.  Our goal is to find out the probability of arriving at a finite tree with exactly $j$ infections given the offspring distribution.

This tree has certain constraints on it.  The first constraint is that it must have exactly $j-1$ transmissions from the $j$ infected individuals.  So we look at the probability of having a sum of $j-1$ when we choose $j$ numbers from the offspring distribution.  This is given by the coefficient of $y^{j-1}$ in $[\mu(y)]^j$.

Next we have to make sure that the sequence is consistent with an outbreak that did not die out sooner.  For example, if an outbreak has exactly two infections, we cannot assume that the first individual infected no-one and then the second individual infected $1$ because the outbreak would have died out without the second individual having the chance to transmit.  So it is not enough for the sequence to add to $j-1$, the order must be consistent with an outbreak of size $j$.

It turns out that we can find a one-to-one mapping between trees on $j$ individuals and ``valid'' sequences summing to $j-1$.  When doing this, we discover that if a sequence sums to $j-1$, there is exactly one cyclic permutation of that sequence which is valid.\footnote{A \emph{cyclic permutation} of a sequence is formed by thinking of the sequence as a loop, and then choosing a different starting point.}   Thus of all sequences of $j$ values chosen from the offspring distribution that sum to $j-1$, a fraction $1/j$ are ``valid'', that is they yield a complete transmission tree.  So the probability is $(1/j)$ times the coefficient of $y^{j-1}$ in $[\mu(y)]^j$.

We now go through the proof in detail.

\ifsolns
\else
\subsection{Proof of Theorem~\ref{thm:power_magic}}
\fi
We take as given a probability distribution so that $p_i$ is the probability of $i$ offspring.

We will first show a way to represent a (finite) transmission tree as a sequence of integers representing the number of offspring of each node.  Additionally we show that the possible sequences coming from a tree can be characterized by a few specific properties.  Then the probability of such a sequence corresponds to the probability of the corresponding tree.

\begin{figure}
\begin{center}
\framebox{\scalebox{0.8}{\parbox{0.3\textwidth}{\begin{tikzpicture}
\node[BasicNode, minimum size = 4mm, fill = colorb!50] (node0) at (2.5,0) {$A$};
\node[BasicNode, minimum size = 4mm, fill = colorb!50] (node1) at (0.5,-0.8) {$B$};
\node[BasicNode, minimum size = 4mm, fill = colorb!50] (node2) at (2.5,-0.8) {$C$};
\node[BasicNode, minimum size = 4mm, fill = colorb!50] (node3) at (2.0,-1.6) {$D$};
\node[BasicNode, minimum size = 4mm, fill = colorb!50] (node4) at (1.75,-2.4000000000000004) {$E$};
\node[BasicNode, minimum size = 4mm, fill = colorb!50] (node5) at (2.25,-2.4000000000000004) {$F$};
\node[BasicNode, minimum size = 4mm] (node6) at (3.0,-1.6) {$G$};
\node[BasicNode, minimum size = 4mm] (node7) at (4.5,-0.8) {$H$};
\node[BasicNode, minimum size = 4mm] (node8) at (4.5,-1.6) {$I$};
\draw [-{Latex[length=2mm,width=2mm,angle'=30]}] (node0) -- (node1);
\draw [-{Latex[length=2mm,width=2mm,angle'=30]}] (node7) -- (node8);
\draw [-{Latex[length=2mm,width=2mm,angle'=30]}] (node2) -- (node6);
\draw [-{Latex[length=2mm,width=2mm,angle'=30]}] (node0) -- (node7);
\draw [-{Latex[length=2mm,width=2mm,angle'=30]}] (node2) -- (node3);
\draw [-{Latex[length=2mm,width=2mm,angle'=30]}] (node3) -- (node4);
\draw [-{Latex[length=2mm,width=2mm,angle'=30]}] (node0) -- (node2);
\draw [-{Latex[length=2mm,width=2mm,angle'=30]}] (node3) -- (node5);
\end{tikzpicture}\\[20pt]
$\mathcal{S} = (3, 0, 2, 2, 0, 0,\ldots)$}}}
\qquad 
\framebox{\scalebox{0.8}{\parbox{0.3\textwidth}{\begin{tikzpicture}
\node[BasicNode, minimum size = 4mm, fill = colorb!50] (node0) at (2.5,0) {$A$};
\node[BasicNode, minimum size = 4mm, fill = colorb!50] (node1) at (0.5,-0.8) {$B$};
\node[BasicNode, minimum size = 4mm, fill = colorb!50] (node2) at (2.5,-0.8) {$C$};
\node[BasicNode, minimum size = 4mm, fill = colorb!50] (node3) at (2.0,-1.6) {$D$};
\node[BasicNode, minimum size = 4mm, fill = colorb!50] (node4) at (1.75,-2.4000000000000004) {$E$};
\node[BasicNode, minimum size = 4mm, fill = colorb!50] (node5) at (2.25,-2.4000000000000004) {$F$};
\node[BasicNode, minimum size = 4mm, fill = colorb!50] (node6) at (3.0,-1.6) {$G$};
\node[BasicNode, minimum size = 4mm, fill = colorb!50] (node7) at (4.5,-0.8) {$H$};
\node[BasicNode, minimum size = 4mm, fill = colorb!50] (node8) at (4.5,-1.6) {$I$};
\draw [-{Latex[length=2mm,width=2mm,angle'=30]}] (node0) -- (node1);
\draw [-{Latex[length=2mm,width=2mm,angle'=30]}] (node7) -- (node8);
\draw [-{Latex[length=2mm,width=2mm,angle'=30]}] (node2) -- (node6);
\draw [-{Latex[length=2mm,width=2mm,angle'=30]}] (node0) -- (node7);
\draw [-{Latex[length=2mm,width=2mm,angle'=30]}] (node2) -- (node3);
\draw [-{Latex[length=2mm,width=2mm,angle'=30]}] (node3) -- (node4);
\draw [-{Latex[length=2mm,width=2mm,angle'=30]}] (node0) -- (node2);
\draw [-{Latex[length=2mm,width=2mm,angle'=30]}] (node3) -- (node5);
\end{tikzpicture}\\[20pt]
$\mathcal{S} = (3, 0, 2, 2, 0, 0, 0, 1, 0)$}}}
\end{center}
\caption{Demonstration of the steps mapping the tree $\Tree$ to the
  sequence $\Seq$.  The nodes are traced in a depth-first traversal
  and their number of offspring is recorded.  For the labeling given,
  a depth-first traversal traces the nodes in alphabetical order.  At
  an intermediate stage (left) the traversal has not finished the
  sequence.  The final sequence (right) is uniquely determined once
  the order of a node's offspring is (randomly) chosen.}
\label{fig:tree2seq}
\end{figure}
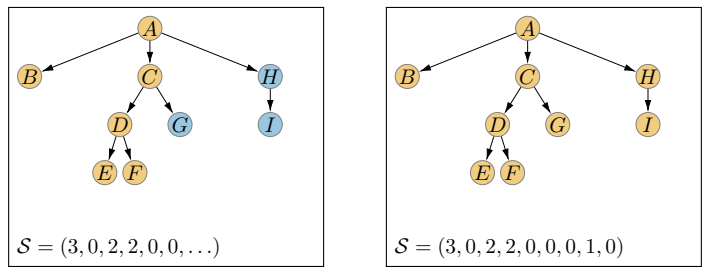

Given a finite transmission tree $\Tree$, we first order the offspring of any individual (randomly) from ``left'' to ``right''.  We then construct a sequence $\Seq$ by performing a depth-first traversal of the tree and recording the number of offspring as we visit the nodes of the tree, as shown in Fig.~\ref{fig:tree2seq}.  A sequence constructed in this way is called a \emph{\L{}ukasiewicz word}~\cite{stanleyEC2}.

It is straightforward to see that if we are given a \L{}ukasiewicz
word $\Seq_\Tree$, we can uniquely reconstruct the (ordered) tree $\Tree$
from which it came.  



\begin{figure}
\begin{center}
\input{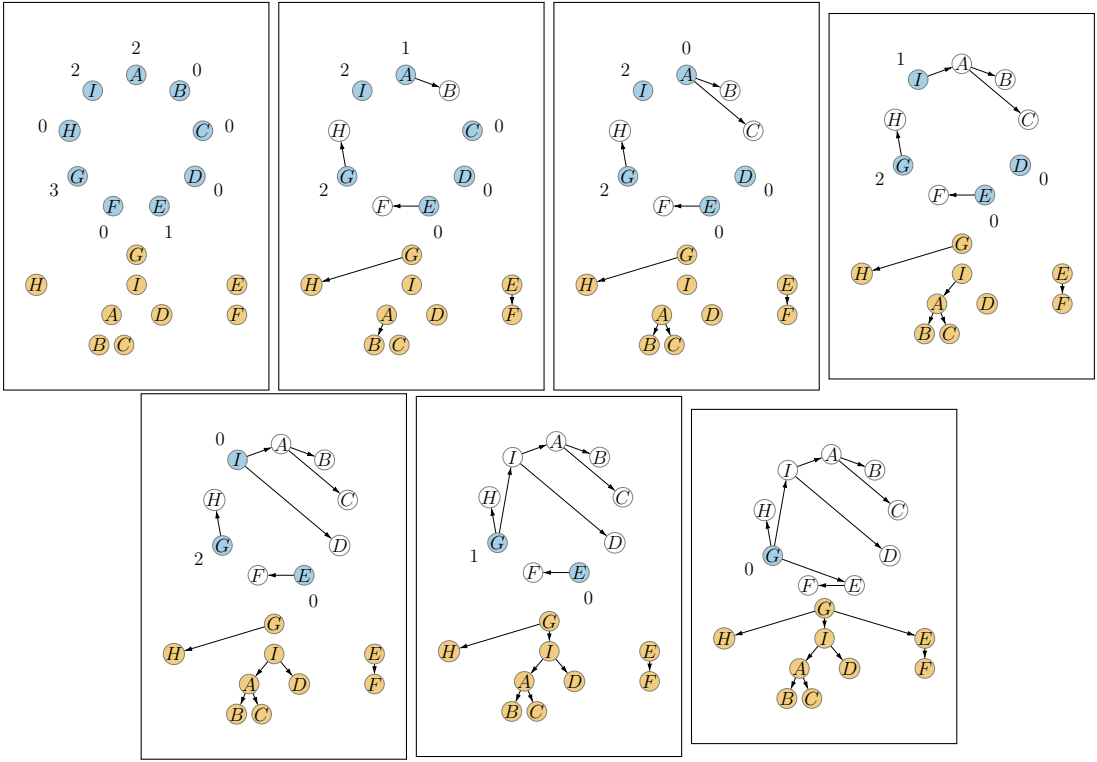}
\end{center}
\caption{The steps of the construction of a tree with  $\hat{\Seq} = (2, 0, 0, 0, 1, 0, 3, 0, 2)$ [note that this is a cyclic permutation of the previous $\Seq$].  Each frame shows next step in building a tree on a ring.  The resulting tree is not rooted at the top.  The names of the nodes in the tree are a cyclic permutation of the original.}
  \label{fig:cyclic}
\end{figure}

We now note that the probability of observing a given length-$j$ sequence $\hat{\Seq}$ by choosing $j$ numbers from the offspring distribution is simply $\pi_{\hat{\Seq}} = \prod_{s_i\in \hat{\Seq}} p_{s_i}$.  

Similarly, as infection spreads, each infected individual infects some
number $s_i$ with probability $p_{s_i}$.  If we calculate each $s_i$ in the order of a depth-first search, it is clear that the probability of observing a given
tree $\Tree$ with \L{}ukasiewicz word $\Seq_\Tree$ is exactly $\pi_{\Seq_\Tree}$.


Now we look for the probability that a random length-$j$ sequence $\hat{\Seq}$
created by choosing numbers from the offspring distribution is a \L{}ukasiewicz
word.\footnote{If the sequence is not a \L{}ukasiewicz word, then
  either it is the start of a sequence corresponding to a larger
  (possibly infinite tree), or some initial subsequence corresponds to
  a completed tree.}

To be a \L{}ukasiewicz word, $\hat{\Seq}$ must satisfy that $\sum_{s_i \in \hat{\Seq}} s_i = j-1$ because the
sum is the total number of transmissions occurring which is one less
than the total number of infections.  By repeated application
of Property~\ref{property:sum2product}, the probability a sequence of $j$ numbers chosen from the offspring distribution sums to $j-1$ is the coefficient of $y^{j-1}$ in $[\mu(y)]^j$.  So the probability that a random sequence $\hat{\Seq}$ satisfies this constraint is the coefficient of $y^{j-1}$ in $[\mu(y)]^j$.

Momentarily we will show that
given a length-$j$ sequence $\hat{\Seq}$ which sums to $j-1$, exactly one of its
$j$ cyclic permutations is a \L{}ukasiewicz word, but let us for now assume this result is true.

Consider the $j$ distinct sequences that are cyclic permutations of a sequence $\hat{\Seq}$ which sums to $j-1$.  Since each of these is a sequence of exactly the same values they have the same probability.  Our assumption that exactly one of them is a \L{}ukasiewicz word means that if $\hat{\Seq}$ satisfies the constraint that it sums to $j-1$ then with probability $1/j$ it is a \L{}ukasiewicz word.  So the probability that a random sequence is a \L{}ukasiewicz word would be $1/j$ times the probability it sums to $j-1$.  That is it would be $1/j$ times the coefficient of $y^{j-1}$ in $[\mu(y)]^j$.  This is the claim of Theorem~\ref{thm:power_magic}.

However, our earlier assumption must still be proven: if $\hat{\Seq}$ sums to $j-1$ then exactly one of its $j$ permutations is a \L{}ukasiewicz word.

Given a length $j$ sequence $\hat{\Seq}$ of non-negative integers that sum to $j-1$, we place $j$ nodes on a ring starting at the top and ordered clockwise, following the example in Figure~\ref{fig:cyclic}.  We label each $i$th node with $s_i$.  If a node $v$ is labeled with $0$ and the adjacent position in the counter-clockwise direction has node $u$ with a positive label, we place an edge from $u$ to $v$ (with $v$ to the right of any previous edge from $u$ to another node) and remove $v$.  We decrease $u$'s label by one.  Note that at a given step there may be multiple pairs eligible to have edges placed between them, in which case we do all of them.  If we did one at a time, the final outcome would be the same.

Each edge addid in this process reduces both the number of nodes and their sums by one, leaving all numbers as non-negative integers.  So
the sum remains one less than the remaining number of nodes.  This guarantees at
least one zero and at least one nonzero value until only one node
remains.  Thus we can always find an appropriate pair $u$ and $v$
until only a single node remains.  The process constructs a directed tree (there are $j$ nodes with $j-1$ edges and the fact that a node is removed from the algorithm once an edge is added pointing to it guarantees no cycles).  Fig.~\ref{fig:cyclic} demonstrates the steps.  

If the tree is rooted at the node that began at the top of the ring, then $\hat{\Seq}$ corresponds to a depth-first traversal of that tree.  It is a \L{}ukasiewicz word.  Each cyclic permutation of $\hat{\Seq}$ rotates the location of the root to be one of the $j$ nodes.  Only the case when the root is at the top will result in a \L{}ukasiewicz word.  Thus $\hat{\Seq}$ has exactly $j$ distinct cyclic permutations, and exactly one of them is \L{}ukasiewicz word.  This completes the final detail of the proof.

So we finally conclude that the probability of a tree of $j$ infected nodes is
equal to $1/j$ times the probability that $j$ randomly-chosen values
from the offspring distribution sum to $j-1$.  This is $1/j$ times the coefficient of $y^{j-1}$ in $[\mu(y)]^j$ as Theorem~\ref{thm:power_magic} claims.

\ifsolns
\else
\subsection{Theorem~\ref{thm:cts_power_magic}}
\fi
We can prove Theorem~\ref{thm:cts_power_magic} as a special case of
Theorem~\ref{thm:power_magic} by calculating the offspring
distribution (Exercise~\ref{exercise:cts_power_magic}).  However, a more illuminating proof is by noting that
if we treat a transmission event as a node disappearing and being replaced by two infected nodes and a
recovery event as a node disappearing with no offspring, then we have a tree
where each node has $2$ or $0$ offspring.  The total number of actual
individuals infected in the outbreak is equal to the number of nodes with $0$ offspring in the tree.

Following the arguments above, we are looking for sequences of length
$2j-1$ in which $2$ appears $j-1$ times and $0$ appears $j$
times. There are $\binom{2j-1}{j-1}$ such sequences.  The probability
of each is $\beta^{j-1}\gamma^j/(\beta+\gamma)^{2j-1}$ and a fraction
$1/(2j-1)$ of these correspond to trees.  Thus, the probability a
length $2j-1$ sequence is a \L{}ukasiewicz word is
\[
\frac{1}{2j-1} \frac{\beta^{j-1}\gamma^j}{(\beta+\gamma)^{2j-1}}
\binom{2j-1}{j-1} = \frac{1}{j}
\frac{\beta^{j-1}\gamma^j}{(\beta+\gamma)^{2j-1}} \binom{2j-2}{j-1}
\]
Using the same approach as before, we conclude that this is the probability of exactly
$j$ infections.

\begin{exercise}
\label{exercise:cts_power_magic}
If we do not think of an infected individual as disappearing and being
replaced by two infected individuals when a transmission happens, but
rather, we count up all of the transmissions the individual causes, we
get a geometric distribution with $q = \beta/(\beta+\gamma)$.  The
details are in Exercise~\ref{exercise:continuous_vs_discrete}.  Use
this along with Theorem~\ref{thm:power_magic} and
Table~\ref{table:final_probs} (which was derived in
exercise~\ref{exercise:final_probs}) to give a different proof of Theorem~\ref{thm:cts_power_magic}.
\end{exercise}
\begin{solution}
The offspring distribution is geometric with $q =
\beta/(\beta+\gamma)$.

The result in Table~\ref{table:final_probs}
predicts $j$ infections with probability
\[
\frac{1}{j} \binom{2j-2}{j-1}
\left(\frac{\gamma}{\beta+\gamma}\right)^j
\left(\frac{\beta}{\beta+\gamma}\right)^{j-1} = \frac{1}{j}
\frac{\gamma^j\beta^{j-1}}{(\beta+\gamma)^{2j-1}} \binom{2j-2}{j-1}
\]
\end{solution}
\ifsolns
\else
\section{Software}
\fi
\label{app:software}
We have produced a python package, \texttt{Invasion\_PGF} which can be used to solve the equations of Section~\ref{sec:discrete} or Section~\ref{sec:cts} once the PGF of the offspring distribution or $\beta$ and $\gamma$ are determined.  Because the numerical method involves solving differential equations in the complex plane, it requires an integration routine that can handle complex values.  For this we use \texttt{odeintw}~\cite{odeintw}.

Table~\ref{tab:software} briefly summarizes the commands available in \texttt{Invasion\_PGF}.

\begin{table}
\begin{center}
\begin{tabular}{|Sc|Sc|}
\hline
\textbf{Command} & \textbf{Output}\\
\hline\hline
\pbox{0.38\textwidth}{\texttt{R0($\mu$)}} & Approximation of $\Ro$.\\
\hline
\pbox{0.38\textwidth}{\texttt{extinction\_prob($\mu$, gen)}}
& \parbox{0.57\textwidth}{Probability $\alpha_{\texttt{gen}}$ of extinction by generation \texttt{gen} given offspring PGF $\mu$.}\\
\hline
\pbox{0.38\textwidth}{\texttt{cts\_time\_extinction\_prob($\beta$, $\gamma$, $T$)}}
                 & \parbox{0.57\textwidth}{Probability $\alpha(T)$ of
                   extinction by time $T$ given transmission and
                   recovery rates $\beta$ and $\gamma$.}\\
\hline
\pbox{0.38\textwidth}{\texttt{active\_infections($\mu$, gen, $M$)}}
                 & \parbox{0.57\textwidth}{Array containing
                   probabilities $\phi_0, \ldots, \phi_j, \ldots,
                   \phi_{M-1}$ of having $j$ active infections in generation \texttt{gen} given offspring PGF $\mu$.}\\
\hline
\pbox{0.38\textwidth}{\texttt{cts\_time\_active\_infections($\beta$, $\gamma$, $T$)}}
& \parbox{0.57\textwidth}{Array containing
                   probabilities $\phi_0, \ldots, \phi_j, \ldots,
                   \phi_{M-1}$ of having $j$ active infections at time $T$ given transmission and recovery rates $\beta$ and $\gamma$.}\\
\hline
\pbox{0.38\textwidth}{\texttt{completed\_infections($\mu$, gen, $M$)}}
                 & \parbox{0.57\textwidth}{Array containing
                   probabilities $\omega_0, \ldots, \omega_j, \ldots,
                   \omega_{M-1}$ of having $j$ completed infections in
                   generation \texttt{gen} given offspring PGF $\mu$.}\\
\hline
\pbox{0.38\textwidth}{\texttt{cts\_time\_completed\_infections
  ($\beta$, $\gamma$, $T$)}}
& \parbox{0.57\textwidth}{Array containing probabilities $\omega_0, \ldots, \omega_j, \ldots, \omega_{M-1}$ of having $j$ completed infections at time $T$ given transmission and recovery rates $\beta$ and $\gamma$.}\\
\hline
\pbox{0.38\textwidth}{\texttt{active\_and\_completed($\mu$, gen, $M_1$, $M_2$)}}
& \parbox{0.57\textwidth}{$M_1\times M_2$ array containing probabilities $\pi_{i,r}$ of $i$ active infections and $r$ completed infections in generation \texttt{gen} given offspring PGF $\mu$.}\\
\hline
\pbox{0.38\textwidth}{\texttt{cts\_time\_active\_and\_completed
  ($\beta$, $\gamma$, $T$)}}
 & \parbox{0.57\textwidth}{$M_1\times M_2$ array containing probabilities $\pi_{i,r}$ of $i$ active infections and $r$ completed infections at time $T$ given transmission and recovery rates $\beta$ and $\gamma$.}\\
\hline
\pbox{0.38\textwidth}{\texttt{final\_sizes($\mu$, $M$)}}
 & \parbox{0.57\textwidth}{Array
                                    containing probabilities
                                    $\omega_0, \ldots, \omega_j,
                                    \ldots, \omega_{M-1}$ of having
                                    $j$ total infections in an
                                    outbreak given offspring PGF $\mu$.}\\
\hline
\pbox{0.38\textwidth}{\texttt{cts\_time\_final\_sizes($\beta$,
  $\gamma$, $T$)}}
 & \parbox{0.57\textwidth}{Array containing probabilities $\omega_0, \ldots, \omega_j, \ldots, \omega_{M-1}$ of having $j$ total infections in an outbreak given transmission and
                   recovery rates $\beta$ and $\gamma$.}\\
\hline
\end{tabular}
\end{center}
\caption{Commands of \texttt{Invasion\_PGF}.  Many of these have an
  optional boolean argument \texttt{intermediate\_values} which, if
  \texttt{True}, will result in returning values from generation $0$ to
  generation \texttt{gen} in the discrete-time case or at some
  intermediate times in the continuous-time case.  For the discrete-time results, the input $\mu$ is the
  offspring distribution PGF.  For the continuous-time version, $\beta$
  and $\gamma$ are the transmission and recovery rates respectively.}
\label{tab:software}
\end{table}


%
%
%
%

\begin{pyconcode}
\end{pyconcode}

We now demonstrate a sample session with these commands.
\begin{pyconsole}
\end{pyconsole}

\ifsolns
\else
\section*{Acknowledgments}
\fi
This work was funded by Global Good.

I thank Linda Allen for useful discussion about the Kolmogorov
equations.  Hao Hu played an important role in inspiring this work and
testing the methods.  Hil Lyons and Monique Ambrose provided valuable feedback on the
discussion of inference.  Amelia Bertozzi-Villa and Monique Ambrose read over drafts and recommended a number of changes that have significantly improved the presentation.

The python code and output in Appendix~\ref{app:software} was
incorporated using Pythontex~\cite{poore2015pythontex}.  I relied
heavily on \url{https://tex.stackexchange.com/a/355343/70067} by ``touhami'' in setting up the solutions to the exercises.

\ifsolns
\else
\bibliographystyle{plain}
\bibliography{abbreviations,pgf}
\fi
\end{document}